\documentclass[pdftex,12pt]{article}
\usepackage{amsmath,amsfonts,amsbsy,latexsym,amssymb,amscd,amstext}
\usepackage{pst-node}
\usepackage{dsfont}
\usepackage{cool}
\usepackage{overpic,youngtab}
\pdfoutput=1
\usepackage{rotating}
\usepackage{amsmath}
\usepackage{amssymb}
\usepackage{amsthm}

\usepackage{tikz}
\usepackage{slashed}
\def\bpm{\begin{pmatrix}}
	\def\epm{\end{pmatrix}}
\def\be{\begin{equation}}
\def\cO{\cal O}
\def\ee{\end{equation}}
\def\bea{\begin{eqnarray}}
\def\bA{\overline{A}}
\def\ap{\a^\prime}
\def\eea{\end{eqnarray}}
\def\pd{\partial}
\def\a{\alpha}

\def\tp{t^\prime}

\def\b{\beta}

\def\g{\gamma}
\def\d{\delta}
\def\m{\mu}

\def\n{\nu}
\def\t{\tau}

\def\l{\lambda}

\def\r{\rho}

\def\bR{\bar{R}}

\def\cD{\cal D}

\def\bg{\bar{g}}

\def\xp{x^\prime}
\def\bD{\bar{D}}
\def\bG{\bar{G}}
\def\bp{\bar{\phi}}
\def\s{\sigma}
\def\e{\epsilon}

\def\cD{\cal D}

\def\bma{\begin{pmatrix}}
	\def\ema{\end{pmatrix}}

\def\bp{\bar{\phi}}

\def\bg{\bar{g}}
\def\bi{\begin{itemize}}
	\def\ei{\end{itemize}}
\def\bn{\bar{\nabla}}

\def\bp{\bar{\phi}}
\def\tr{{\rm tr\,}}

\def\bph{\bar{\phi}}

\begin{document}

		\vspace*{-1cm}
		\phantom{hep-ph/***} 
		{\flushleft
			{{FTUAM-21-xx}}
			\hfill{{ IFT-UAM/CSIC-22-30}}}
		\vskip 1.5cm
		\begin{center}
		{\LARGE\bfseries   Covariant techniques in Quantum Field Theory.}\\[3mm]
			\vskip .3cm
		
		\end{center}

		\vskip 0.5  cm
		\begin{center}
			{\large Enrique \'Alvarez and  Jes\'us Anero}
			\\
			\vskip .7cm
			{
				Departamento de F\'isica Te\'orica and Instituto de F\'{\i}sica Te\'orica, 
				IFT-UAM/CSIC,\\
				Universidad Aut\'onoma de Madrid, Cantoblanco, 28049, Madrid, Spain\\
				\vskip .1cm

				\vskip .5cm
				
				\begin{minipage}[l]{.9\textwidth}
					\begin{center} 
							\textit{E-mail:} 
						\tt{enrique.alvarez@uam.es},
						\tt{jesusanero@gmail.com}
					\end{center}
				\end{minipage}
			}
		\end{center}
	\thispagestyle{empty}
	
\begin{abstract}
	\noindent
In these lectures some techniques useful to perform quantum field theory computations in a covariant manner are reviewed. In particular the background field gauge, the zeta function regularization and the heat kernel approach are highlighted. Some detailed calculations of the Schwinger-de Witt coefficients of the small proper time expansion of the heat kernel are also repeated in detail. This work reports lectures given by Enrique \'Alvarez at the IFT-UAM-CSIC in Madrid.
\end{abstract}

\newpage
\tableofcontents
	\thispagestyle{empty}
\flushbottom

\newpage

%%%%%%%%%%%%%%%%%%%%%%%%%%%%%%%%%%%%%%%%%%%%%%%%%%%%%%%%%%%%%%%%%%%%%%%%%%%%%%%%%%%%%%%%%%%%%%%%%%%%%%%%%%%%%%%%%%%%%%%%%%%%%%%
 \section{Introduction. }
 %%%%%%%%%%%%%%%%%%%%%%%%%%%%%%%%%%%%%%%%%%%%%%%%%%%%%%%%%%%%%%%%%%%%%%%%%%%%%%%%%%%%%%%%%%%%%%%%%%%%%%%%%%%%%%%%%%%%%%%%%%%%%
We are living exciting times in theoretical physics. Polchinski's famous question "What is string theory?"  \cite{Polchinski} can perhaps be turned inside out to ask the  question "What is quantum field theory?" (QFT). As happens often in physics, what was believed to be understood is not really so once we are able to ask deeper questions. In recent years, duality relationships between different (supersymmetric) gauge theories, and the AdS/CFT duality between some conformal quantum field theories (CFT) and some  specific  string theories suggest  to raise the question of whether string theory is not really a part of quantum field theory, in some unknown sense.
\par
It is still true that the main problem that prevents further progress is the inability to perform nontrivial computations in the strong coupling regime of QFT. Perturbative approaches are not enough. This is true even in the theory of strong interactions, QCD, where there is a plethora of experimental data in the strong coupling regime. The only theoretical grasp of this is through effective field theories, and of course, through the lattice approach.
\par
 What we are able to do beyond perturbation theory is only a partial subset of all interesting computations, and this only in  theories with certain amount of supersymmetry. It could be that all those marvelous conjectures and/or facts mentioned above are just an artifact of supersymmetry, and disappear as soon as supersymmetry is broken and maybe they would disappear  in an uncontrollable way. 
 \par
 On the other hand, it is  known that there are some quantum field theories that do not admit a Lagrangian formulation, like little string theories \cite{Losev}, which do not even have any local operators.  In QFT in non-commutative spaces \cite{Matusis} it is not possible to disentangle the infrared (IR) and the ultraviolet (UV) sectors of the theory; those appear linked in an intricate way. In the same vein, there are also QFT \cite{Seiberg} that cannot be latticized. This means that any framework that makes use of a Lagrangian, like the path integral, or the action principle, is bound not to be applicable to all QFTs. Similar observations can obviously be applied to the lattice approach. Any general treatment should be able to encompass all those non-Lagrangian theories as well. This is not possible at the moment, and we limit ourselves in these notes to the Lagrangian case, but we wanted to call attention to the fact that we are not able to treat all known QFTs in an unified way.
\par
In these notes we shall dwell on the most conservative approach about quantum effects in gravity, namely assuming that the geometric variables (metric tensor and/or the connection one-form) are the natural variables to be quantized.
In spite of the fact that  space-time  is naturally endowed in General Relativity (GR) with a semi-Riemannian metric tensor (that is, one with Lorentzian signature), we shall perform our calculations in a Riemannian space. Even with this simplification, the Einstein-Hilbert Lagrangian fails to be positive definite, which implies that most of the usual arguments used in QFT to argue for an expansion around saddle points of the action are actually not compelling.
\par
Furthermore, given a Riemannian space, there is no an unique way to define a semi-Riemannian one that is in some sense its {\em analytic continuation}. (The flat space trick of defining a complex time coordinate does not work in general). Sometimes there is a complex manifold of which both the Riemannian and the semi-Riemannian manifolds are real sections \cite{Woodhouse}, but this is not true in general.
\par
It has been suggested by Hawking to make the analytic continuation at the end of the whole computation, namely on the transition amplitudes themselves, but we do not know how to do this in detail. For example, what is the semi-Riemannian meaning of the transition amplitude from a certain Riemannian  space $\Sigma$ to another one, $\Sigma^\prime$?  We insist that given a Riemannian manifold is not defined (much less uniquely) a semi-Riemannian one which is in some sense  its analytic continuation. This fact sheds some doubts on the whole {\em Euclidean quantum gravity program}. On the other hand working directly with Lorentzian signature is much more difficult and not many general results have been obtained.
\par
There is a {\em vanilla problem}, which is the study of quantum field theories defined in an external classical background gravitational field. Some beautiful results have been obtained here, and this is probably the only instance in which people have been able to tackle the problems inherent to the Lorentzian signature. Hawking radiation and the Unruh effect are probably the milestones here. Once nontrivial time dependence is introduced, lots of ambiguities appear, like in the (Poincar\'e patch of the) de Sitter background \cite{WittenS}. In spite of the interest of this topic, it is not clear \cite{HarlowJ} in what sense this is a consistent approximation to a full quantum gravity treatment.
\par
What we do want to present in some detail in this course is the simplest instance of all the above, namely a set of beautiful techniques useful in order to do QFT computations (in particular divergent pieces) in Riemannian spaces in a covariant way. The purpose is pedagogical; this is why most  calculations are done in gory detail. A general very good elementary reference is Mukhanov and Winitzki's book \cite{Mukhanov}. Full of gems, but not an easy read is the two-volume DeWitt's book \cite{DeWitt}. Somewhat more mathematical books  are \cite{Gilkey}\cite{Blumenhagen}\cite{Fursaev}\cite{Avramidi}\cite{Kirsten}\cite{Kiefer}\cite{Elizalde}\cite{Mottola}. Quite good reviews are also available \cite{Alvarez-Gaume}\cite{Barvinsky}\cite{Gracia-Bondia}\cite{Vassilevich}\cite{Vilkovisky}. 
\newpage
%%%%%%%%%%%%%%%%%%%%%%%%%%%%%%%%%%%%%%%%%%%%%%%%%%%%%%%%%%%%%%%%%%%%%%%%%%%%%%%%%%%%%%%%%%%%%%%%%%%%%%%%%%%%%%%%%%%%%%%%%%%%%
\section{Gravitation and quantum field theory: The Big Picture}
%%%%%%%%%%%%%%%%%%%%%%%%%%%%%%%%%%%%%%%%%%%%%%%%%%%%%%%%%%%%%%%%%%%%%%%%%%%%%%%%%%%%%%%%%%%%%%%%%%%%%%%%%%%%%%%%%%%%%%%%%%%%
There are many obvious issues when considering quantum gravity, by which we mean some unknown quantum theory that in the classical limit reduces to GR. For example, one of the basis of quantum field theory \cite{Symanzik} is {\em microcausality}, the statement that field variables defined at points spacelike separated should commute. Also the canonical commutators are defined at {\em equal time}. It is plain that these concepts make sense in a fixed gravitational background at best; and even then with caveats when horizons are present.
\par
In a similar vein, any attempt to write a Schr\"odinger equation for the gravitational field must face the fact that there is no natural notion of time in GR, even classically. The Wheeler-DeWitt equation is obtained by  interpreting the Hamiltonian constraint as an operator equation by substituting the canonical momenta by functional derivatives. It is similar to the Schr\"odinger equation, except precisely for the absence of time. It has been repeteadly conjectured (\cite{Alvarez} for a review) that such a time can appear when  a WKB type of semiclassical approximation is performed on the Wheeler-DeWitt equation, but this has not been properly substantiated.
\par
Some people try to apply the canonical approach to a clever set of variables introduced by Ashtekar \cite{Rovelli}. Those variables are related to the spacetime metric in a complicated way. It is unclear how this approach is related to the classical regime at all. This whole approach is dubbed {\em loop quantum gravity}, because a loop representation is useful to understand some aspects of the corresponding Hilbert space.
\par
We think it is fair to say that the results obtained so far from the canonical approach are quite modest.
\par
A natural first guess would be to use the functional integral to define transition amplitudes from one three-dimensional metric on a given three-dimensional manifold $\Sigma_i$, say $h_i$ to another three dimensional surface $\Sigma_f$ with its corresponding metric $h_f$, something like
\be
K\left[h_i,h_f\right]\equiv\int {\cal D}g e^{i S_{EH}\left(g\right)}
\ee
where the integration is performed over all metrics defined on a four-dimensional  domain $D$ such that
\be
\pd D=\Sigma_f-\Sigma_i
\ee
\par

Nevertheless, before we are even able to understand this conjecture, we have to face several ugly facts. First of all, the gravitational action is not positive definite, even with the Euclidean signature. As we have already suggested before, the loop expansion is not then justified by any sort of saddle point expansion. 
Even worse, the set of four dimensional manifolds is a complicated one. It has been shown by Markov \cite{Markov} that the problem of classifying four-dimensional geometries is an undecidable one. Given any two four-dimensional manifolds, there is no set of topological invariants that can discriminate when those two manifolds are diffeomorphic. The problem lies mainly with the first fundamental group, $\pi_1(M)$. It does not seem the case that there exists any justification for restricting the functional integral to any subset of manifolds (i.e. simply connected).
\par
Were the spacetime geometry to fluctuate we would have to build a new all our ideas about QFT, which we understand when defined in flat space only, and even there as mentioned above, we miss non-perturbative effects known to be important.
\par
Another issue is the following. Assuming that the symmetry group of the quantum theory is still diffeomorphism (Diff) invariance, what are the observables? For a {\em fixed} manifold, integrals of n-forms are Diff invariant objects, but there are not many of those.
\par
The preceding difficulties did not deter physicist to think on quantum aspects of gravitation. Besides many long and inconclusive discussions of the basic foundational points, to be mentioned later, such as what are the observables of the theory, the main breakthrough was made by 't Hooft and Veltman employing techniques invented by DeWitt and Feynman. What is computed are the quantum fluctuations around an {\em arbitrary} background, $\bg_{\a\b}(x)$, which can be any solution of Einstein's equations of motion (EM). General Relativity is considered in this treatment as an ordinary gauge theory, forgetting about all questions of principle. Actually the calculation is usually done with Euclidean signature, making an appropiate analytical continuation at the end of the procedure. Particularly easy is the computation of the divergences of the effective action, which must be eliminated in the renormalized theory. In this computation beautiful mathematical techniques can be employed. The propagator is assumed to be however well-defined for a generic background metric, which is a delicate assumption in the presence of horizons and/or singularities.

It is doubtful whether we can assert any  proposition about quantum \footnote{
In order to understand the sequel, the reader is assumed a working knowledge of quantum field theory at a graduate level, up to and including, Feynman's path integral.
}
gravity with some confidence. 
\par
The tree level estimate for the cross section for  production of gravitons in particle-antiparticle annihilation is of the order of the inverse of the mass scale associated to this problem which just by sheer dimensional analysis is
Planck's mass, which is given in terms of Newton's constant, $G$ by
\be
m_p\equiv\sqrt{\hbar c\over 8\pi G}\sim 10^{19}GeV
\ee 
 if we remember that $1\,GeV(=10^3\, MeV)$ is the rough scale of hadronic physics (the mass and inverse Comptom wavelength of a proton, for example), this means that quantum gravity effects will only be apparent when we are able to explore concentrated energy  $10^{19}$ times bigger (or a scale distance correspondingly smaller; these two statements are supposed to be equivalent owing to Heisenberg's principle). To set the scale, the Large Hadron Collider works roughly at the $TeV(=10^3\,GeV)$ scale, so there is a long way to go before reaching expected quantum gravity effects in accelerators.
\par
In terms of the cross section, this yields up to numerical factors of order unity
\be
\s\sim l_p^2\sim 10^{-66}~cm^2\sim 10^{-40}~ fm^2
\ee
This is more or less 40 orders of magnitude smaller than typical nuclear reactions.
\par
There are however some interesting experimental facts such as the ones studied in \cite{Colella}. Free fall of neutrons has been reported there.
Also there interference effects due to the Earth's  classical gravitational field on a neutron's wave function are analyzed. The experimental apparatus is a neutron interferometer. The phase shift between the two different paths is given by
\be
\Delta \phi\sim {2\pi m_p^2 l g \Delta h \over h^2}
\ee 
where $l$ is the common horizontal span of the paths and  $\Delta h$ is the difference in height. There are some more contributions in the actual experiment and the precision is not too big. Nevertheless the effect seems clear. It is not clear however what is its meaning with respect to the relationship between gravitation and quantum mechanics. More recently \cite{Nesvizhevsky} experimental evidence for gravitational quantum bound states of neutrons has been claimed.
\par
If we want to get direct experimental information  of quantum gravitational effects, we have to turn our attention towards Cosmology, or perhaps look for some clever precision experiment in the laboratory. 
Lacking any experimental clue, the only thing we can do is to think and try to look for logical (in)consistencies.  
\par
Many people believe that those should stem from Einstein's equations themselves. There is evidence that the second member (the energy momentum tensor) is quantized, and as such, is subject to quantum interference effects. It seems logical to assume that the same thing should happen with the first member of Einstein's equations; that is, the geometry.
\par 
It has been repeatedly argued by many particle physicists that the practical utility of the answer to this question will not presumably be great. How would we know for sure beforehand?. There has always been a recurrent dream, exposed vehemently by Salam \cite{Salam} that the inclusion of the gravitational interaction would cure many of the diseases and divergences of quantum field theory, through the inclusion in the propagator of terms of the type
\be
e^{-{1\over m_p r}}
\ee
So that for example, the sum of tree graphs that leads to the Schwarzschild solution as worked out by Duff \cite{Duff}
\be
{1\over r}+ {2M\over m_p^2 r^2}+{4 M^2\over m_p^4 r^3}+\ldots={1\over r\left(1-{2M\over m_P^2 r}\right)}
\ee
would get modified to
\be
{1\over r}e^{-{1\over m_p r}}+ {2M\over m_p^2 r^2}e^{-{ 2 \over m_p r}}+\ldots\sim {1\over r}e^{-{1\over m_p r}}\frac{1}{\Big[ {1 	-{2M\over m_p^2r}\,e^{-{1\over m_p r}}}\Big]}
\ee
shifting the location of the horizon and eliminating the singularity at $r=0$. Nobody has been able to substantiate this dream so far.
\par
On the other hand, it has been speculated that quantum gravitational effects can tame the infinities that appear in QFT yielding a finite theory eventually. Some arguments in favor of this (first proposed by  Arnowitt, Deser and Misner \cite{ADM}, the inventors of ADM) are as follows.
The self-energy of a body of radius $\e$ and mass $m$ and charge $e$ which in newtonian theory reads
\be
m_\e=m+{e^2\over 8\pi \e}-{G m^2\over 2\e}
\ee
It diverges in the pointlike limit $\e\rightarrow 0$. The only modification borne out by General Relativity 
was shown by ADM to be the replacement in the second member of $m_0$ by $m_\e$ (because all energy gravitates)
\be
m_\e=m+{e^2\over 8\pi \e}-{G m_\e^2\over 2\e}
\ee
solving the quadratic equation yields
\be
m_\e={\e\over G}\left[-1\pm \sqrt{1+{2 G\over \e}~\left(m+{e^2\over 8\pi\e}\right)}\right]
\ee
which has a finite limit when $\e\rightarrow 0$ namely
\be
m_0=\sqrt{e^2\over 4\pi G}
\ee

\par
Ay ant rate quantum gravity is nevertheless a topic which has fascinated whole generations of physicists, just because it is a natural boundary of two of the most successful physical theories mankind has discovered.
There seems to be a strong tension between the beautiful, geometrical world of General Relativity and the no less marvelous, less geometrical, somewhat mysterious, but very well tested experimentally, world of Quantum Mechanics.

As with all matters of principle  we can hope to better understand both quantum mechanics and gravitation if we are able to clarify the issue.
\par
The most conservative approach is of course to start from what is already known with great precision about the standard model of elementary particles associated to the names of Glashow, Weinberg and Salam. This can be called the {\em bottom-up approach} to the problem.
In this way of thinking Wilson taught us that there is a working low energy effective theory, and some quantum effects in gravity can be reliably computed for energies much smaller than Planck's mass.
\par
 There are two caveats to this. First of all, we do not understand why the observed cosmological constant is so small: the natural value from the low energy effective Lagrangian point of view ought to be much bigger.  The second point is that one has to rethink again the lore of effective theories in the presence of horizons. 
We shall comment on both issues in due time.
\par
There is not a universal consensus even on the most promising avenues of research from the opposite {\em top-down} viewpoint. Many people think that strings are the best
buy (we sort of agree with this); but it is true that after more than two decades of intense effort nothing substantial has come out of them on the particular problem of our interest here. Others \cite{Rovelli} try to quantize directly the Einstein-Hilbert Lagrangian, something that is at variance with our experience in effective field theories. But it is also true that as we have already remarked, the smallish value of the observed cosmological constant also cries out of the standard effective theories lore.
\par
It is generally accepted that General Relativity, a generally covariant theory, is akin to a gauge theory, in the sense that the diffeomorphism group of the apace-time manifold, $Diff(M)$ plays a role similar to the compact gauge group in the standard model of particle physics. There are some differences though. To begin with, the group, $ Diff(M)$ is too large; is not even a Lie group. Besides, its detailed structure depends on the manifold, which is a dynamical object not given a priori. Other distinguished subgroups (such as the volume-preserving diffeomorphisms, like in unimodular gravity, already discussed in an appendix of Pauli's  wonderful book on Relativity \cite{Pauli}) are perhaps also arguable for. Those leave invariant a given measure, such as the Lebesgue measure.
\par
It also seems clear that when there is a boundary of space-time, then the gauge group is restricted to the subgroup consisting on those diffeomorphisms that act trivially on the boundary. The subgroup that act not-trivially is related to the set of conserved charges. In the asymptotically flat case this is precisely the Poincar\'e group, $SO(1,3)$ that gives rise to the ADM mass. 
\par
In the asymptotically anti-de Sitter case, this is related to the conformal group $SO(4,2)$. 
\par
It is nevertheless not clear what is the physical meaning of keeping constant the boundary of spacetime (or keeping constant some set of boundary conditions) in a functional integral of some sort.
\par
A related issue which we have already mentioned is that
 it is very difficult to define what could be {\em observables} in a diffeomorphism invariant theory, other than global ones defined as integrals of scalar composite operators $O(\phi_a(x))$ (where $\phi_a, a=1\ldots N$ parametrizes all physical fields) with the peudo-Riemannian measure
\be
{\cal O}\equiv \int \sqrt{|g|}d^4 x O(\phi_a(x))
\ee
 Some people claim that there are no local observables whatsoever, but only {\em pseudolocal } ones; the fact is that we do not know.
 Again, the exception to this stems from keeping the boundary conditions fixed; in this case it is possible to define an $S$-matrix in the asymptotically flat case, and a conformal quantum field theory in the asymptotically anti-de Sitter case. 
 Unfortunatelly, the most interesting case from the cosmological point of view, which is when the space-time is asymptotically de Sitter is not well understood.
\par
It has been already mentioned  that the equivalence problem in four-dimensional geometries is undecidable \cite{Markov}. In three dimensions  Thurston's geometrization conjecture has recently been put on a firmer basis by Hamilton and Perelman, but it is still  not clear whether it can be somehow implemented in a functional integral without some drastic restrictions. Those caveats should be kept in mind when reading the sequel. 
\par
A radically different viewpoint has recently been advocated by Gerardus 't Hooft by insisting in causality to be well-defined, so that the conformal class of the space-time metric should be determined by the physics, but not necessarily the precise point in a given conformal orbit. If we write the spacetime metric in terms of a unimodular metric and a conformal factor
\be
g_{\m\n}=\omega^2(x)\hat{g}_{\m\n}
\ee
with
\be
det\,\hat{g}_{\m\n}=1
\ee
then the unimodular metric is in some sense intrisic and determines causality, whereas the conformal factor depends on the observer in a way dictated by {\em black hole complementarity}.
\par
Finally, there is always the (in a sense, opposite) possibility that space-time (and thus diffeomorphism invariance) is not a fundamental physical entity in such a way that the appropiate variables for studying short distances are non geometrical. Some recent references are \cite{Verlinde}. Something like that could happen in string theory, but our understanding of it is still in its infancy.
\par

\newpage
%%%%%%%%%%%%%%%%%%%%%%%%%%%%%%%%%%%%%%%%%%%%%%%%%%%%%%%%%%%%%%%%%%%%%%%%%%%%%%%%%%%%%%%%%%%%%%%%%%
\section{Schwinger's action principle and\\Peierls'brackets.}
%%%%%%%%%%%%%%%%%%%%%%%%%%%%%%%%%%%%%%%%%%%%%%%%%%%%%%%%%%%%%%%%%%%%%%%%%%%%%%%%%%%%%%%%%%%%%%%%%%
 {\em Schwinger's action principle} \cite{Schwinger}\cite{Symanzik}\cite{Toms} expresses the variation of a transition amplitude $\langle A| B\rangle$ between two states $\left|B\right\rangle$ and $\left|A\right\rangle$ in terms of the expectation value of the variation of the action, $\d S$. Symbollically
\be
\d \langle A(t_1)| B(t_2)\rangle= i \langle A(t_1) | \d S_{12} | B(t_2)\rangle
\ee
This principle  is the starting point for all dynamical laws in QFT for Schwinger's school. 
\par
Let us elaborate. In Heisenberg's representation
\be
\dot{\hat{q}}=-i \left[\hat{q},\hat{H}\right]
\ee
and in particular,
\be
\dot{\hat{H}}=0
\ee

This means that the time dependence of the operators is given by
\be
\hat{q}(t)=e^{i \hat{H}(t-t_0)}\hat{q}(t_0) e^{-i \hat{H}(t-t_0)}
\ee
let us denote by 
\be
|q\, t\rangle
\ee
 a state defined by measurements or preparations at time $t$ (i.e., eigenstates of $q(t)$). Also
\be
| q \,t^\prime\rangle
\ee
means that we have to replace $t$ by $t^\prime$; but otherwise keep the same preparation as before.
 We have 
 \bea
 &&\hat{q}(t)|q \,t\rangle=q|q \,t\rangle\nonumber\\
 &&\hat{q}(t^\prime)| q \,t^\prime\rangle=q| q\, t^\prime\rangle=e^{iH(t^\prime-t)}\,\hat{q}(t)\,e^{-i H(t^\prime-t)}|q\, t^\prime\rangle
 \eea
it follows that
\be
| q \,t^\prime\rangle=e^{i H(t^\prime-t)}|q\,t\rangle
\ee
so that
\be
\langle q\,t|q\, t^\prime\rangle=\langle q,t|e^{i H(t^\prime-t)}|q\,t\rangle
\ee
assuming that the Hamiltonian $\hat{H}$ depends on some external parameter, say $\l$,
\be
\d_\l \langle q\,t|q\, t^\prime\rangle=i\left\langle q,t\left|\int_{t^\prime}^t d\t\d_\l H(\t)\,e^{i H(t^\prime-t)}\right|q\,t\right\rangle=i\left\langle q\,t\left|\int_{t^\prime}^t d\t \d_\l H(\t)\right| q\, t^\prime\right\rangle
\ee
Schwinger's principle can easily be generalized \cite{Symanzik} to the case when an operator insertion is included at some intermediate time $t_2\leq t\leq t_1$
\bea
&\d_\l\left\langle q\,t_1\left|{\cO} (q(t),\l) \right|q\,t_2\right\rangle=\left\langle q\,t_1\left|\d_\l {\cO}(q(t),\l) \right|q\,t_2\right\rangle\    +\nonumber\\
&+\left.i\left\langle q\,t_1\left|\int_t^{t_1} \d_\l L(q(\t),\l) d\t\right|_{q(t)\,\text{fixed}}\,{\cO}(q(t),\l) \right|q\,t_2\right\rangle+\nonumber\\
	&+\left.i\left\langle q\,t_1\left|{\cO}(q(t),\l)\int_{t_2}^t \d L(q(\t),\l) d\t\right|_{q(\t)\, \text{fixed}} \right|q\,t_2\right\rangle
\eea

When either the initial or the final time grows to infinity past or future, we write
\be
\left.\left.\d^{\text{ret}}{\cal O}(q(t))=\d{\cal O}(q(t)\right|_{q\,\text{fixed}}+i \int_{-\infty}^t d\t \bigg[ {\cal O}(q(t))\, ,\,\d L(q(\t))\right|_{q\,\text{fixed}}\bigg]
\ee

\be
\left.\left.\d^{\text{adv}}{\cal O}(q(t))=\d{\cal O}(q(t)\right|_{q\,\text{fixed}}+i \int^t_{\infty} d\t \bigg[ {\cal O}(q(t))\, ,\,\d L(q(\t))\right|_{q\,\text{fixed}}\bigg]
\ee
if follows that
\be
\left.\left(\d^{\text{ret}}-\d^{\text{adv}}\right)\,{\cal O}(q(t))=i \int_{-\infty}^{\infty} d\t \bigg[ {\cal O}(q(t))\, ,\,\d L(q(\t))\right|_{q\,\text{fixed}}\bigg]
\ee
this formula defines  the {\em Peierls' bracket}, which is a generalization of Poisson's one, and that helps to determine QFT commutators in an elegant way. 

Consider the simplest example of a quantum mechanical harmonic oscillator. The retarded Green function is given by
\be
G_R(t-\tp)=\theta(t-\tp){\sin\,\omega t\over \omega}
\ee
and the advanced one
\be
G_A(t-\tp)=-\theta(\tp-t){\sin\,\omega t\over \omega}
\ee
now, take a perturbation
\be
\d L= j(t) q(t)
\ee
then
\bea
&&p=\dot{q}={d\over dt}\int_{-\infty}^t d\tp\,G_R(t-\tp)\,j(\tp)=\int_{-\infty}^t d\tp\,\cos\,\omega(t-\tp)\,j(\tp)=\nonumber\\
&&={d\over dt}\int^{\infty}_t d\tp\,G_A(t-\tp)\,j(\tp)=-\int^{\infty}_t d\tp\,\cos\,\omega(t-\tp)\,j(\tp)
\eea

It follows that
\be
\left(\d^R-\d^A\right)\,p=\int_{-\infty}^\infty d\tp\,\cos\,\omega(t-\tp)\,j(\tp)
\ee
besides
\be
\d L=j(\t) q(\t)
\ee
it follows that
\be
\left[p,q\right]=-i \hbar
\ee
In the general case the {\em Peierls' bracket}, is defined by computing  the change of some function of all field variables, ${\cO}_1$ under a perturbation  of the Lagrangian with some other operator, $\d L= j {\cO}_2$ 
\be
\left\{{\cO}_2,{\cO}_1\right\}\equiv \left(\d^R-\d^A\right)_{{\cO}_2} {\cO}_1
\ee

\par
Schwinger's principle is in some sense  a functional differential form of Feynman's path integral, where
\be
\langle A| B\rangle=\int_B^A{\cal D}\phi\, e^{i S}
\ee
the paths over which we are to integrate in the path integral are usually characterized by the initial and final points: $(x,x^\prime)$ meaning the path that obeys the boundary conditions
\bea
&x_c(t^\prime)=x^\prime\nonumber\\
&x_c(t)=x
\eea
but that could be equally well be characterized  by the initial position and momentum $(x^\prime,p^\prime)$.
\par
Hamilton's equations do tell us that
\be
\dot{x}_c^i={\pd H_c\over \pd p_i}
\ee
The jacobian mapping these two specificacions is called the {\em van Vleck-Morette} determinant
\be
{\pd(p,x)\over \pd(x,x^\prime)}\equiv \det\,D_{ij}
\ee
\be
D_{ij}\left(x\, t|x^\prime\, t^\prime\right)={\pd p_j\over \pd x^\prime_i}=-{\pd^2\over \pd x_i \pd x^\prime_j} S\left(x\, t|x^\prime\, t^\prime\right)
\ee
The Hamilton-Jacobi equation
\be
{\pd S\over \pd t}+ H_c\left(x,{\pd S\over \pd x},t\right)=0
\ee
leads to an equation for the van Vleck-Morette determinant
\be
{\pd D\over \pd t}+\pd_i\left( D \dot{x}_c^i\right)=0
\ee

This in turn leads to a representation of the path integral in the WKB (one-loop) approximation 
\be
K\left(x\, t|x^\prime\, t^\prime\right)= \tilde{N}(x,x^\prime) \,D^{1/2}\left(x\, t|x^\prime\, t^\prime\right)\, e^{i S\left(x\, t|x^\prime\, t^\prime\right)}
\ee
 Let us elaborate.  As in many other instances, Pauli's field theory book  \cite{Pauli} gives the simplest explanation of the necessity of the van Vleck determinant. To be specific, consider the simplest Hamiltonian
\be
H=\sum_k {p_k^2\over 2 m_k}+V(q)
\ee
then the solution of Schr\"odinger's equation
\be
i \hbar {\pd \over \pd t}-H=0
\ee
to $O(\hbar^2)$ is given by
\be
K_c\equiv \left(2\pi i \hbar\right)^{-{n\over 2}}\,D^{1/2}\, e^{i{S\over \hbar}}
\ee

The Hamilton-Jacobi equation reads
\be
{\pd S\over \pd t}+\sum_k {1\over 2 m_k} S_k^2+V(q)=0
\ee
It follows that we derive
\be
\pd_\t S_{i}+\sum_k {1\over m_k} S_k S_{k i}=0
\ee
and derive again we get
\be
\pd_\t S_{ij}+\sum_k {1\over m_k}\left( S_{k j} S_{k i} +S_k S_{k i j}\right)=0
\ee

Remembering that $D_{ij}\equiv -S_{ij}$ multiplying by $(D^{-1})^{ij}$ we get
\be
D^{-1}\pd_\t D+\sum_k{1\over m_k}\left( S_{kk}+ S_k D^{-1}\pd_k D\right)=0
\ee
Now it is simple exercise to show that
\be
i\hbar\pd_t K_c=\left(-\pd_t S+ {i\hbar\over 2} D^{-1}\pd_t D\right)\, K_c
\ee
as well as 
\be
i\hbar\pd_k K_c=\left( -S_k+{i\hbar\over 2} D^{-1}D_k\right)\, K_c
\ee
Derive again
\bea
-\hbar^2\pd_k\pd_k K_c&&=\bigg\{\left(-S_k+ {i\hbar\over 2} D^{-1}D_k\right)^2 -i\hbar S_{kk}+\nonumber\\
&&+{\hbar^2\over 2}D^{-2} D_k^2-{\hbar^2\over 2}D^{-1}D_{kk}\bigg\} K_c\nonumber\\
\eea
Collecting all results
\bea
&&\pd_t S- {i\hbar\over 2} D^{-1}\pd_t D+\sum_{k}\frac{1}{2m_k}\bigg\{\left(-S_k+ {i\hbar\over 2} D^{-1}D_k\right)^2 -i\hbar S_{kk}+\nonumber\\
&&+{\hbar^2\over 2}D^{-2} D_k^2-{\hbar^2\over 2}D^{-1}D_{kk}\bigg\}+V=0
\eea
\bi
\item To $O(\hbar^0)$
\be
{\pd S\over \pd t}+\sum {S_k^2\over 2 m_k}+V=0
\ee
is just the Hamilton-Jacobi equation.
\item To $O(\hbar)$
\be
D^{-1}\pd_t D+\sum_k {1\over m_k}\left(S_{kk}+S_kD^{-1} D_k\right)=0
\ee
the equation we got a while ago.

\item To $O(\hbar^2)$ appear {\em Pauli's false terms} and the ansatz needs to be corrected.

\ei

\newpage
%%%%%%%%%%%%%%%%%%%%%%%%%%%%%%%%%%%%%%%%%%%%%%%%%%%%%%%%%%%%%%%%%%%%%%%%%%%%%%%%%%%%%%%%%%%%%%%%%%%%%%%%%%%%%%%%%%%%%%%%%%%%%
\section{Gravitation and Quantum Field Theory:\\Poor man's approach.}
%%%%%%%%%%%%%%%%%%%%%%%%%%%%%%%%%%%%%%%%%%%%%%%%%%%%%%%%%%%%%%%%%%%%%%%%%%%%%%%%%%%%%%%%%%%%%%%%%%

%%%%%%%%%%%%%%%%%%%%%%%%%%%%%%%%%%%%%%%%%%%%%%%%%%%%%%%%%%%%%%%%%%%%%%%%%%%%%%%%%%%%%%%%%%%%%%%%%%%%%%%%%%%%%%%%%%%%%%%%%%%%%%
%\subsection{The Effective Lagrangian Approach to Quantum Gravity}
%%%%%%%%%%%%%%%%%%%%%%%%%%%%%%%%%%%%%%%%%%%%%%%%%%%%%%%%%%%%%%%%%%%%%%%%%%%%%%%%%%%%%%%%%%%%%%%%%%%%%%%%%%%%%%%%%%%%%%%%%%%%%
Following Wilson's effective Lagrangian approach, and to the extent that our previous experience with the other fundamental interactions is to be of any relevance here, there ought to be a regime, experimentally accessible in the not too distant future, in which gravitons propagating in flat spacetime can be isolated. This is more or less unavoidable, once gravitational waves have been observed \cite{LIGO}, and the road towards gravitons should not be too different from the road that lead from the discovery of electromagnetic waves to the identifications of photons as the quanta of the corresponding interaction, a road that led from Hertz to Planck. 
\par
Any quantum gravity theory that avoids identifying gravitational radiation as consisting of large numbers of gravitons in a semiclassical state would be at variance with all we believe to know about quantum mechanics. 
\par
What we expect instead to be confirmed by observations somewhere in the future is that the number of gravitons per unit volume with frequencies between 
$\omega$ and $\omega+d\omega$ is given by Planck's formula
\be
n(\omega)d\omega={\omega^2 \over \pi^2}{1\over e^{\hbar \omega\over k T}-1} d\omega
\ee

It is natural to keep an open mind for surprises here, because it can be argued that gravitational interaction is not alike any other fundamental interaction in the sense that the whole structure of space-time ought presumably be affected, but it cannot be denied that this is the most conservative approach and as such it should be explored first, up to its very limits, which could hopefully indicate further avenues of research.

From our experience then with the standard model of elementary particles, and assuming we have full knowledge of the fundamental symmetries of our problem, we know that we can parametrize our ignorance on the
{\em fundamental} ultraviolet physics by writing down all local operators in the low energy fields $\phi_i(x)$ compatible with the basic symmetries we have assumed.
\be
L=\sum_{n=0}^\infty{\l_n(\Lambda)^n\over \Lambda^n}{\cal O}^{(n+4)}\left(\phi_i\right)
\ee
Here $\Lambda$ is an ultraviolet cutoff, which restricts the contributions of large Euclidean momenta (or small Euclidean distances) and $\l_n(\Lambda)$ is an infinite set of dimensionless bare couplings. Two caveats. First, all this is done in a {\em flat} background. There are almost no experimental clues on what happens when the background is curved. Second, there is some contention on what is exactly the symmetry group of General Relativity. After all, {\em any} theory can be written in a covariant form. We shall be conservative in that respect.
\par
Standard Wilsonian arguments imply that {\em irrelevant operators}, corresponding to $n > 4$, are less and less important as we are interested in deeper and deeper infrared ({\em low energy}) variables.  The opposite occurs with {\em relevant operators}, corresponding to $n<4$, like the masses, that become more and more important as we approach the IR. 
The intermediate role is played by the {\em marginal operators}, corresponding to precisely $n=4$, and whose relevance in the IR is not determined solely by dimensional analysis, but rather by quantum corrections. 
The range of validity of any finite number of terms in the expansion is roughly
\be
{E\over\Lambda}\ll 1
\ee
where $E$ is a characteristic energy of the process under consideration.
\par
In the case of gravitation, we assume that general covariance (or diffeomorphism invariance) is the basic symmetry characterizing the interaction. We can then write
\bea
&&L_{eff}=\l_0 \Lambda^4 \sqrt{|g|}+\l_1 \Lambda^2 R \sqrt{|g|}+\l_2 R^2+ {1\over 2}g^{\a\b}\nabla_\a\phi\nabla_\b\phi\sqrt{|g|}+\nonumber\\
&&+\l_3 {1\over \Lambda^2}R^{\a\b}\nabla_\a\phi\nabla_\b\phi\sqrt{|g|}+\l_4 {1\over \Lambda^2}R^3\sqrt{|g|}+\l_5 \phi^4\sqrt{|g|}+\nonumber\\
&&+\bar{\psi}\left(e^\m_a \g^a \left(\pd_\m-\omega_\m\right)\psi-m\right)\psi+{\l_5\over \Lambda^2}\bar{\psi}e^\m_a \g^a R\left(\pd_\m-\omega_\m\right)\psi+\ldots
\eea
where $e_a^\m$ is the tetrad, such that
\be
e_a^\m e_\b^\n \eta^{\a\b}=g^{\m\n}
\ee
$\eta^{\a\b}$ being Minkowski's metric. The quantities $\omega_\m$ are the spin connection.
\par
The need to recover General Relativity in the classical IR limit means 
\be
\l_1\Lambda^2=-{c^3\over 16\pi G}\equiv -2 M_p^2
\ee
this in turn, means that if
\be
\l_0\Lambda^4
\ee
is to yield the observed value for the cosmological constant (which is of the order of magnitude of Hubble's constant, $H_0^4$, which is a very tiny figure when  expressed in particle  physics units, $H_0\sim 10^{-33}\,eV$) then
\be
\l_0\sim 10^{-244}
\ee
this is one aspect of the cosmological constant problem; it seems most unnatural that the cosmological constant is observationally so small from the effective Lagrangian point of view. We do not have anything new to say on this.
\par
This expansion is fine as long as it is considered a low energy expansion. As Donoghue \cite{Donoghue} has emphasized, even if it is true that each time that a renormalization is made there is a finite arbitrariness, there are physical predictions stemming from the non-local finite parts.
\par
The problem is when energies are reached that are comparable to Planck's mass,
\be
E\sim M_p.
\ee
then all couplings  in the effective Lagrangian become of order unity, and there is no {\em decoupling limit} in which gravitation can be considered by itself in isolation from all other interactions.
This then seems the minimum prize one has to pay for being interested in quantum gravity; all couplings in the derivative expansion become important simultaneously. No significant differences appear when supergravity is considered.
\par 
In conclusion, it does not seem likely that much progress can be made by somehow quantizing Einstein-Hilbert's Lagrangian in isolation. To study quantum gravity means to study all other interactions as well.
\par
On the other hand, are there any reasons to go beyond the standard model (SM)? 
\par
Yes there are some, both theoretical, and experimental. From the latter, and most important, side, both the existence of neutrino masses and  dark matter do not fit into the SM. And from the former, abelian sectors suffer from Landau poles and are not believed to be UV complete; likewise the self-interactions in the Higgs sector appear to be a trivial theory. Also the experimental values of the particle masses in the SM are not natural from the effective Lagrangian point of view.
\par
The particle physics community has looked thoroughly for such extensions since the eighties: extra dimensions (Kaluza-Klein), supersymmetry and supergravity, technicolor, etc. From a given point of view, the natural culmination of this road is string theory 
\par
A related issue is the understanding of the so-called {\em semiclassical gravity}, in which the second member of Einstein's equations is taken as the expectation value of some quantum energy-momentum operator. It can be proved that this is the dominant $1/N$ approximation in case there are $N$ identical matter fields (confer \cite{Hartle}). In spite of tremendous effort, there is not yet a full understanding of Hawking's emission of a black hole from the effective theory point of view . Another topic in which this approach has been extensively studied is Cosmology. Novel effects (or rather old ones on which no emphasis was put until recently) came from lack of momentum conservation and seem to point towards some sort of inestability \cite{Polyakov}; again the low energy theory is not fully understood; this could perhaps have something to do with the presence of horizons.

\par
Coming back to our theme, and  closing the loop, what are the prospects to make progress in quantum gravity? 
\par

Insofar as effective Lagrangians are a good guide to the physics there are only two doors open: either there is a ultraviolet attractive fixed point in coupling space, such as in Weinberg's {\em asymptotic safety} or else new degrees of freedom, like in string theory  exist in the UV.
Even if Weinberg's approach is vindicated, the fact that the putative fixed point most likely lies at strong coupling combined with our present inability to perform analitically other than perturbative computations, means that our only means to get physical information on that regime should come from lattice simulations assuming they will be able to cope with the integration over (a subclass of) geometries before physical predictions could be made with the techniques at hand at the present moment.
\par

 It is to be remarked that sometimes theories harbor the seeds of their own destruction. Strings for example, begin as theories in flat spacetime, but there are indications that space itself should be a derived, not fundamental concept. It is hoped that a simpler formulation of string theory exists bypassing the roundabouts of its historical development. This is far from being the case at present.

\par

 Finally, it is perhaps worth pointing out that to the extent that a purely gravitational canonical approach, as the ones based upon the use of Ashtekar variables makes contact with the classical limit (which is an open problem from this point of view)  the preceding line of argument should  still  carry on. 
 \par
 It seems {\em unavoidable} with our present understanding, that any theory of quantum gravity should recover, for example, the prediction that there are quantum corrections to the gravitational potential given by
 \cite{BjerrumBohr}
 \be
 V(r)=-{G m_1 m_2 \over r}\left(1+3 {G\left(m_1+m_2\right)\over r}+{41 \over 10\pi}{G\hbar\over r^2}\right)
 \ee
(the second term is also a loop effect, in spite of the conspicuous absence of $\hbar$.)
Similarly, and although this has been the subject of some controversy, it seems now established that there are gravitational corrections to the running of gauge couplings, first uncovered by Robinson and Wilczek  and given in standard notation by
\be
\b(g,E)=-{b_0\over (4\pi)^2}g^3 -3 {16\pi G\over (4\pi)^2 \hbar c^3}g E^2
\ee
sometimes these effects are dismissed as perturbative, and therefore trivial. This is not a healthy attitude.
\par
Something that can be done is to ignore most of the conceptual problems of quantum gravity, and treat it as a gauge theory. This is possible because the action of diffeomorphisms is formally similar to the one of the symmetry group of an ordinary gauge theory. Locally the fact that the group of diffeomorphisms of a given manifold, $\text{Diff}(M)$ is not a fixed entity, but rather depends in a complicated way on the specific manifold considered, this problem we say if of no concern for our perturbative analysis. All we aim at is to compute the quantum corrections to the gravitational action to first order in the coupling constant, $\kappa$. This was first done in a classic paper by 't Hooft and Veltman in 1973 \cite{tHooft}, as a byproduct of their analysis of one-loop amplitudes in non-abelian gauge theories. An essential tool of their analysis is the background field technique, first devised by DeWitt, to which we now turn.

\newpage
%%%%%%%%%%%%%%%%%%%%%%%%%%%%%%%%%%%%%%%%%%%%%%%%%%%%%%%%%%%%%%%%%%%%%%%%%%%%%%%%%%%%%%%%%%%%%%%%%%%%%%%%%%%%%%%%%%%%%%%%%%%%%
\section{Exact symmetries in quantum gravity.}
%%%%%%%%%%%%%%%%%%%%%%%%%%%%%%%%%%%%%%%%%%%%%%%%%%%%%%%%%%%%%%%%%%%%%%%%%%%%%%%%%%%%%%%%%%%%%%%%%%%%%%%%%%%%%%%%%%%%%%%%%%%%%%
There are some {\em folk theorems} nicely summarized in \cite{Banks} ({\em confer} a detailed discussion in \cite{Harlow} and \cite{Misner} for a pre-stringy approach) on what are the consistent symmetries allowed in quantum gravity. The arguments for this theorem stem mostly from consistency of the statistical interpretation of the black hole thermodynamics, although in \cite{Harlow} the theorem is argued to be a consequence of AdS/CFT.
\par
In a nutshell, this theorem asserts that there are no global symmetries. Only gauge symmetries with compact gauge group are possible. Besides, given such a gauge group, the Hilbert space must include states transforming with every possible finite dimensional irreducible representation of such a gauge group.
\par
The (simplified) argument goes as follows.
\par
 Imagine that there is a global symmetry. Then there must me some states of mass $m$ transforming with some representation $R$ of $G$. This implies the possible existence of black holes made of matter transforming with $\otimes^n R$, with $n$ arbitrarily large. The long-distance physics of this black hole will be independent of the representation $R$. This black hole cannot lose its global charge, so this implies a stable remmant once full Hawking evaporation gas taken place. Stability is a consequence of the fact that any state with such a large representation of $G$ must be heavier than the remmant. Similar arguments still hold in the massless case, $m=0$ \cite{Banks}. This leads to an infinite number of remmant states.
\par
 This in turn contradicts the {\em covariant entropy bound} (CEB) \cite{Fischler}\cite{Bousso}. The CEB  conjecture that given any two-dimensional surface of area $ A$, consider   $L$ to be the  hypersurface generated by surface-orthogonal null geodesics with non-positive expansion. The conjecture then asserts that  the entropy on $L$, $S$  does not exceed $A/4$.
 \be
 S\leq {A\over 4}
 \ee
 
 \par
 It is indeed easy to write dowm $QFT$ models that violate this theorem. Consider, for example \cite{Harlow}, Einstein-Hilbert Lagrangian coupled to two abelian $U(1)$ fields. This theory has a $\mathbb{Z}_2$ global symmetry exchanging both abelian gauge fields. Also it does not have charged matter fields.
 \par
  According to the theorem what happens is that this theory (and similar ones) cannot appear as the low energy limit of a consistent theory of quantum gravity; those theories are in the {\em swampland}, in Vafa's language \cite{Vafa}.
  \par
  All this approach relies heavily on the assumption that when we learn more about quantum gravity, we are not going to change our conception of how general are the (Schwarzschild) black hole states, as well as the esential characteristics of their Hawking evaporation. This may or may not be true. Time will say. 
\newpage
%%%%%%%%%%%%%%%%%%%%%%%%%%%%%%%%%%%%%%%%%%%%%%%%%%%%%%%%%%%%%%%%%%%%%%%%%%%%%%%%%%%%%%%%%%%%%%%%%%%%%%%%%%%%%%%%%%%%%%%%%%%%%
\section{The background field approach in quantum field theory.}
%%%%%%%%%%%%%%%%%%%%%%%%%%%%%%%%%%%%%%%%%%%%%%%%%%%%%%%%%%%%%%%%%%%%%%%%%%%%%%%%%%%%%%%%%%%%%%%%%%%%%%%%%%%%%%%%%%%%%%%%%%%%%
The only problem in quantum field theory that will concern us in these lectures is the computation of the partition function, which is nothing else than Schwinger's {\em vacuum persistence amplitude} in the presence of an external source, $J(x)$. It is useful to represent it as a functional integral
\be
Z[J]\equiv e^{i W[J]}\equiv \int {\cal D} \phi~ e^{i S[\phi]+ i \int J(x)\phi(x)}
\ee
Where in this formal analysis we represent all fields (including the gravitational field) by $\phi(x)$, and we add a coupling to an arbitrary external source as a technical device to compute Green functions out of it by taking functional derivatives of $Z[J]$ and then putting the sources equal to zero. This trick was also invented by Schwinger. The partition funtion generates all Green functions, connected and disconnected. Its logarithm, $W[J]$ sometimes dubbed the {\em free energy} generates connected functions only. These names  come from a direct analogy with similar quantities in statistical physics.
\par
It is possible to give an intuitive meaning to the path integral in quantum mechanics as a transition amplitude from an initial state to a final state. This is actually the way Feynman derived it, and it is also the way of connecting it with {\em Schwinger's action principle}. It has already been pointed out that, in some sense, Feynman's approach is an integral version of Schwinger's differential approach to the quantum dynamics.
\par
In QFT the integration measure is not mathematically well-defined.
For loop calculations, however, it is enough to {\em formally  define} the gaussian path integral as a functional determinant, that is
\be
\int {\cal D}\phi\, e^{i \left(\phi K \phi\right)}=\left(\text{det}~K\right)^{-{1\over 2}}
\ee
where the scalar product is defined as 
\be
\left(\phi, K \phi\right)\equiv \int d(vol)~\phi~ K~\phi
\ee
where $d(vol)$ is the appropiate measure (often $d(vol)\equiv \sqrt{|g|}\,dx^0\wedge\ldots\wedge dx^{n-1}$), and $K$ is a differential operator, usually
\be
K=\Box+\text{something}
\ee
there are implicit indices in the operator to pair the (also implicit) components of the field $\phi$.
\par
The only extra postulate needed is translation invariance of the measure, in the sense that
\be
\int {\cal D}\phi\, e^{i ~\left(\left(\phi+\chi\right) K \left(\phi+\chi\right)\right)}=\int {\cal D}\phi\, e^{i \left(\phi K \phi\right)}
\ee
this is the crucial property that allows the computation of integrals in the presence of external sources by completing the square.
\par
It is quite useful to introduce a generating function for one-particle irreducible (1-PI)  Green functions. This is usually called the {\em effective action} and is obtained through a Legendre transform, quite analogous to the one  performed when passing from the Lagrangian to the Hamiltonian in classical mechanics. 
\par
One defines the {\em classical field} as a functional of the external current by
\be
\phi_c[J]\equiv {1\over i}~{\d W[J]\over \d J(x)}
\ee
the Legendre transform then reads
\be
\Gamma[\phi_c]\equiv W[J]-i \int d^n x  J(x)\phi_c(x)
\ee
it is a fact that
\be
{\d \Gamma\over \d \phi_c(x)}=\int d^n z ~{\d W\over \d J(z)}~{\d J(z)\over \d \phi_c(x)}-i J(x)-i \int d^n z \phi_c (z)~{\d J(z)\over \d \phi_c(x)}=- i J(x)
\ee

The background field technique was invented by Bryce Dewitt as a clever device to keep track of divergent terms in theories (such as gravity) with a complicated algebraic structure.
The main idea is to  split the integration fields into a {\em classical} and a {\em quantum} piece:
\be
W_\m\equiv \bar{A}_\m+ A_\m
\ee
\par
The gauge transformations are
\be
\left(\bA_\m+A_\m\right)^\prime=g\left(\bA_\m+A_\m\right)g^{-1}+g\pd_\m g^{-1}
\ee
there is a subset of those, that we shall call {\em quantum gauge transformations}, which are those under which the background field remains inert 
\bea\label{quantum}
&&\bar{A}_\m^\prime=\bar{A}_\m\nonumber\\
&&A^\prime_\m=g\left(\bar{A}_\m+A_\m+\pd_\m\right)g^{-1}-\bar{A}_\m
\eea
that is
\bea
&&\d \bA_\m=0\nonumber\\
&&\d A^a_\m= i f^a_{~ b c}~\omega^b~\left(\bA^c_\m+ A^c_\m\right)-\pd_\m \omega^a=-\bD_\m \omega^a + i f^a_{~bc} \omega^b A^c_\m
\eea
those are the gauge transformations that we have got to gauge fix. The thing is that there is another, {\em background gauge} transformation, which can be kept even when gauge fixing \eqref{quantum}. Namely
\bea\label{classical}
&&\bar{A}_\m^\prime=g\left(\bar{A}_\m+\pd_\m\right) g^{-1}\nonumber\\
&&A^\prime_\m=g~A_\m~g^{-1}
\eea
under which the quantum fields rotate in the adjoint. 
This is
\bea
&&\d \bA^a_\m= i f^a_{~ b c}~\omega^b~\bA^c_\m-\pd_\m \omega^a\nonumber\\
&&\d A^a_\m= i f^a_{~ b c}~\omega^b~ A^c_\m
\eea
\par
Currents transform in such a way that
\be
\d_C \int J_a^\m A^a_\m=0
\ee
that is
\be
\d J^a_\m=if^a_{~bc} \omega^b J^c_\m
\ee

We insist that the beauty of the background field method is that it is possible to gauge fix the quantum symmetry while preserving the classical gauge symmetry. All computations are then invariant under gauge transformations of the clasical field, and so are the counterterms. This simplifies the heavy work involved in computing with gravity.
\par

The simplest background field gauge is
\be
\bar{F}^a[A]\equiv \left(\bar{D}_\m A^\m \right)^a
\ee
where $\bar{D}_\m$ represents the covariant derivative with respect to the classical field.
\par
L. Abbott \cite{Abbott} was able to prove a beautiful theorem to the effect that the effective action computed by the background field method is simply related to the ordinary effective action
\be
\Gamma_{BF}[A^{BF}_c,\bar{A}]=\left.\Gamma[A_c]\right|_{A_c=A_c^{BF}+\bar{A}}
\ee
this means in particular, that
\be
\Gamma[A_c]=\Gamma_{BF}[0,\bar{A}=A_c]
\ee
at the one loop order all this simplifies enomoursly. Working in Euclidean space
\bea
&&e^{-W[\bA]}\equiv \int {\cal D}A\, e^{-S[\bA]-\int A K[\bA]A-\int JA}=\nonumber\\
&&=e^{-S[\bA]-{1\over 2}\text{log~det}~K[\bA]-{1\over 2} \int J K^{-1}[\bA] J}
\eea
where the operator $K$ incorporates the contributions of $L_{gauge}$ as well as $L_{gf}$.
This means that
\be
A_c=- \int K^{-1}~[\bA] J
\ee
so that
\be
J= -\int K[\bA]~A_c
\ee
and
\bea
&&\Gamma^{BF}[A_c,\bA]=W[J(A_c)]-\int JA_c=\nonumber\\
&&=S[\bA]+{1\over 2}\text{log~det}~K[\bA]+{1\over 2} \int K A_c K^{-1}[\bA] K A_c-\int K A_c A_c=\nonumber\\
&& =S[\bA]+{1\over 2}\text{log~det}~K[\bA]-{1\over 2}\int A_c K A_c
\eea
finally by Abbott's theorem
\be
\Gamma(A_c)=\Gamma^{BF}[0,\bA=A_c]=W[\bA]\equiv S[\bA]+{1\over 2}\text{log~det}~K[\bA]
\ee

In order to compute the counterterm at  one-loop order, we need to take the effective action
\be
e^{iW}=\int {\cal D }A\, e^{\frac{i}{\hbar}S[A]}
\ee
with the background technique $A\rightarrow \bar{A}+A$, 
\bea
e^{iW}&&=\int {\cal D }A\, e^{\frac{i}{\hbar}S[\bar{A}+A]}\nonumber\\
&&=\int {\cal D }A\, e^{\frac{i}{\hbar}\left(S[\bar{A}]+\int S_1[\bar{A}]A+\frac{1}{2}\int S_2[\bar{A}]A^2\right)}
\eea
where
\be \left.S_n\left[\bar{A}\right]=\frac{\partial^n S}{\partial A^n}\right|_{\bar{A}}\ee
rescale now
\be
A\rightarrow \hbar^{1\over 2}\,A
\ee
\bea
e^{iW}&&=\int {\cal D }A\, e^{\left(\frac{i}{\hbar}S[\bar{A}]+\frac{i}{\hbar^{1/2}}\int S_1[\bar{A}]A+\frac{1}{2}\int S_2[\bar{A}]A^2\right)}
\eea
It can be proved (cf. \cite{Buchbinder}) that only even powers of $\hbar$ appear in the expansion; and also that only 1PI diagrams need to be considered. The linear term vanishes whenever the classical field is a solution of the equations of motion
\be
S_1[\bar{A}]=0
\ee
the first nontrivial order is the one-loop contribution
\be
e^{iW}=\int {\cal D}A\,e^{ {i\over 2} \int S_2[\bar{A}]A^2}=\left(\det\, S_2[\bar{A}]\right)^{- 1/2}
\ee
when the field is complex,
\be
e^{iW}=\int {\cal D}\varphi{\cal D}\bar{\varphi}\,e^{ {i\over 2}\bar{\varphi} S_2[\bar{A}]\varphi}=\left(\det\, S_2[\bar{A}]\right)^{-1}
\ee
finally, for fermionic fields
\be
e^{iW}=\int {\cal D}\psi{\cal D}\bar{\psi}\,e^{ {i\over 2}\bar{\psi} S_2[\bar{A}]\psi}=\left(\det\, S_2[\bar{A}]\right)
\ee

%%%%%%%%%%%%%%%%%%%%%%%%%%%%%%%%%%%%%%%%%%%%%%%%%%%%%%%%%%%%%%%%%%%%%%%%%%%%%%%%%%%%%%%%%%%%%%%%%%%%%%%%%%%%%%%%%%%%%%%%%%%%%
\subsection{Gauge invariance of the one loop effective action.}
%%%%%%%%%%%%%%%%%%%%%%%%%%%%%%%%%%%%%%%%%%%%%%%%%%%%%%%%%%%%%%%%%%%%%%%%%%%%%%%%%%%%%%%%%%%%%%%%%%%%%%%%%%%%%%%%%%%%%%%%%%%%%
We have just proved that when an appropiate gauge fixing term is used, background gauge invariance is maintained all the way except for the source term. It is then plausible that when $J=0$, that is, when
\be
\overline{S}_1\equiv{\d S[\bA]\over \d \bA}=0
\ee
the effective action is background gauge invariant.
\par
In fact the gauge dependence of the effective action has been discussed extensively by Kallosh \cite{Kallosh} who proved that not only this is indeed the case; but also that when $J=0$; that is
\be
{\d \Gamma[\bA]\over \d \bA}=0
\ee
the background field effective action is independent of the gauge fixing term. This is a nontrivial statement.
\par
Let us show a simplified proof of this fact, following \cite{Buchbinder}. We begin with an action
\be S[A]=-\frac{1}{4}\int d^nx F_{\a\b}F^{\a\b}\ee
with
\be F_{\a\b}=\partial_\a A_\b-\partial_\b A_\a-ig\left[A_\a,A_\b\right]\ee 

We shall  restrict ourselves for simplicity to linear gauge fixing
\be
\chi^\a\equiv t^\a_\b A^\b
\ee
with $\pd_\m t^\a_\b\equiv \pd_{A^\m}t^\a_\b=0$, then the gauge fixing action will be
\be
S_{gf}\equiv\frac{1}{2}\int d^nx g_{\a\b} \chi^\a \chi^\b
\ee
and the corresponding  ghost
\be
S_{gh}\equiv\int d^nx\bar{c}^\a M^\b_{~a} c_\b \equiv \int d^nx\bar{c}^\a t^\m_\a R^\b_\m c_\b
\ee
where the generator of gauge transformations is
\be R_\m^\a=\left(\d^\a_\b\partial_\m+igf^\a_{~\b\g}A^\g_\m\right)\omega^\b\ee

The partition function, with a source $J$,  will then depend upon the gauge fixing through both $g_{\a\b}$ and $\chi^\a$
\be
Z[J]\equiv \int {\cD} A{\cD}  c{\cD} \bar{c}\,e^{i\left( S[A]+S_{gf}+S_{gh}-{i\over 2}\tr\log\,g_{\a\b}+J A\right)}\equiv\langle 1\rangle
\ee
then under an arbitrary variation $\d g_{\a\b}$ and $\d t^\a_\b$
\bea
\d Z&&=i\bigg\langle -{1\over 2} g^{\a\b} \d g_{\a\b}+{1\over 2} \d g_{\a\b}\chi^\a \chi^\b+\frac{1}{2}g_{\a\b}\left( \d t^\a_\m  t^\b_\n+  t^\a_\m  \d t^\b_\n\right)A^\m A^\n-\nonumber\\
&&- (M^{-1})_\b^\a \d t^\m_\a R^\b_\m+J_\a\d  A^\a\bigg\rangle
\eea

Next, we perform a gauge transformation on the fields
\be
A^\a=A^\a+ R^\a_\m\, \xi^\m
\ee
with parameter
\be
\xi^\m=-(M^{-1})^\m_\n\left(\d t^\n_\a+{1\over 2} g^{\n\l}\d g_{\l\t} t^\t_\a\right)A^\a
\ee
Then
\bea\label{Z}
&&\d Z=i\bigg\langle -{1\over 2} g^{\a\b} \d g_{\a\b}+{1\over 2} \d g_{\a\b}\chi^\a \chi^\b+\frac{1}{2}g_{\a\b}\left( \d t^\a_\m  t^\b_\n+  t^\a_\m  \d t^\b_\n\right)A^\m A^\n- \nonumber\\
&&-(M^{-1})_\b^\a \d t^\m_\a R^\b_\m+\frac{1}{2}g_{\a\b}t^\a_\m t^\b_\n \left(A^\m R^\n_\l+A^\n R^\m_\l\right)\xi^\l-R^\a_\m\xi^\m_{,\a}-\nonumber\\
&&-(M^{-1})^\a_\b t^\b_ \m R^\m_{\a,\l}R^\l_\t\xi^\t+J_\a A^\a+J_\a R^\a_\m\xi^\m\bigg\rangle
\eea
but
\bea
&&\frac{1}{2}g_{\a\b}t^\a_\m t^\b_\n \left(A^\m R^\n_\l+A^\n R^\m_\l\right)\xi^\l=-\frac{1}{2}g_{\a\b}\left( \d t^\a_\m  t^\b_\n+  t^\a_\m  \d t^\b_\n\right)A^\m A^\n-{1\over 2} \d g_{\a\b}\chi^\a \chi^\b\nonumber\\
\eea
and the rest of terms, except the sources one
\bea
&&-{1\over 2} g^{\a\b} \d g_{\a\b}- (M^{-1})_\b^\a \d t^\m_\a R^\b_\m-R^\a_\m\xi^\m_{,\a}-(M^{-1})^\a_\b t^\b_ \m R^\m_{\a,\n}R^\l_\t\xi^\t=\nonumber\\
&&=-f^\a_{\a\b}(M^{-1})^\b_\t\left(\d t^\t_\a+\frac{1}{2}g^{\t\l}\d g_{\l\m} t^\m_\a\right)A^\a=0
\eea
taking in to account that the gauge algebra implies that 
\be
R^j_\g \pd_j R^i_\b-R^j_\b \pd_j R^i_\g= f^\d\,_{\g\b} R^i_\d
\ee
and that in dimensional  regularization (where by definition $\left.{d^n\over dx^n} \d(x)\right|_{x=0}=0\quad \forall n$) 
\be
f^\b\,_{\b\g}=0
\ee
then only the term proportional to the external source 
\be
\d Z=i\left\langle J_\a A^\a+J_\a R^\a_\m\xi^\m\right\rangle
\ee
survives. 

In conclusion, if there is not source
\be
0=J={\pd \Gamma[A]\over \pd A}
\ee
 it suffices for gauge fixing independence. Note that this argument (which seems to have its origin in DeWitt \cite{DeWitt}) is independent of the background field approach.

\par
In \cite{Kallosh}, Kallosh was also able to show that when a counterterm vanishes owing to the equations of motion, there is always a different gauge fixing term where the divergences vanish even off shell.

\newpage
%%%%%%%%%%%%%%%%%%%%%%%%%%%%%%%%%%%%%%%%%%%%%%%%%%%%%%%%%%%%%%%%%%%%%%%%%%%%%%%%%%%%%%%%%%%%%%%%%%%%%%%%%%%%%%%%%%%%%%%%%%%%%
\section{Geometric computation of the one loop effective action.}
%%%%%%%%%%%%%%%%%%%%%%%%%%%%%%%%%%%%%%%%%%%%%%%%%%%%%%%%%%%%%%%%%%%%%%%%%%%%%%%%%%%%%%%%%%%%%%%%%%%%%%%%%%%%%%%%%%%%%%%%%%%%%
 We have just proved that to one loop order all functional integral computations reduce to gaussian integrals, which can in turn be formally represented as functional determinants. This is hardly of any advantage when computing finite parts of correlators. Contrasting with that, a geometric approach for computing the {\em divergent piece} of the effective action exists. This approach was pioneered by Julian Schwinger and Bryce DeWitt (a former student of Schwinger's).
 \par
 When breaking the total gravitational field $g_{\m\n}(x)$ into a {\em background part}, $\bg_{\m\n}(x)$ and a quantum fluctuation, $h_{\m\n}(x)$, we are working in a {\em background manifold}, $\overline{M}$, with metric $\bg_{\m\n}(x)$, and thereby avoiding most of the problems of principle of quantum gravity. Quantum gravitational fluctuations are treated as ordinary gauge fluctuations. This approach was culminated by the brilliant work of 't Hooft and Veltman \cite{tHooft}, where it was shown that pure quantum gravity is one loop finite on shell. This is not true any more as soon as some matter is added. Even pure quantum gravity at two-loops is divergent on shell, as was shown by Goroff and Sagnotti \cite{Goroff}.
 \par
 The formalism is such that in order to  compute the divergent piece of the effective action, background gauge invariance can be maintained, so that we do not commit to any specific background, although we assume that some such background always exists. 
 \par
 Were we to compute correlators, then the particular Green function appropiate to each background is needed, and then all subtle points associated with background  horizons and singularities will reappear. The Unruh radiation is the simplest manifestation of these.
 \par
 It is to be emphasized that quantum Diff invariance is spontaneously broken in this approach.
 The background gauge transformations read
 \bea
&& \d \bg_{\m\n}=\xi^\l\pd_\l \bg_{\m\n}+\bg_{\l\n}\pd_\m \xi^\l +\bg_{\m\l}\pd_\n\xi^\l =\bn_\m\xi_\n+\bn_\n\xi_\m\nonumber\\
&&\d h_{\m\n}=\xi^\l\pd_\l h_{\m\n}+h_{\l\n}\pd_\m \xi^\l +h_{\m\l} \pd_\n\xi^\l 
\eea
 and the quantum gauge transformations read
 \bea
 &&\d \bg_{\m\n}=\xi^\l\pd_\l \bg_{\m\n}\nonumber\\
 &&\d h_{\m\n}=\xi^\l\pd_\l h_{\m\n}+\pd_\m \xi^\l \left(\bg_{\l\n}+h_{\l\n}\right)+\pd_\n\xi^\l \left(\bg_{\m\l}+h_{\m\l}\right)
 \eea
 
 Working to one loop order, they simplify to
  \bea
&& \d \bg_{\m\n}=\xi^\l\pd_\l \bg_{\m\n}+\bg_{\l\n}\pd_\m \xi^\l +\bg_{\m\l}\pd_\n\xi^\l =\bn_\m\xi_\n+\bn_\n\xi_\m\nonumber\\
&&\d h_{\m\n}=\xi^\l\pd_\l h_{\m\n}
\eea
 and to
 \bea
 &&\d \bg_{\m\n}=\xi^\l\pd_\l \bg_{\m\n}\nonumber\\
 &&\d h_{\m\n}=\xi^\l\pd_\l h_{\m\n}+\bg_{\l\n}\pd_\m \xi^\l +\bg_{\m\l}\pd_\n\xi^\l 
 \eea
 they still act nonlinearly of the quantum fluctuations owing to the inhomogeneous terms. This physically means that the quantum fluctuations  behave as goldstone bosons of broken Diff invariance.
 \par
 To study the Diff invariant phase would mean to compute with 
 \be
 \overline{g}_{\m\n}=0
 \ee
 which is not possible, bacause there is then no background geometry. For starters, it is not possible to define even the inverse metrix, $\bg^{\m\n}$, and consequently neither can the Christoffels be computed.
 \par
 In some cases, and using the first order formalism, it is possible to functionally integrate without the restriction that the determinant of the metric does not vanish $\bg\neq 0$.
 An example is Witten's treatment \cite{Witten3} of three-dimensional quantum gravity as a gauge theory.
 \par
 It is not clear what are the conclusions to draw for the four-dimensional case.

\newpage 
%%%%%%%%%%%%%%%%%%%%%%%%%%%%%%%%%%%%%%%%%%%%%%%%%%%%%%%%%%%%%%%%%%%%%%%%%%%%%%%%%%%%%%%%%%%%%%%%%%%%%%%%%%%%%%%%%%%%%%%%%%%%%
\section{Zeta function}
%%%%%%%%%%%%%%%%%%%%%%%%%%%%%%%%%%%%%%%%%%%%%%%%%%%%%%%%%%%%%%%%%%%%%%%%%%%%%%%%%%%%%%%%%%%%%%%%%%%%%%%%%%%%%%%%%%%%%%%%%%%%%
Consider the partition function in euclidean signature
\be 
Z\equiv\int{\cal D}\phi~e^{-{1\over 2}\int \sqrt{|g|}d^n x~\phi A\phi}
\ee
this means that the dimension of the fields $\phi$ must be ${n-d_A\over 2}$, where $d_A$ is the mass dimension of the operator $A$; usually $d_A=2$. The eigenvalues equation for this operator is
\be
A \phi_n=\l_n\phi_n
\ee
the dimension of $\l_n$ must necessarily be that of the operator $A$. We can fool around with the dimension of $\phi_n$, or fix it through normalization:
\be
\langle\phi_n|\phi_m\rangle\equiv \int \sqrt{|g|}~d^n x~ \phi^*_n~ \phi_m=\d_{mn}
\ee
the dimension of $\phi_m$ is then ${n\over 2}$ in the Kronecker case, or $0$ in the continuous case when the Kronecker delta is replaced by a Dirac delta of momentum  $\d^n(k)$. 
\par
If the set of eigenfunctions is complete in the functional space, it is possible to formally expand
\be
\phi\equiv\sum a_n~\phi_n
\ee
the dimensions of the expansion coefficients $a_n$ is ${n-d_A\over 2}-{n\over 2}=-{d_A\over 2}$ with the discrete normalization.
\par
It is tempting to {\em define the functional measure} as the dimensionless quantity
\be
{\cal D}\phi\equiv \prod_n \m^{d_A\over 2}~da_n
\ee
then the gaussian integral is represented by the infinite product
\bea
&&Z=\prod_n \m^{d_A\over 2}\int da_n~e^{-{1\over 2}\int \sqrt{|g|}d^n x~a_n\phi_n Aa_n\phi_n}=\nonumber\\
&&=\prod_n \m^{d_A\over 2}\int da_n~e^{-{a_n^2\over 2}\l_n}=\prod_n \m^{d_A\over 2}\sqrt{2\pi\over \l_n}
\eea

The zeta-function associated to the operator $A$ is now defined by analogy with Riemann's zeta function
\be
\zeta(s)\equiv \sum_n \left({\l_n \over \m^{d_A}}\right)^{-s}
\ee
and find
\be
\zeta^\prime(s)=-\sum_n~\text{log}~\left({\l_n \over \m^{d_A}}\right)~\left({\l_n \over \m^{d_A}}\right)^{-s}
\ee
so that
\be
-\zeta^\prime(0)=\sum_n~\text{log}~\left({\l_n \over \m^{d_A}}\right)~=\text{log}~\text{det}~A
\ee
then the determinant of the operator itself is defined by analytic continuation as
\be
\text{det}~ A\equiv e^{-\zeta^\prime\left(0\right)}
\ee
this definition was first proposed in the mathematical literature by Ray and Singer \cite{Ray}; in physics was first used by Dowker and Critchley \cite{Dowker} and Hawking \cite{Hawking} which studied its conformal properties and rederived the conformal anomaly.
\par
It would seem that this definition immediatly implies
\be
\det\,(AB)=\det\,A\det\,B
\ee
this is in fact obvious in the finite case, because
\be
\zeta_{AB}(s)=\sum_{m,n} \l^A_m \l^B_n
\ee
in such a way that
\be
\zeta^\prime_{AB}(0)=-\sum_{mn}\log\,\left(\l^A_m \l^B_n\right)=-\sum_m \log\,\l^A_\m-\sum_n\,\log\,\l^B_n=\zeta^\prime_A(0)+\zeta^\prime_B(0)
\ee

In general this is not so, and the {\em multiplicative anomaly} \cite{Elizalde}\cite{Kontsevich}\cite{McKenzie-Smith} is defined as
\be
a_{AB}\equiv \log\,\det\,(AB)-\log \det\,A-\log\det\,B
\ee

Let us work in detail the most basic of all determinants, the one of the flat space d'Alembertian.
The dimensionless eigenfunctions are plane waves
\be
\phi_k\equiv {1\over (2\pi)^{n\over 2}}~e^{i k x}
\ee
and are normalized in such a way that
\be
\int d^n x~\phi_k^*(x) \phi_{k^\prime}(x)=\d^n~\left(k-k^\prime\right)
\ee
the eigenvectors are simply
\be
\l_k=-k^2
\ee
the continuum normalization means that fields are expanded as 
\be
\phi(x)=\int d^n k ~a_k~ \phi_k(x)
\ee
this means that the dimension of the expansion coefficients is now
\be
\left[a_k\right]=-{n+d_A\over 2}
\ee
the zeta function is given by
\be
\zeta(s)=\int {d^n k\over (2\pi)^n}~\left({-{k^2\over \m^2}}\right)^{- s}=\int {d^n k\over (2\pi)^n}~e^{-s~\text{log}~\left({-{k^2\over \m^2}}\right)}
\ee

This leads to the expression for the determinant of the ordinary d'Alembert operator
\be\label{flat}
\text{log~det}~\Box=\int {d^n~ k\over (2\pi)^n} \text{log}\left({-{k^2\over \m^2}}\right)
\ee

\newpage
%%%%%%%%%%%%%%%%%%%%%%%%%%%%%%%%%%%%%%%%%%%%%%%%%%%%%%%%%%%%%%%%%%%%%%%%%%%%%%%%%%%%%%%%%%%%%%%%%%%%%%%%%%%%%%%%%%%%%%%%%%%%%
\section{Heat kernel}
%%%%%%%%%%%%%%%%%%%%%%%%%%%%%%%%%%%%%%%%%%%%%%%%%%%%%%%%%%%%%%%%%%%%%%%%%%%%%%%%%%%%%%%%%%%%%%%%%%%%%%%%%%%%%%%%%%%%%%%%%%%%%%
Let us now follow a slightly different route which is however intimately related. We begin, following Schwinger,  by considering 
the divergent integral which naively is independent of $\l$
\be
I(\lambda)\equiv\int_0^{\infty} \frac{dx}{x}~e^{- x\lambda}
\ee
the integral is actually divergent, so before we begin speaking about it, it has to be regularized. It
can be defined through
\be
I(\lambda)\equiv \lim_{\epsilon\rightarrow 0} I(\epsilon,\lambda)\equiv 
\lim_{\epsilon\rightarrow 0}\int_\epsilon^{\infty} 
\frac{dx}{x}e^{- x\lambda}
\ee
such that
\be
\lim_{\epsilon\rightarrow 0}\frac{\pd I(\epsilon,\lambda)}{\pd\lambda}=-\int_\e^\infty dx\,e^{-x\l}=\left.{e^{-x\l}\over \l}\right|_\e^\infty=
-\frac{1}{\lambda}
\ee
it follows
\be
I(\lambda)=-\log{\lambda}+C
\ee
\par
It is natural to define (for trace class \footnote{
In the physical Lorentzian signature, all quantities will be computed from analytic continuations from Riemannian configurations where they are better defined. This procedure is not always unambiguous when gravity is present.})  operators
\be
\log\det \Delta= \text{tr}\log\Delta\equiv \sum_n \log\lambda_n
\ee
\par
Now given an operator (with purely discrete, positive spectrum) we could generalize the above idea (Schwinger)
\be
\log\det\Delta\equiv -\int_0^\infty \frac{d\tau}{\tau} \text{tr}~ e^{-\tau\Delta}
\ee
the trace here encompasses not only discrete indices, but also includes an space-time integral.
Let is define now the {\em heat kernel} associated to that operator as the operator
\be
K(\tau)\equiv e^{-\tau \Delta}
\ee
formally the inverse operator is given through
\be
\Delta^{-1}\equiv \int_0^\infty d\t~K(\t)
\ee
where the kernel obeys the heat equation
\be
\left(\frac{\pd}{\pd\tau}+\Delta\right)K(\tau)=0
\ee
in all case that will interest us, the operator $\Delta$ will be a differential operator. Then the heat equation is a parabolic equation
\be
\left(\frac{\pd}{\pd\tau}+\Delta\right)K(\tau;x,y)=0
\ee
which need to be solved with the boundary condition
\be
K(x,y,0)=\delta^{(n)}(x-y)
\ee
\par
The mathematicians have studied operators which are deformations of the laplacian of the type
\be
\Delta\equiv -D^{\m}D_{\m}+Y
\ee
where $D_\m$ is a gauge covariant derivative
\be
D_{\m}\equiv \nabla_{\m}+X_{\m}
\ee
and $\nabla_\m$ is the usual covariant space-time derivative.
\par
In the simplest  case $X=Y=0$ and $\nabla_\m=\pd_\m$, the flat space solution corresponding to the Euclidean Laplacian is given by
\be
K_0(x,y;\tau)=\frac{1}{(4\pi \tau)^{n/2}}e^{-\frac{\sigma(x,y)}{2\tau}}
\ee
where $\s(x,y)$ is Synge's world function \cite{Synge}, which in flat space is simply given by
\be
\sigma(x,y)\equiv {1\over 2}(x-y)^2
\ee
This can be easily checked by direct computation.
\bea
\langle x \left|K_0(\t)\right| \xp\rangle&&=\langle x \left|e^{\t\Box}\right| \xp\rangle=e^{\t\Box}\langle x| \xp\rangle=\nonumber\\
&&=e^{\t\Box}\d^{(n)}\left(x-\xp\right)=\int{d^n k\over (2\pi)^n}\,e^{-\t k^2+i k(x-\xp)}=\nonumber\\
&&=\int {d^n k\over (2\pi)^n} e^{-\left(k\sqrt{\t}-i{x-\xp\over 2\sqrt{\t}}\right)^2}e^{-{(x-\xp)^2\over 4 \t}}={1\over (4 \pi\t)^{n\over 2}}\,e^{-{(x-\xp)^2\over 4\t}}\nonumber\\
\eea

\par
 Lorentzian signature leads to the replacement of the heat equation by Schr\"odinger's one (cf\cite{HartleHawking})
 \be
 S\equiv {1\over 2}\int d^4 x \phi(-\Box)\phi
 \ee
 the one-loop operator is $-\Box$. Let us define
 \be
 K(s)=e^{-is \Box}
 \ee
 so that
 \be
 i{\pd \over \pd s} K(s)= \Box K(s)
 \ee
then
 \bea
 \Delta&&\equiv \int_0^\infty ds \int \frac{d^4 k}{(2\pi)^4} e^{i s k^2}e^{i k x}=\nonumber\\
&&=\int_0^\infty ds \int \frac{d^4 k}{(2\pi)^4} e^{is(k+{x\over 2s})^2-i {x^2\over 4 s}}=\int_0^\infty ds e^{-{i \s\over 2 s}}
 \eea
it is possible to regularize the UV divergence at the coincidence limit $x=y$ by substituting
\be
\s(x,y)\rightarrow \lim_{\e\rightarrow 0^+}\,\s(x,y)- i\e
\ee

 It is unfortunately quite difficult to get explicit solutions of the heat equation except in very simple cases. This limits the applicability of the method for computing finite determinants. These determinants are however divergent in all cases of interest in QFT, and their divergence is due to the lower limit of the proper time integral. It we were able to know the solution close to the lower limit, we could get at least some information on the structure of the divergences. This is exactly how far it is possible to go.
\par
The small proper time expansion of Schwinger and DeWitt is given by a Taylor expansion
\be
K\left(\tau;x,y\right)=K_0 \left(\tau;x,y\right)~\sum_{p=0}^\infty~a_p \left(x,y\right)\tau^p
\ee
with
\be
a_0(x,x)=1
\ee
the coefficients $a_p\left(x,y\right)$ characterize the operator whose determinant is to be computed. They are {\em universal} in the sense that they are independent on the dimension of the space-time manifold, as well as on the non-invariant characteristics of the gauge fields; they only depend on {\em local} geometrical and gauge invariants \cite{Gilkey}. Actually, for the purpose at hand, only their diagonal part, $\tr\,a_n\left(x,x\right)$ is relevant.
\par
The integrated diagonal coefficients will be denoted by capital letters
\be
A_n\equiv \int \sqrt{|g|}~d^n x~ a_n(x,x)
\ee
in such a way that
\be
A_0=vol\equiv \int_M \sqrt{|g|}~d^n x
\ee
the determinant of the operator is then given by an still divergent integral. The short time expansion did not arrange anything in that respect. This integral has to be regularized by some procedure. One of the possibilities is to keep $x\neq y$ in the exponent, so that
\bea
&&\text{log~det}~\Delta\equiv -\int_0^\infty \frac{d\tau}{\tau} \text{tr}~K(\tau)\equiv \nonumber\\
&&\equiv -\lim_{\s\rightarrow 0}\int_0^{\infty}\frac{d\tau}{\tau}\frac{1}{(4\pi \tau)^{n/2}}\sum_{p=0}^\infty\tau^p \text{tr}~ a_p(x,y)~ e^{-\frac{\sigma}{2\tau}}
\eea
we have regularized the determinant by point-splitting. For consistency, also  the off-diagonal part of the short-time coefficient ought to be kept.
\par
All ultraviolet divergences are given by the behavior in the $\t\sim 0$ endpoint. Changing the order of integration, and performing first the proper time integral, the Schwinger-de Witt expansion leads to
\be
\log\det~\Delta=-\frac{1}{(4\pi)^{n/2}}\int \sqrt{|g|}~d^n x~\lim_{\s\rightarrow 0}\sum_{p=0}^\infty \frac{ \sigma^{p-{n\over 2}}}{2^{p-{n\over 2}} }~\Gamma\left({n\over 2}-p\right)~\tr\, a_p (x,y)
\ee
here it has not been not included the $\sigma$ dependence of
\be
\lim_{\s\rightarrow 0}~a_n\left(x,y\right)
\ee
in flat space this corresponds to
\be
\left(x-y\right)^2=2 \s\rightarrow 0
\ee
assuming this dependence is analytic, this could only yield higher powers of $\s$, as will become plain in a moment.

The term $p=0$ diverges in four dimensions when $\s\rightarrow 0$ as
\be
{1\over \s^2}
\ee
but this divergence  is common to all operators and can be absorbed by a counterterm proportional to the total volume of the space-time manifold. This renormalizes the  the cosmological constant. 
\par
The next term corresponds to $p=2$, and is independent on $\s$. In order to pinpoint the divergences,  When $n=4-\e$ it is given by
\be
\log\det\left.\Delta\right|_{n=4}\equiv\frac{1}{(4\pi)^2}\frac{2}{\e}~ A_2
\ee
from this term on, the limit $\s\rightarrow 0$ kills everything.

A different way to proceed is to take $\s=0$ from the beginning and put explicit IR ($\m$) and UV ($\Lambda$) proper time cutoffs, such that ${\Lambda\over\m}>>1$. It should be emphasized that these cutoffs are not cutoffs in momentum space; they respect in particular all gauge symmetries the theory may enjoy.
\be
\log\det\Delta\equiv -\int\frac{d\tau}{\tau} \text{tr}~K(\tau)\equiv -\int_{1\over \Lambda^2}^{1\over \m^2}\frac{d\tau}{\tau}\frac{1}{(4\pi \tau)^{n/2}}\sum_{p=0}\tau^p \text{tr}~ A_p\left[\Delta\right]~ 
\ee
this yields, for example in $n=4$ dimensions
\be
\log\det\Delta=-{1\over (4\pi)^2}\Big[{1\over 2}\left(\Lambda^4-\m^4\right)+ A_1\left[\Delta\right]~\left(\Lambda^2-\m^2\right) +A_2\left[\Delta\right]~ \text{log}~{\Lambda^2\over \m^2}\Big]
\ee
there are finite contributions that are not captured by the small proper time expansion; those are much more difficult to compute and, as has been already pointed out,  the heat kernel method is not particularly helpful in that respect.

\par 
 It is possible to relate the $\zeta$ function to the heat kernel, namely
 \be
 \zeta(s)={1\over \Gamma(s)}\,\tr\,\int_0^\infty d\t\, \t^{s-1}\,K(\t)
 \ee
 in such a way that
 \be
 K(\t)={1\over 2 \pi i}\oint {d\t\over \t}\Gamma(s)\zeta(s)
 \ee
 
Another possible covariant regulator in the effective action by 
 \be
 W\equiv\lim_{s\rightarrow 0}W_s\equiv  \lim_{s\rightarrow 0}\int_0^\infty\,{d\t\over \t^{1-s}} K(\t)= \lim_{s\rightarrow 0} \Gamma(s) \zeta(s)=
\ee
\be
 =\lim_{s\rightarrow 0}\left({1\over s}-\g_E\right)\left(\zeta(0)+ s \zeta^\prime(0)\right)=\lim_{s\rightarrow 0} {\zeta(0)\over s}+\zeta^\prime(0)-\g_E\zeta(0)
 \ee
 here it is clearly seen a split between the divergence and the finite part.
 
Introducing the short time expansion of the heat kernel before point splitting
\bea
&&W=\lim_{s\rightarrow 0}\int_0^\infty\,{d\t\over \t^{1-s}} {1\over (4\pi\t)^{n\over 2}} \sum_p \lim_{\s\rightarrow 0}\, e^{-{\s\over 2 \t}} a_p(\s) \t^p=
\nonumber\\
&&={1\over (4\pi)^{n\over 2}}\lim_{s\rightarrow 0}\sum_p {1\over  2^{s +p-{n\over 2}}}\Gamma\left({n\over 2}-p-s\right)\lim_{\s\rightarrow 0}\,a_p(\s)\, \s^{s+p-{n\over 2}}
\eea
this also yields a somewhat symbolic identity valid for $s\sim 0$
\bea
&& \zeta(s)={W_s\over \Gamma(s)}=
{1\over (4\pi)^{n\over 2}}{1\over \Gamma(s)}\sum_p {1\over  2^{s +p-{n\over 2}}}\Gamma\left({n\over 2}-p-s\right)\lim_{\s\rightarrow 0}\,a_p(\s)\, \s^{s+p-{n\over 2}}\nonumber\\
\eea
there is no logarithmic singularity as long as  $s\neq 0$, introducing both IR (${1\over \m^2}$) and UV (${1\over \Lambda^2}$) cutoffs in proper time we are led to
\be
W={1\over (4\pi)^{n/2}}\,\sum_{p=0}^{n\over 2}\,{a_p\over {p-\frac{n}{2}}}\,\Big[\left(\m^2\right)^{{n\over 2}-p}-\left(\Lambda^2\right)^{{n\over 2}-p}\Big]+{1\over (4\pi)^{n/2}}\,\,a_{n\over 2}\,\log\,{\Lambda^2\over \m^2}
\ee

In particular, for Klein-Gordon operator $\Delta_{KG}=-\Box-m^2$ reads
\be
K_{KG}(\t;x-y)={1\over (4\pi\t)^{n\over 2}}\,e^{-{(x-y)^2\over 4 \t}-m^2 \t}
\ee
then the $\zeta$-function can be recovered from the heat kernel through
\bea
\zeta_{KG}(s)&&={1\over \Gamma(s)}\int_0^\infty {d\t\over \t^{1-s}}{1\over (4\pi\t)^{n\over 2}} e^{-m^2\t}=\sum_{p=0}^{\infty}a_p{\left(m^2\right)^{{n\over 2}-p-s}\over (4\pi)^{n\over 2}}\,\frac{\Gamma\left(s+p-{n\over 2}\right)}{\Gamma(s)}\nonumber\\
\eea
in even dimension, $n\in 2 \mathbb{N}$,
\be
\frac{\Gamma\left(s+p-{n\over 2}\right)}{\Gamma(s)}=\frac{1}{(s-1)(s-2)\ldots \left(s+p-{n\over 2}\right)}
\ee
so that
\be
\zeta_{KG}(s)=\sum_{p=0}^{\infty}a_p{\left(m^2\right)^{{n\over 2}-p-s}\over  (4\pi)^{n\over 2}(s-1)(s-2)\ldots \left(s+p-{n\over 2}\right)}
\ee
it follows that
\be
\zeta_{KG}(0)=\sum_{p=0}^{\infty}a_p {m^{n-2p}\over  (4\pi)^{n\over 2} (-1)^{p-\frac{n}{2}} \left(p-{n\over 2}\right)!}
\ee

\newpage 
 %%%%%%%%%%%%%%%%%%%%%%%%%%%%%%%%%%%%%%%%%%%%%%%%%%%%%%%%%%%%%%%%%%%%%%%%%%%%%%%%%%%%%%%%%%%%%%%%%%%%%%%%%%%%%%%%%%%%%%%%%%%%
\section{Covariant perturbation theory.}
%%%%%%%%%%%%%%%%%%%%%%%%%%%%%%%%%%%%%%%%%%%%%%%%%%%%%%%%%%%%%%%%%%%%%%%%%%%%%%%%%%%%%%%%%%%%%%%%%%%%%%vv%%%%%%%%%%%%%%%%%%%%%%%
Let us begin by constructing a perturbation theory for the operator in flat space \cite{Mukhanov}
\be
\Delta\equiv -\Box-E
\ee
if $E$ can be treated as a small correction to the operator $\Box$, we can assume that
\be
K(\t)=\sum_{n=0}^\infty K_n(\t)
\ee
where $K_n(\t)$ is of order $E^n$, then the heat equation 
\be
{\pd \over \pd \t} K(\t)=\left(\Box+E\right)\, K(\t)
\ee
implies that
\be
{\pd K_n\over \pd \t}=\Box K_n+ E K_{n-1}
\ee

We start the recurrence with
\be
K_0(\t)=e^{\t\Box}
\ee
where the {\em parallel displacement operator} is defined as
\be
\phi(x)=a_0(x,\xp)\phi_0(x)
\ee
It is clear that a formal solution to the recurrence reads
\be
K_n(\t)= e^{\t\Box}\int_0^\t ds\,e^{-s\Box} E\,K_{n-1}(s)=K_0(\t)\int_0^\t ds\,K_0(-s)\, E\,K_{n-1}(s)
\ee
we can compute the trace
\bea
&&\langle x\left|K_1(\t)\right| x\rangle=\left\langle x\left|K_0(\t)\int_0^\t ds\,K_0(-s)\, E\,K_0(s)\right| x\right\rangle=\nonumber\\
&&=\int d^n y\,d^n z\int_0^\t ds\,\left\langle x\left|K_0(\t-s)\right|y\right\rangle\langle y\left|E(z)\right| z\rangle\langle z\left| K_0(s)\right|x\rangle=\nonumber\\
&&={1\over (4\pi \t)^{n\over 2}}\int_0^\t\, ds\, e^{{s(\t-s)\over \t}\Box}\,E(x)
\eea
when the background is not flat, the trick introduced by Barvinsky and Vilkovisky \cite{Barvinsky} consists in postulating an auxiliary metric, $g_{\m\n}^a$ such that $R_{\m\n\r\s}(g^a)=0$ as well as $\left[\nabla_\m^a,\nabla_\n^a\right]\,\phi=0$. We already know how to construct a perturbation theory for the auxiliary space with the  operator
\be
\Delta\equiv -\Box^a - E^a
\ee
then we write
\bea
&g_{\a\b}= g_{\a\b}^a+ h_{\a\b}\nonumber\\
&\nabla_\m\phi=\left(\nabla^a_\m+\Gamma_\m\right)\phi
\eea
and expand in powers of $h$, then
\be E(x)=\frac{1}{4}\d_{\m\n}\Box h^{\m\n}(x)-V(x)+{\cal O}(h^2)\ee

The full expression to second order reads \cite{Barvinsky}
\bea
\tr\,K(\t)&&={1\over (4\pi\t)^{n\over 2}}\int d^n x \sqrt{g}\, \tr\bigg\{1+\t \left(\frac{R}{6}-V\right) +\nonumber\\
&&+\frac{\t^2}{2}\left(V-\frac{R}{6}\right) f_1(-\t\Box) V+\t^2 V f_2(-\t\Box) R+\nonumber\\
&&+\t^2R f_3(-\t\Box) R+\t^2R_{\m\n}f_4(-\t\Box)R^{\m\n}\bigg\}
\eea
where the different form factor are given by
\bea
&&f_1(x)=\int_0^1\, d\a\, e^{-\a(1-\a) x}\nonumber\\
&&f_2(x)=-\frac{f_1(x)}{6}-\frac{f_1(x)-1}{2x}\nonumber\\
&&f_3(x)=\frac{f_1(x)}{32}+\frac{f_1(x)-1}{8x}-\frac{f_4(x)}{8}\nonumber\\
&&f_4(x)=\frac{f_1(x)-1+\frac{x}{6}}{x^2}
\eea

\newpage
%%%%%%%%%%%%%%%%%%%%%%%%%%%%%%%%%%%%%%%%%%%%%%%%%%%%%%%%%%%%%%%%%%%%%%%%%%%%%%%%%%%%%%%%%%%%%%%%%%%%%%%%%%%%%%%%%%%%%%%%%%%%%
\section{Flat space determinants}
%%%%%%%%%%%%%%%%%%%%%%%%%%%%%%%%%%%%%%%%%%%%%%%%%%%%%%%%%%%%%%%%%%%%%%%%%%%%%%%%%%%%%%%%%%%%%%%%%%%%%%%%%%%%%%%%%%%%%%%%%%%%%
Let us see in detail  how the heat equation can be iterated to get the coefficients of the short time expansion for operators pertaining to flat space gauge theories. 
\par
Consider the operator
\be
\Delta\equiv -D_\m D^\m+Y
\ee
where
\be
D_\m\equiv \pd_\m+A_\m
\ee
The small proper time expansion of the heat kernel
\bea
&&K\left(\t;x,y\right)\equiv\frac{1}{(4\pi \tau)^{n/2}}e^{-\frac{\sigma}{2\tau}}\sum_{p=0}^\infty\tau^p a_p(x,y)
\eea
when substituted into the heat equation leads to
\bea
&&{\pd \over \pd \t}~K\left(\t;x,y\right) = {1\over (4\pi)^{n\over 2}}~e^{-\frac{\sigma}{2\tau}}\sum_{p=0}^\infty\left(a_p{\s\over 2} +(p-{n\over 2}-1)a_{p-1}\right)\t^{p-2-{n\over 2}}\nonumber\\
\eea
\bea
&&D_\m K\left(\t;x,y\right)= {1\over  (4 \pi \t)^{n\over 2}}e^{-\frac{\sigma}{2\tau}}\sum_{p=0}^\infty \left(-{\s_\m\over 2\t}a_p+D_\m a_p\right)\t^p\eea
\bea
&&\sum_\m D_\m^2 K\left(\t;x,y\right)=\nonumber\\
&&={1\over (4\pi)^{n\over 2}}~e^{-\frac{\sigma}{2\tau}}~\sum_{p=0}^\infty\left(- {n\over 2}~a_{p-1}+{\s\over 2}a_p-\sum_\m~\s^\m~ D_\m a_{p-1}+ D^2 a_{p-2}\right)\t^{p-2-{n\over 2}}\nonumber\\\eea
\bea
&&-\Delta K\left(\t;x,y\right)=\left(D_\m^2-Y\right) K=\nonumber\\
&&={1\over (4\pi)^{n\over 2}}~e^{-\frac{\sigma}{2\tau}}~\sum_{p=0}^\infty\left(- {n\over 2}~a_{p-1}+{\s\over 2}a_p-\sum_\m~\s^\m~ D_\m a_{p-1}- \Delta a_{p-2}\right)\t^{p-2-{n\over 2}}\nonumber\\
\eea
where $\s_\m\equiv (x_\m-y_\m)$.
\par
The more divergent terms are those in $\t^{-2-{n\over 2}}$, but they do not give anything new
\be
{a_0\over 4}(x-y)^2={a_0\over 4}(x-y)^2
\ee
The next divergent term (only even p contribute to the expansion in a manifold without boundaries) is $\t^{1-{n\over 2}}$
\be
-{n\over 2} a_0=-{n\over 2} a_0-\s^\m D_\m a_0
\ee
so that we learn that
\be\label{cero}
\s^\m D_\m a_0=0.
\ee
Generically,
\be
\left(p-{n\over 2}-1\right)a_{p-1}=-{n\over 2} a_{p-1}-\s^\m D_\m a_{p-1}-\Delta a_{p-2}
\ee
which is equivalent to
\be\label{iter}
(p+1)a_{p+1}+\s^\m D_\m a_{p+1}+\Delta a_p=0
\ee

Taking the covariant derivative of \eqref{cero}
\be\label{uno}
D_{\lambda}(\sigma^{\m}D_{\m}a_0)=D_\l a_0+\s^\m D_\l D_\m a_0=0
\ee
 The first {\em coincidence limit} ($x\rightarrow y$) follows
\be
\left[D_{\m}a_0\right]\equiv \lim_{x\rightarrow y}D_{\m}a_0(x,y) =0
\ee
please note that $[a_0]=1$ which we knew already, does not imply the result. Taking a further derivative of \eqref{uno}, we get 
\be
\left[(D_{\m}D_{\n}+D_{\n}D_{\m})a_0\right]=0
\ee
whose trace reads
\be
\left[D^2 a_0\right]=0
\ee
With the usual definition of gauge field strength
\be
F_{\m\n}\equiv \left[D_{\m},D_{\n}\right]
\ee
implies
\be
\left[D_{\m}D_{\n}~a_0\right]=\frac{1}{2}\left[([D_{\m},D_{\n}]+\{D_{\m},D_{\n}\})a_0\right]=\frac{1}{2}F_{\m\n}
\ee
where the fact  has been used that
\be
\left[a_0\right]=1
\ee

Taking $p=0$ in (\ref{iter})
\be
a_1=-\s^\m D_\m a_1-\Delta a_0
\ee
so that
\be
\left[a_1\right]=-\left[\Delta  a_0\right]=-Y
\ee
since $\Delta=-D^2+Y$.

When $p=1$ in (\ref{iter})
\be
-2 a_2 =\Delta a_1 +\s^\m D_\m a_2
\ee
so that
\be
\left[a_2\right]=-\frac{1}{2}\left[\Delta a_1\right]
\ee

Let us derive again the $p=0$ expression before the coincidence limit:
\be
D_{\m}a_1=-D_{\m}D^2 a_0-D_\m\left(\s^\n D_\n a_1\right)= -D_\m D^2 a_0-D_{\m}a_1-\sigma^{\lambda}D_{\m}D_{\lambda}a_1
\ee
then
\be
2 D_\m a_1=-D_\m\Delta a_0-\s^\l D_\m D_\l a_1
\ee
which implies at the coincidence limit
\be
\left[D^2 a_1\right]=-\frac{1}{3}\left[D^2 \Delta a_0\right]
\ee
that is
\be
\left[\Delta a_1\right]\equiv \left[D^2 a_1\right]+\left[Y a_1\right]=-\frac{1}{3}\left[D^2 D^2 a_0\right]-Y^2 +\frac{1}{3}D^2 Y
\ee

Now deriving four times the equation (\ref{cero})
\be
(D_{\delta}D_{\sigma}D_{\rho}D_{\m}+D_{\delta}D_{\sigma}D_{\m}D_{\rho}+
D_{\delta}D_{\rho}D_{\m}D_{\sigma}+D_{\sigma}D_{\rho}D_{\m}D_{\delta}+
\s^{\lambda}D_{\delta}D_{\sigma}D_{\rho}D_{\m}D_{\lambda})a_0=0
\ee
contracting with $\eta^{\delta\sigma}\eta^{\rho\m}$
\be
\left[(D^2 D^2+D^{\m}D^2 D_{\m})a_0\right]=0
\ee
and contracting instead with $\eta^{\delta\rho}\eta^{\sigma\m}$
\be
\left[(D^{\m}D^{\n}D_{\m}D_{\n})a_0\right]=0
\ee
now
\be
\left[(D^{\sigma}D^{\m}D_{\m}D_{\sigma})a_0\right]=\left[(D^{\m}D^{\sigma}D_{\m}D_{\sigma}a_0+
F^{\s\m}D_\m D_\s a_0\right]
\ee
it follows that
\be
\left[D^\a D^2 D_\a a_0\right]=0+F^{\s\m}\left[D_\m D_\s a_0\right]=-{1\over 2} F^2_{\m\n}
\ee
so that
\be
\left[D^2 D^2 a_0\right]= \frac{1}{2} F^2_{\m\n}
\ee
and finally
\be
\left[a_2\right]=-\frac{1}{2}\left[\Delta a_1\right]=\frac{1}{6}\left[D^2 D^2 a_0\right]+\frac{1}{2}Y^2-
\frac{1}{6}D^2 Y=\frac{1}{12}F^2_{\m\n} +\frac{1}{2}Y^2-\frac{1}{6}\Box Y	
\ee

The final expression for the divergent piece of the determinant of the flat space gauge operator reads
\be\label{ados}
\text{log~ det}~ \Delta =\frac{1}{ (4\pi)^2}\frac{2}{4-n}\int d^n x~ \text{tr}~\left(\frac{1}{12}\,F^2_{\m\n}+\frac{1}{2}Y^2\right)
\ee
the term in $\Box Y$ is a surface term which does not contribute in the absence of boundaries.
%%%%%%%%%%%%%%%%%%%%%%%%%%%%%%%%%%%%%%%%%%%%%%%%%%%%%%%%%%%%%%%%%%%%%%%%%%%%%%%%%%%%%%%%%%%%%%%%%%%%%%%%%%%%%%%%%%%%%%%%%

This computation is immediatly applicable to gauge theories in flat space \cite{Osborn}. The simplest gauge invariant action is
\be
S\equiv -{1\over 4 g^2}\int d^4 x\, F_{\m\n}F^{\m\n}
\ee
the gauge fixing term reads
\be
S_{gf}\equiv-{1\over 2 g^2\xi}\int d^4 x\,\partial^\m A_\m \partial^\n A_\n
\ee
and the ghost action is
\be
S_{gh}\equiv -{1\over 2 g^2}\int  d^4 x\,\partial^\m\bar{c} \partial_\m c
\ee
the expansion up to quadratic order reads
\be S_2=-\int d^n x\left(\frac{1}{2}A_\m\Delta^{\m\n}A_\n-\bar{c}\Box c\right)\ee
where
\be
\Delta_{\m\n}=-\eta_{\m\n}\Box+\left(1-\frac{1}{\xi}\right)\partial_\m\partial_\n
\ee
and we have removed the factor $1/g^2$ by redefining
\bea 
&&A_\m\rightarrow gA_\m\nonumber\\
&&c\rightarrow gc
\eea

Then the one-loop effective action in Minkowski space is given by
\be
\Gamma(\bA)\equiv {1\over 2}\left(\log\det\,\Delta_{\m\n}- 2\log\det \,\Box\right)
\ee
Please note the factor and the sign of the ghost contribution. The divergent contributions can be read directly from our analysis:
\be
\log\det\,\Delta_{\m\n}=-{2\over 4-n}{C\over (4\pi)^2}{20\over 3}\int d^n x F_{\m\n}^2
\ee
\be
\log\det \,\Box={2\over 4-n}{C\over (4\pi)^2}{1\over 3}\int d^nx F_{\m\n}^2
\ee
Collecting both results
\be
\Gamma(\bA)=-{1\over \e}{C\over (4\pi)^2}{22\over 3}\int d^nx F_{\m\n}^2
\ee
where $\e=4-n$.

In order to renormalize the gauge theory to this order we have to write
\be
{1\over g_0^2}=\m^{-\e}\left({1\over g^2}+{22\over 3}{C\over (4\pi)^2}\frac{1}{\e}\right)
\ee
The origin of the essential term $\m^{-\e}$ is that in $n$ complex dimensions the coupling constant has dimensions of
\be
\left[ g^2_0\right]=4-n
\ee
(the gauge field has mass dimension one in any spacetime dimension).
This means that the beta function (more on this momentarily) is given by
\be
\left. \m{\pd \over \pd \m} g\right|_{g_0}=-\frac{1}{2}\e g+\b(g)
\ee
differentiating $g_0^{-2}$ with respect to $\m$ then gives
\be \e\left({1\over g^2}+{22\over 3}{C\over (4\pi)^2}\frac{1}{\e}\right)=\left[-\frac{1}{2}\e g+\b(g)\right]\frac{\partial}{\partial g}\frac{1}{g^2}\ee
finally
\be
\b(g)=-{11\over 3}{C\over (4\pi)^2}
\ee
in the usual case when $G=SU(n)$ the index $C=n$.

\newpage

%%%%%%%%%%%%%%%%%%%%%%%%%%%%%%%%%%%%%%%%%%%%%%%%%%%%%%%%%%%%%%%%%%%%%%%%%%%%%%%%%%%%%%%%%%%%%%%%%%%%%%%%%%%%%%%%%%%%%%%%%%%%%
\section{The conformal anomaly}
%%%%%%%%%%%%%%%%%%%%%%%%%%%%%%%%%%%%%%%%%%%%%%%%%%%%%%%%%%%%%%%%%%%%%%%%%%%%%%%%%%%%%%%%%%%%%%%%%%%%%%vv%%%%%%%%%%%%%%%%%%%%%
Every renormalizable massless QFT without dimensionful coupling constants (masses count as coupling constants) enjoys scale invariance in flat space under
\be
x^\m\rightarrow \l x^\m
\ee
the conformal group $SO(4,2)$ \cite{Blumenhagen} is an extension of the Lorentz group (which has six generators: three {\em rotations} and three  {\em boosts}). This extension has 15 generators; besides the ones of the Lorentz group, one {\em dilatation}, four {\em translations} and four {\em special conformal transformations}
\bea
&&x^\m\rightarrow y^\m
\nonumber\\
&&{y^\m\over y^2}={x^\m\over x^2}-b^\m
\eea
Although the inversion is not connected to the identity, the special conformal transformation is. An elegant way of characterizing the full set of conformal transformations is as generated by {\em conformal Killing vectors}, under which the Minkowski metric is not fully invariant, but rather is proportional to itself.
\be
\eta_{\m\n}\rightarrow \Omega^2(x) \eta_{\m\n}
\ee
this in turn gives the clue as to how to generalize this property to curved spacetimes as Weyl rescalings of the metric tensor, which we shall do in a moment.

Even when a theory is classically scale invariant, it fails to to be so when quantum corrections are considered. This is because the theory has to be regularized, and this introduces a mass scale, say $\m$ as in dimensional regularization. Even after renormalizations has taken rid of all infinities, the coupling constants remember the scale that has been introduced. This is the idea of the {\em beta function}. Somewhat symbollically, under a scale transformation,
\be
\d {\cal L}=\sum_i \b(g_i){\pd  {\cal L}\over \pd g_i}
\ee
where
\be
\b(g_i)\equiv \m\,{\pd g_i\over \pd \m}
\ee
We shall see later on that {\em in addition} to this scale violation by quantum effects, there are other violations once gravitational fields are present; those occur because  those background fields are not scale invariant; there is always a scale associated with them.
\par
Let us consider a generic scalar field theory. Non zero  spin does not substantially change the analysis, so that our results will be quite generic.
\be
S=\int d^4 x~\left({1\over 2}\pd_\m\phi\pd^\m\phi-{m^2\over 2}\phi^2-{g\over 4!}\phi^4+g_1 \phi^6+\ldots\right)
\ee
Let us  perform a  scale transformation
\bea
&&x\equiv \l x^\prime\nonumber\\
&&\phi=\l^{-1}\phi^\prime
\eea
In terms of the new fields
\be
S=\int d^4 x^\prime~ ~\left({1\over 2}\pd_{\m^\prime}\phi^\prime\pd^{\m^\prime}\phi^\prime-{m^2 \l^2\over 2}\phi^2-{g \over 4!}\left(\phi^\prime\right)^4+g_1 \l^{-2} \left(\phi^\prime\right)^6+\ldots\right)
\ee
Consider the theory in the vicinity of the gaussian fixed point, which correspond to a free scalar field, that is
\be
S=\int d^4 x^\prime~ ~\left({1\over 2}\pd_{\m^\prime}\phi^\prime\pd^{\m^\prime}\phi^\prime\right)
\ee
When we run towards the infrared ($\l\rightarrow \infty$)
we see that
\bea
&&m^\prime\rightarrow\infty\nonumber\\
&&g_1\rightarrow 0\nonumber\\
&&g\rightarrow g
\eea
Operators like $m^2 \phi^2$ of dimension less than four are called {\em relevant}. Operators like $g_2\phi^6$ of scaling dimension bigger than four are called {\em irrelevant}. and operators like $\l \phi^4$ of scaling dimension exactly equal to four  are called {\em marginal}.
\par
Any physical observable should be independent of the scale $\m$, which has been introduced as an intermediate step in the regularization and is thus completely arbitrary. Observables then  obey
\bea
&&0=\left.{d\over d\m}S\left[p_i,g_0,m_0\right]\right|_{g_0,m_0}=\left.{d\over d\m}S\left[p_i,g_R,m_R,\m\right]\right|_{g_0,m_0}=\nonumber\\
&&=\bigg[{\pd\over\pd ~\text{log}~\m}+\left.\b(g_R,m_R)~{\pd\over\pd g_R}\right|_{\m,m_R}-\nonumber\\
&&-\left.\g_m(g_R,m_R){\pd\over \pd\text{log}~m_R}\right|_{g_R,\m}\bigg]S\left[p_i,g_R,m_R,\m\right]\nonumber\\
\eea
this is dubbed the {\em renormalization group equation} (RGE). We have defined
\bea
&&\b(g_R,m_R)\equiv \left.\m{\pd\over \pd\m}g_R(\m)\right|_{g_0,m_0}\nonumber\\
&&\g_m(g_r,m_R)\equiv \left.-{\m\over m_R}{\pd\over\pd \m}~m_R(\m)\right|_{g_0,m_0}
\eea
since the function $S\left[p_i,g_R,m_R,\m\right]$ is analytic at $n=4$, it is natural to expect that both functions $\b$ and $\g_m$ are analytic as well. In order to compute these universal functions, and remembering that our renormalization counterterms are defined as
\be
m_0=Z_m^{1\over 2}~m_R
\ee
and
\be
g_0=Z_g~ g_R~ \m^{\e\over 2}
\ee
where $\e=4-n$

\bea
&&\b(g_R,m_R)\equiv \left.g_0 \m{\pd\over \pd\m}{1\over Z_g \m^{\e\over 2}}  \right|_{g_0,m_0}\nonumber\\
&&\g_m(g_r,m_R)\equiv \left.-{m_0\over m_R}\m{\pd\over\pd \m}~Z_m^{-{1\over 2}}\right|_{g_0,m_0}
\eea
All this is much simpler in a mass independent renormalization scheme, where the renormalization constants are independent of $m_R$ and $\m$. In fact {\em  minimal substraction schemes} MS (or $\overline{MS}$) are  such schemes.

\par
It is plain that
\be
\m~{\pd \over \pd \m}~g_0=0=\m^{\e\over 2}\left({\e\over 2} Z_g g_R+\m~{\pd\over \pd \m }\left(Z_g g_R\right)\right)
\ee
now
\be
g_R~ Z_g=g_R+\sum_{p=1}^\infty a_p(g_R) \left({2\over \e}\right)^p
\ee
let us make the Laurent-type ansatz (we shall see later that it is actually necessary)
\be
\beta(g_R)\equiv \b_0(g_R)+\b_1(g_R)\e+\ldots
\ee
we get
\bea
&&{\e\over 2}\left(g_R+a_1(g_R) {2\over \e}+\ldots\right)+\b_0(g_R)+\b_1(g_R)\left(\e\right)+\ldots+\nonumber\\
&&+ {da_1(g_R)\over d g_R}\left(\b_0(g_R)+\b_1(g_R)\e+\ldots\right){2\over \e}+\ldots=0
\eea
terms of ${\cal O}(\e^0)$ (which are now seen to be neccessary) yield
\be a_1(g_R)+\b_0(g_R)+2\b_1(g_R) {da_1(g_R)\over d g_R}=0\ee
and the terms of ${\cal O}(\e^1)$ 
\be
{g_R\over 2}+\b_1(g_R)+2\b_0(g_R){da_1(g_R)\over d g_R}=0
\ee
there are recursion relations worked out by 't Hooft \cite{tHooftS} to compute all $a_p, p>1$ from the knowledge of $a_1$.
\par
For example, in the theory $\phi^4_4$ the result is
\be
\b=3~{\l^2\over (4\pi)^2}
\ee
whereas for the (also renormalizable) six-dimensional theory $\phi^3_6$
\be
\b=-3~{\l^3\over 4 (4\pi)^3}
\ee
in QED
\be
\b={e^3\over 12\pi^2}
\ee
and for a nonabelian $SU(N)$ gauge theory with $n_f$ fermion flavors in the fundamental representation,
\be
\b=-{g^3\over (4\pi)^2}~\left({11\over 3} N-{4\over 3} n_f T(F)\right)
\ee

We take a general beta function to assert
\be
\b\equiv b \l^3
\ee
with the definition of $\b$
\be \b=\frac{d\l}{d \text{log}~\m}\ee
means that
\be
{d\l\over \l^3}= b~d~\text{log}~\m
\ee
integrating with the boundary conditions that
\be
\l=\l_0
\ee
when
\be
\m=\m_0
\ee
yields the dependence of the coupling constant on the RG scale, $\m$
\be
\l^2={\l^2_0\over 1-2b \l_0^2~\text{log}~{\m\over\m_0}}
\ee

When $b$ is positive (like in $\phi^4_4$ or QED), there is a {\em Landau pole} at
\be
\Lambda\equiv\m=\m_0~e^{\frac{1}{2b~\l_0^2}}
\ee
those theories are {\em infrared safe}, but they do not enjoy an UV consistent limit.
\par
When $b<0$ (this is precisely what happens for $\phi^3_6$ and also for ordinary gauge theories) there is a pole at
\be
\Lambda\equiv\m= \m_0~e^{-{1\over 2\left|b\right|~\l_0^2}}
\ee
the paradigm of these theories is QCD. They are {\em asymptotically free} but {\em infrared slave}. The Landau pole is now located in the infrared region. Its scale is also denoted by $\Lambda$ is obviously renormalization group invariant and experimentally its value is
\be
\Lambda\sim 217 MeV
\ee
and it signals the scale at which QCD starts being strongly coupled.

\par
Green functions also obey some different renormalization group equations, because they are multiplicatively renormalized. The starting point is that
\be
\m~{\pd \over \pd \m}~ \Gamma^0=0
\ee
then
\bea
&&\bigg\{\m~\left.{\pd\over \pd\m}\right|_{g_R,m_R}+\left.\b(g_R){\pd\over \pd g_R}\right|_{\m,m_R}-\g_m(g_R)\left.{\pd\over \pd m_R}\right|_{\m,g_R }-\nonumber\\
&&-n \g_\phi(g_R)\bigg\}~\Gamma_R\left(p_i, g_R,m_R,\m\right)=0\nonumber\\
\eea
where 
\be \g_m(g_R)=-\mu\frac{\pd m_R}{\pd\m}\ee
and we have defined the {\em anomalous dimension}
\be
\g_\phi(g_R)\equiv {1\over 2}~\m{\pd \over \pd \m} \text{log}~Z_\phi={1\over 2}~\b(g_R)~{\pd \over \pd g_R} \text{log}~Z_\phi
\ee
for example, for the theory $\phi^3_6$
\be
\g_\phi(g)={1\over 12}{\l^3\over (4\pi)^3}+{13\over 432}~\left({\l^2\over (4\pi)^3} 
\right)^2
\ee

The RE equations for 1PI in gauge theories are best writing by first defining the operator
\be
{\cal D}\equiv \m\frac{\pd}{\pd \m}+\b(g_R)\frac{\pd}{\pd g_R}-\g_m(g_R)\frac{\pd}{\pd m_R}+\d(g_R,\l_R)\frac{\pd}{\pd \a_R}-n_A\g_A-n_f \g_\psi -n_c \g_c
\ee
the 1PI equation itself reads
\be
{\cal D}~ \Gamma_{R,n}(g_R,m_R,\zeta_R)=0
\ee
the number of external gauge fields is  ($n_A$), external fermions by ($n_f$) and ghosts ($n_c$), and their corresponding anomalous dimensions  $\g_A$,$\g_f$,$\g_c$, where the generic definition of the {\em anomalous dimension} reads
\be
\g\equiv \frac{1}{2}\m\frac{\pd}{\pd \m}log\,Z
\ee
these objects are in general  gauge dependent.

\par

We have already mentioned the fact that when a nontrivial gravitational field is present, the generalization of the conformal group $SO(4,2)$ in flat space is given by Weyl rescalings of the spacetime metric
\be
g_{\m\n}\rightarrow \Omega^2(x) g_{\m\n}
\ee
when $\Omega$ is constant, a given operator $\Delta$ scales under Weyl transformations as
\be
\Delta\rightarrow \Omega^{-\l} \Delta
\ee
where $\l$ is the so called {\em conformal weight} of the operator in question. Recalling that the energy momentum tensor is defined as
\be
\d W={1\over 2} \int d(vol) T^{\m\n}\d g_{\m\n}
\ee
under a Weyl transformation $\d g_{\m\n}=2 \Omega \,g_{\m\n}\d \Omega$ the change in the effective action is proportional to the trace of the energy momentum tensor.
\be
\d_\Omega W={1\over 2} \int d(vol)2 \Omega \, T\d \Omega
\ee
This means that if the action is Weyl invariant, the trace of the energy momentum tensor vanishes. Let us now compute in an explicit way the change in the effective action following \cite{Hawking}
\par
The  zeta function associated to $\Delta$ transforms under such a global rescaling as
\be
\zeta(s)\rightarrow \Omega^{-\l s} \zeta(s)
\ee
then
\be \zeta^{'}(s)\rightarrow -\l\Omega^{-\l s-1} \zeta(s)+\Omega^{-\l s} \zeta^{'}(s)\ee
which conveys the fact that
\be
\log\,\det\, \Delta \rightarrow \frac{\l}{\Omega}\zeta(0)+\log\,\det\,\Delta
\ee
At the level of the effective action
\be
\d W={1\over 2} \d \log\,\det\,\Delta={1\over 2} \d \log\,\det\,\Delta+{1\over 2}\l \zeta(0)\d \Omega
\ee
This can be computed from the heat kernel through
\be
\zeta(0)=\lim_{s\rightarrow 0} \frac{1}{\Gamma(s)}\,\int_0^\infty d\t \t^{s-1} K(\t)
\ee
This result means that the {\em conformal} (or {\em trace}) anomaly is proportional to   the divergent part of the effective action when computed in dimensional regularization. In order to calculate the one loop conformal anomaly it is enough to  determine  the corresponding counterterm.

\newpage
%%%v%%%%%%%%%%%%%%%%%%%%%%%%%%%%%%%%%%%%%%%%%%%%%%%%%%%%%%%%%%%%%%%%%%%%%%%%%%%%%%%%%%%%%%%%%%%%%%%%%%%%%%%%%%%%%%%%%%%%%%%%%
\section{Vacuum energy}
%%%%%%%%%%%%%%%%%%%%%%%%%%%%%%%%%%%%%%%%%%%%%%%%%%%%%%%%%%%%%%%%%%%%%%%%%%%%%%%%%%%%%%%%%%%%%%%%%%%%%%%%%%%%%%%%%%%%%%%%%%%%%
It is well known that when gravitational effects are neglected, the value of the energy is defined only up to an additive constant.  The zero point of energy is arbitrary, and can be selected at will. When gravitation is taken into account this is not so, and vacuum energy gravitates (owing to the $\sqrt{|g|}$ in the volume element).  The corresponding operator is
\be
{\cO}\equiv \int \sqrt{|g|} d^n x\,V_0
\ee
\par
Even in flat space, it is possible to compute the difference of energies between two different vacua. The simplest instance occurs when this difference between vacua is due to the presence of boundary conditions. In the Casimir effect, those boundary conditions  are due to the presence of conducting plates, where the electromagnetic field should vanish.
\par
In the few instances where the vacuum energy can be computed explicitly, \cite{Parker} zeta function reveals itself useful.
Consider, as a vanilla example a one-dimensional free  scalar field with two Dirichlet boundary conditions in two hyper surfaces 
\be 
\phi(x,y)=\phi(x,y+2\pi R)
\ee
the heat kernel can be expanded as 
\be K(\t, x,x',y,y')=\sum_{p=-\infty}^{\infty}\frac{1}{2\pi R}K_p(\t,x,x')e^{\frac{i}{R}(y-y')p}
\ee
Where each Fourier component must satisfy
\be 
\frac{\partial}{\partial\t}K_p(\t,x,x')=-\left(\Box^x_{n}+m^2\right)K_p(\t,x,x')
\ee

The heat equation reads 
\be 
\frac{\partial}{\partial\t}K_p(\t,x,x')=-\left(\Box^x_{n-1}+m^2+\frac{p^2}{R^2}\right)K_p(\t,x,x')
\ee
whose solution is
\be 
K_p(\t,x,x')=(4\pi\t)^{-(n-1)/2}e^{-\frac{(x-x')^2}{4\t}-\left(m^2+p^2/R^2\right)\t}
\ee
The effective action is easily computed
\bea
W&&=-\hslash\int\frac{d\t}{\t}K(\t, x,x,y,y)=\nonumber\\
&&=-\hslash\int\frac{d\t}{\t}\sum_{p=-\infty}^{\infty}\frac{1}{2\pi R}(4\pi\t)^{-(n-1)/2}e^{-\left(m^2+p^2/R^2\right)\t}=\nonumber\\
&&=-\frac{\hslash}{2\pi R}(4\pi)^{-(n-1)/2}\sum_{p=-\infty}^{\infty}\left(m^2+\frac{p^2}{R^2}\right)^{(n-1)/2}\int d\t \t^{-1-(n-1)/2}e^{-\t}=\nonumber\\
&&=-\frac{\hslash}{2\pi R}(4\pi)^{-(n-1)/2}\Gamma\left(\frac{1-n}{2}\right)\sum_{p=-\infty}^{\infty}\left(m^2+\frac{p^2}{R^2}\right)^{(n-1)/2}\eea

For a massless scalars
\bea
 W&&=-\frac{\hslash}{2\pi R^n}(4\pi)^{-(n-1)/2}\Gamma\left(\frac{1-n}{2}\right)\sum_{p=-\infty}^{\infty}p^{(n-1)}=\nonumber\\
&&=-\frac{\hslash}{2\pi R^n}(4\pi)^{-(n-1)/2}\Gamma\left(\frac{1-n}{2}\right)2\zeta(1-n)
\eea
which in four dimensions, in $n=4$ reduces to
\bea
 W&&=-\frac{\hslash}{720\pi^2 R^4}
 \eea
Quite often it is expressed in terms of  $L=2\pi R$  
\bea 
W&&=-\frac{\hslash\pi^2}{45 L^4}
\eea

\newpage
%%%%%%%%%%%%%%%%%%%%%%%%%%%%%%%%%%%%%%%%%%%%%%%%%%%%%%%%%%%%%%%%%%%%%%%%%%%%%%%%%%%%%%%%%%%%%%%%%%%%%%%%%%%%%%%%%%%%%%%%%%%%
\section{The DeWitt computation of gravitational determinants}
%%%%%%%%%%%%%%%%%%%%%%%%%%%%%%%%%%%%%%%%%%%%%%%%%%%%%%%%%%%%%%%%%%%%%%%%%%%%%%%%%%%%%%%%%%%%%%%%%%%%%%%%%%%%%%%%%%%%%%%%%%%%%
When the gravitational interaction is physically relevant, things are much more complicated, \cite{DeWitt}. First of all, the space-time manifold is not flat, so that the flat space free solution has got to be generalized. All computations should be covariant. It is precisely at this point that the world function will become handy. On the other hand it is when dealing with this sort of problems that the real power of the heat kernel technique is visible.
We shall keep denoting the {\em coincidence limit} of any bi-scalar function by
\be
\left[W\right]\equiv \lim_{x\rightarrow x^\prime}W(x,x^\prime)
\ee
There is a general rule, called {\em Synge's rule} for computing such limits. The rule as applied to the world function states that
\be
\left[\nabla_{\ap} \s_{\ldots}\right]=\nabla_{\ap}\left[\s_{\ldots}\right]-\left[\nabla_\a \s_{\ldots}\right]
\ee
where the dots indicate further derivations. Let us prove it. 
\begin{proof}
	Consider some  bi-scalar (that is, an scalar both at $x$ and $\xp$)
	\be
	\Omega_{A B^\prime}(x,x^\prime)
	\ee
	where  $A,B,\ldots$ are multi-indexes. Further consider a physical quantity $P^A (x)$ with the same multi-index structure as $A$; and another one $Q^{B^\prime} (x^\prime)$ with the same multi-index structure as $B^\prime$. Both objects are parallel propagated
	\be
	u^\a \nabla_\a P^A(x)= u^{\a^\prime}\nabla_{\a^\prime} Q^{B^\prime}(x^\prime)=0
	\ee
	the bi-scalar
	\be
	H(x,x^\prime)\equiv \Omega_{A B^\prime}(x,x^\prime) P^A(x) Q^{B^\prime}(x^\prime)
	\ee
	can be Taylor expanded in two different ways
	\bea
	&&H(\l_1,\l_0)=H(\l_0,\l_0)+(\l_1-\l_0)\left.{\pd H\over \pd\l_1}\right|_{\l_1=\l_0}+\ldots=\nonumber\\
	&&=H(\l_1,\l_1)-(\l_1-\l_0)\left.{\pd H\over \pd \l_0}\right|_{\l_0=\l_1}+\ldots
	\eea
	Therefore
	\be
	H(\l_0,\l_0)\equiv \left[\Omega_{A B^\prime}\right] P^A Q^{B^\prime}
	\ee
	obeys
	\bea
	&&{d\over d \l_0} H(\l_0,\l_0)\equiv \lim_{\l_1\rightarrow \l_0}{H(\l_1,\l_1)-H(\l_0,\l_0)\over \l_1-\l_0}=\nonumber\\
	&&=\left.{\pd H\over \pd \l_0}\right|_{\l_0=\l_1}+\left.{\pd H\over \pd \l_1}\right|_{\l_1=\l_0}=\nonumber\\
	&&=u^{\ap}\left[\nabla_{\ap}\Omega_{A B^\prime}\right] P^A Q^{B^\prime}+ u^\m \left[\nabla_\m \Omega_{A B^\prime}\right] P^A Q^{B^\prime}
	\eea
	Finally
	\be
	\nabla_{\ap}\left[ \Omega_{A B^\prime}\right]=\left[\nabla_{\ap}\Omega_{A B^\prime}\right]+\left[\nabla_\a \Omega_{A B^\prime}\right]
	\ee
	it is the desired result
\end{proof}
%%%%%%%%%%%%%%%%%%%%%%%%%%%%%%%%%%%%%%%%%%%%%%%%%%%%%%%%%%%%%%%%%%%%%%%%%%%%%%%%%%%%%%%%%%%%%%%%%%%%%%%%%%%%%%%%%%%%%%%%%%%%%
\subsection{Coincidence limits of covariant derivatives of the world function}
%%%%%%%%%%%%%%%%%%%%%%%%%%%%%%%%%%%%%%%%%%%%%%%%%%%%%%%%%%%%%%%%%%%%%%%%%%%%%%%%%%%%%%%%%%%%%%%%%%%%%%%%%%%%%%%%%%%%%%%%%%%%%
A basic structural unit in the heat kernel technique is the world function $\s(x,x^\prime)$, which obeys
\be \label{basic}\s_\m \s^\m-2\s=0\ee 
with
\be
\s_\m\equiv \pd_\m \s
\ee
Let us prove it.
\begin{proof}
	Start for the action for a free particle
	\be
	S\equiv\int_{x,\t}^{x^\prime,\t^\prime}~d\t~ {1\over 2} g_{\m\n} \dot{x}^\m \dot{x}^\n\equiv{\s(x,x^\prime]\over \t^\prime-\t}
	\ee
	where the integral is taken over the geodesic $x^\m=x^\m(\t)$ that goes from the {\em base point} $x^\prime$ at value $\t^\prime$ of the parameter to the {\em field point}  $x^\prime$ at value $\t^\prime$ of the same parameter. This defines the square of the geodesic distance between the points $x^\prime$ and $x$. It is a scalar for independent Einstein transformations of the base and field points. This is exactly  Synge's {\em world function}. He used the notation $\Omega$ for it, but nowadays the notation $\s$ is much more common. When the geodesic is timelike and parametrized with  the proper time
	\be
	\s(x,x^\prime)={(\t-\t^\prime)^2\over 2}
	\ee
	\par
	The canonical momentum is given by
	\be
	p_\m\equiv \pd_\m S= {\nabla_\m\s \over \t^\prime- \t}
	\ee
	The Hamilton-Jacobi equation for the free particle reads
	\be
	{\pd S\over \pd \t} + H=-{\s\over \left(\t^\prime-\t\right)^2}+{1\over 2}{\s_\m \s^\m\over \left(\t^\prime- \t\right)^2}=0
	\ee
	leads to the basic equation obeyed by the world function
	\be
	\s_\m \s^\m= 2 \s
	\ee
\end{proof}
\par
It is instructive to study a more pedestrian derivation. 
\begin{proof}	
	Consider a variation of the world function
	\be
	\d \s\equiv \s(x+\d x,x^\prime)-\s(x,x^\prime)
	\ee
	where we rescale the parameters in such a way that $(\l_0,\l_1)$ label the ends of the new geodesic.
	\par
	The variation can be computed in an standard way
	\bea
	&&\d \s=(\l_1-\l_0)\int_{\l_0}^{\l_1}d\l\left(g_{\m\n}\dot{z}^\m\d\dot{z}^\n+{1\over 2}\pd_\l g_{\m\n}\dot{z}^\m\dot{z}^\n\d z^\l\right)=\nonumber\\
	&&=(\l_1-\l_0)\left. g_{\a\b}\dot{z}^\a \d z^\b\right|_{\l_0}^{\l_1}-(\l_1-\l_0)\int_{\l_0}^{\l_1}\left(g_{\a\b}\ddot{z}^\b+\Gamma_{\a\b\g}\dot{z}^\b\dot{z}^\g\right)\d z^\a d\l
	\nonumber\\
	\eea
	Inserting the information that the line integral is taken over a geodesic, we learn that
	\be
	\d\s=(\l_1-\l_0) g_{\a\b}u^\a\d x^\b
	\ee
	This means that the derivative of the world function is proportional to the tangent vector
	\be
	\nabla_\a \s\equiv \s_\a=(\l_1-\l_0) u_\a=+\sqrt{2 \s}~u_\a
	\ee
	also
	\be
	\nabla_{\a^\prime}\s\equiv \s_{\a^\prime}=-(\l_1-\l_0) u_{\a^\prime}
	\ee
	and it is now obvious that
	\be
	\s_\m\s^\m=\s_{\m^\prime}\s^{\m^\prime}=2 \s
	\ee
	This implies that the equation of parallel transport of any quantity $T$ can be written as
	\be
	u^\m \nabla_\m T=\s^\m \nabla_\m T=0
	\ee
\end{proof}
Let us proceed carefully to work out coincidence limits of covariant derivatives of the world function
It is plain that
\be
\boxed{\left[\s\right]=\left[\s_\m\right]=0}
\ee
(The second equation is true because there is no prefered vector in the manifold.)
Deriving (covariantly) once the equation \eqref{basic}
\be
\s_\m \s^{\m\a}=\s^{\a}
\ee
Derive again
\be
\s_{\m}^{~\b}\s^{\m\a}+\s_\m\s^{\m\a\b}=\s^{\a\b}
\ee
obviously
\be
\Box \s =\s_{\m\n}\s^{\m\n}+\s_\m\Box \s^\m
\ee

It follows that
\be
\boxed{\left[\s_{\m\n}\right]=g_{\m\n}}
\ee
as well as its trace
\be
\boxed{\left[\Box \s\right]=n}
\ee

If we derive for the third time \eqref{basic}
\bea
&&\s_{\m}^{~\b\g}\s^{\m\a}+\s_{\m}^{~\b}\s^{\m\a\g}+\s_{\m}^{~\g}\s^{\m\a\b}+\s_\m\s^{\m\a\b\g}=\s^{\a\b\g}
\eea
At the coincidence limit
\be \label{c1}[\s_{\b\a\g}]+[\s_{\g\a\b}]=0\ee
Ricci's formula for the commutation of covariant derivatives implies that
\be
\s_{\a\b\g}-\s_{\a\g\b}=-\s_\m R^\m_{~\a\g\b}
\ee
(Our conventions are different  from \cite{Poisson}).
It is easy to prove that  the coincidence limit of three derivatives vanishes
\be
\boxed{\left[\s_{\m\n\l}\right]=0}
\ee 

The fourth derivative reads
\bea
&&\s_{\m}^{~\b\g\d}\s^{\m\a}+\s_{\m}^{~\b\g}\s^{\m\a\d}+\s_{\m}^{~\b\d}\s^{\m\a\g}+\s_{\m}^{~\b}\s^{\m\a\g\d}+\nonumber\\
&&+\s_{\m}^{~\g\d}\s^{\m\a\b}+\s_{\m}^{~\g}\s^{\m\a\b\d}+\s_{\m}^{~\d}\s^{\m\a\b\g}+\s_\m\s^{\m\a\b\g\d}=\s^{\a\b\g\d}
\eea
at  coincidence 
\be
\left[\s_{\m\n\a\b}\right]+\left[\s_{\b\n\a\m}\right]+\left[\s_{\a\n\b\m}\right]=0
\ee
 Commuting covariant derivatives 
\be
\s_{\m\n\a\b}=\s_{\n\m\a\b}
\ee
and
\be \s_{\a\b\m\n}-\s_{\a\m\b\n}=-R^{\t}_{~\a\m\b}\s_{\t\n}-\s_\t\nabla_\n R^\t_{~\a\m\b}\ee
Then
\be [\s_{\a\b\m\n}]-[\s_{\a\m\b\n}]=-R_{\n\a\m\b}\ee
In the same way 
\be
\left[\s_{\a\b\m\n}\right]-\left[\s_{\a\b\n\m}\right]=-R_{\b\a\n\m}-R_{\a\b\n\m}=0
\ee
Summarizing
\be [\s_{\a\b\m\n}]-[\s_{\m\n\a\b}]=0\ee
At coincidence
\be\label{4s}
\boxed{\left[\s_{\a\b\m\n}\right]=-{1\over 3}\left(R_{\a\m\b\n}+R_{\a\n\b\m}\right)}
\ee
and in particular
\be
\boxed{\left[\nabla_{\a}\nabla_{\b} \Box\s\right]=-{2\over 3}~R_{\a\b}}
\ee

The expression with five dervatives of \eqref{basic} is also needed
\bea
&&\s_{\m}^{~\b\g\d\e}\s^{\m\a}+\s_{\m}^{~\b\g\d}\s^{\m\a\e}+\s_{\m}^{~\b\g\e}\s^{\m\a\d}+\s_{\m}^{~\b\g}\s^{\m\a\d\e}+\s_{\m}^{~\b\d\e}\s^{\m\a\g}+\nonumber\\
&&+\s_{\m}^{~\b\d}\s^{\m\a\g\e}+\s_{\m}^{~\b\e}\s^{\m\a\g\d}+\s_{\m}^{~\b}\s^{\m\a\g\d\e}+\s_{\m}^{~\g\d\e}\s^{\m\a\b}+\s_{\m}^{~\g\d}\s^{\m\a\b\e}+\nonumber\\
&&+\s_{\m}^{~\g\e}\s^{\m\a\b\d}+\s_{\m}^{~\g}\s^{\m\a\b\d\e}+\s_{\m}^{~\d\e}\s^{\m\a\b\g}+\s_{\m}^{~\d}\s^{\m\a\b\g\e}+\s_{\m}^{~\e}\s^{\m\a\b\g\d}+\nonumber\\
&&+\s_\m\s^{\m\a\b\g\d\e}=\s^{\a\b\g\d\e}
\eea
At the coincidence limit
\be
\left[\s_{\b\a\g\d\e}\right]+\left[\s_{\g\a\b\d\e}\right]+\left[\s_{\d\a\b\g\e}\right]+\left[\s_{\e\a\b\g\d}\right]=0
\ee
Using again Ricci's identity
\bea
&&\left[\s_{\b\a\g\d\e}\right]=\left[\s_{\a\b\g\d\e}\right]
\nonumber\\
&&\left[\s_{\g\a\b\d\e}\right]=\left[\s_{\a\b\g\d\e}\right]-\nabla_\d R_{\g\b\a\e}-\nabla_\e R_{\g\b\a\d}
\nonumber\\
&&\left[\s_{\d\a\b\g\e}\right]=\left[\s_{\a\b\g\d\e}\right]-\nabla_\g R_{\d\b\a\e}-\nabla_\e R_{\d\b\a\g}
\nonumber\\
&&\left[\s_{\e\a\b\g\d}\right]=\left[\s_{\a\b\g\d\e}\right]-\nabla_\g R_{\e\b\a\d}-\nabla_\d R_{\e\b\a\g}
\eea
Putting all together,
\bea
&&4\left[\s_{\a\b\g\d\e}\right]-\nabla_\d R_{\g\b\a\e}-\nabla_\e R_{\g\b\a\d}-\nabla_\g R_{\d\b\a\e}-\nabla_\e R_{\d\b\a\g}-\nonumber\\
&&-\nabla_\g R_{\e\b\a\d}-\nabla_\d R_{\e\b\a\g}=0\nonumber\\
\eea

 Contracting $\a=\b$
\be
\boxed{\left[\s^\a_{~\a\g\d\e}\right]=-{1\over 2}\left(\nabla_\d R_{\g\e}+\nabla_\e R_{\g\d}+\nabla_\g R_{\e\d}\right)}
\ee
and using Bianchi's identity $\nabla_\a R^{\a\b}=\frac{1}{2}\nabla^\b R$
\be
\boxed{\left[\s^{\a~\b}_{~\a~\b\l}\right]=-\nabla_\l R}
\ee

Finally the expression of six derivatives of \eqref{basic}
\bea
&&\s_{\m}^{~\b\g\d\e\s}\s^{\m\a}+\s_{\m}^{~\b\g\d\e}\s^{\m\a\s}+\s_{\m}^{~\b\g\d\s}\s^{\m\a\e}+\s_{\m}^{~\b\g\d}\s^{\m\a\e\s}+\s_{\m}^{~\b\g\e\s}\s^{\m\a\d}+\nonumber\\
&&+\s_{\m}^{~\b\g\e}\s^{\m\a\d\s}+\s_{\m}^{~\b\g\s}\s^{\m\a\d\e}+\s_{\m}^{~\b\g}\s^{\m\a\d\e\s}+\s_{\m}^{~\b\d\e\s}\s^{\m\a\g}+\s_{\m}^{~\b\d\e}\s^{\m\a\g\s}+\nonumber\\
&&+\s_{\m}^{~\b\d\s}\s^{\m\a\g\e}+\s_{\m}^{~\b\d}\s^{\m\a\g\e\s}+\s_{\m}^{~\b\e\s}\s^{\m\a\g\d}+\s_{\m}^{~\b\e}\s^{\m\a\g\d\s}+\s_{\m}^{~\b\s}\s^{\m\a\g\d\e}+\nonumber\\
&&+\s_{\m}^{~\b}\s^{\m\a\g\d\e\s}+\s_{\m}^{~\g\d\e\s}\s^{\m\a\b}+\s_{\m}^{~\g\d\e}\s^{\m\a\b\s}+\s_{\m}^{~\g\d\s}\s^{\m\a\b\e}+\s_{\m}^{~\g\d}\s^{\m\a\b\e\s}+\nonumber\\
&&+\s_{\m}^{~\g\e\s}\s^{\m\a\b\d}+\s_{\m}^{~\g\e}\s^{\m\a\b\d\s}+\s_{\m}^{~\g\s}\s^{\m\a\b\d\e}+\s_{\m}^{~\g}\s^{\m\a\b\d\e\s}+\s_{\m}^{~\d\e\s}\s^{\m\a\b\g}+\nonumber\\
&&+\s_{\m}^{~\d\e}\s^{\m\a\b\g\s}+\s_{\m}^{~\d\s}\s^{\m\a\b\g\e}+\s_{\m}^{~\d}\s^{\m\a\b\g\e\s}+\s_{\m}^{~\e\s}\s^{\m\a\b\g\d}+\s_{\m}^{~\e}\s^{\m\a\b\g\d\s}+\nonumber\\
&&+\s_{\m}^{~\s}\s^{\m\a\b\g\d\e}+\s_\m\s^{\m\a\b\g\d\e\s}=\s^{\a\b\g\d\e\s}
\eea
 yields at coincidence
\bea
&&[\s_{\m}^{~\b\g\d}\s^{\m\a\e\s}]+[\s_{\m}^{~\b\g\e}\s^{\m\a\d\s}]+[\s_{\m}^{~\b\g\s}\s^{\m\a\d\e}]+[\s_{\m}^{~\b\d\e}\s^{\m\a\g\s}]+[\s_{\m}^{~\b\d\s}\s^{\m\a\g\e}]+\nonumber\\
&&+[\s_{\m}^{~\b\e\s}\s^{\m\a\g\d}]+[\s^{\b\a\g\d\e\s}]+[\s_{\m}^{~\g\d\e}\s^{\m\a\b\s}]+[\s_{\m}^{~\g\d\s}\s^{\m\a\b\e}]+[\s_{\m}^{~\g\e\s}\s^{\m\a\b\d}]+\nonumber\\
&&+[\s^{\g\a\b\d\e\s}]+[\s_{\m}^{~\d\e\s}\s^{\m\a\b\g}]+[\s^{\d\a\b\g\e\s}]+[\s^{\e\a\b\g\d\s}]+[\s^{\s\a\b\g\d\e}]=0
\eea
Using once more time Ricci's identity we are led to
\bea
&&\left[\s^{\b\a\g\d\e\s}\right]=\left[\s^{\a\b\g\d\e\s}\right]\eea
\bea
&&\left[\s^{\g\a\b\d\e\s}\right]=\left[\s^{\a\b\g\d\e\s}\right]+\nabla^\e\nabla^\d R^{\a\s\b\g}+\nabla^\s\nabla^\d R^{\a\e\b\g}+\nabla^\s\nabla^\e R^{\a\d\b\g}+\nonumber\\
&&+R^{\a\l\b\g}\left[\s_{\l}^{~\d\e\s}\right]\eea
\bea
&&\left[\s^{\d\a\b\g\e\s}\right]=\left[\s^{\a\b\g\d\e\s}\right]-\nabla^\e\nabla^\g R^{\d\b\a\s}-\nabla^\s\nabla^\g R^{\d\b\a\e}-\nabla^\s\nabla^\e R^{\d\b\a\g}-\nonumber\\
&&-R^{\d\g\b\l}\left[\s_{\l}^{~\a\e\s}\right]-R^{\d\g\a\l}\left[\s_{\l}^{~\b\e\s}\right]-R^{\d\b\a\l}\left[\s_{\l}^{~\g\e\s}\right]\eea
\bea
&&\left[\s^{\e\a\b\g\d\s}\right]=\left[\s^{\a\b\g\d\e\s}\right]+\nabla^\d\nabla^\g R^{\a\s\b\e}+\nabla^\s\nabla^\g R^{\a\d\b\e}+\nabla^\s\nabla^\d R^{\a\g\b\e}+\nonumber\\
&&+R^{\g\l\d\e}\left[\s_{~~~\l}^{\a\b~\s}\right]+R^{\b\l\d\e}\left[\s_{\l}^{~\a\g\s}\right]+R^{\a\l\d\e}\left[\s_{\l}^{~\b\g\s}\right]+R^{\b\l\g\e}\left[\s_{\l}^{~\a\d\s}\right]+\nonumber\\
&&+R^{\a\l\g\e}\left[\s_{\l}^{~\b\d\s}\right]+R^{\a\l\b\e}\left[\s_{\l}^{~\g\d\s}\right]\eea
\bea
&&\left[\s^{\s\a\b\g\d\e}\right]=\left[\s^{\a\b\g\d\e\s}\right]+\nabla^\d\nabla^\g R^{\a\e\b\s}+\nabla^\e\nabla^\g R^{\a\d\b\s}+\nabla^\e\nabla^\d R^{\a\g\b\s}+\nonumber\\
&&+R^{\d\l\e\s}\left[\s_{~~~~\l}^{\a\b\g}\right]+R^{\g\l\e\s}\left[\s_{~~\l}^{\a\b~\d}\right]+R^{\b\l\e\s}\left[\s_{\l}^{~\a\g\d}\right]+R^{\a\l\e\s}\left[\s_{\l}^{~\b\g\e}\right]+\nonumber\\
&&+R^{\g\l\d\s}\left[\s_{~~\l}^{\a\b~\e}\right]+R^{\b\l\d\s}\left[\s_{\l}^{~\a\g\e}\right]+R^{\a\l\d\s}\left[\s_{\l}^{~\b\g\e}\right]+R^{\b\l\g\s}\left[\s_{\l}^{~\a\d\e}\right]+\nonumber\\
&&+R^{\a\l\g\s}\left[\s_{\l}^{~\b\d\e}\right]+R^{\a\l\b\s}\left[\s_{\l}^{~\g\d\e}\right]
\eea

Now putting all it together
\bea\label{6D} &&5\left[\s^{\a\b\g\d\e\s}\right]+\nabla^\e\nabla^\d \left(R^{\a\s\b\g}+R^{\a\g\b\s}\right)+\nabla^\s\nabla^\d\left( R^{\a\e\b\g}+R^{\a\g\b\e}\right)+\nonumber\\
&&+\nabla^\s\nabla^\e\left( R^{\a\d\b\g}+R^{\a\g\b\d}\right)+\nabla^\d\nabla^\g\left(R^{\a\e\b\s}+R^{\a\s\b\e}\right)+\nonumber\\
&&+\nabla^\s\nabla^\g \left(R^{\a\d\b\e}+R^{\a\e\b\d}\right)+\nabla^\e\nabla^\g \left(R^{\a\d\b\s}+R^{\a\s\b\d}\right)+\nonumber\\
&&+R^{\a\l\b\g}\left[\s_{\l}^{~\d\e\s}\right]+R^{\g\l\d\e}\left[\s_{~~~\l}^{\a\b~\s}\right]+R^{\b\l\d\e}\left[\s_{\l}^{~\a\g\s}\right]+R^{\a\l\d\e}\left[\s_{\l}^{~\b\g\s}\right]+\nonumber\\
&&+R^{\b\l\g\e}\left[\s_{\l}^{~\a\d\s}\right]+R^{\a\l\g\e}\left[\s_{\l}^{~\b\d\s}\right]+R^{\a\l\b\e}\left[\s_{\l}^{~\g\d\s}\right]+R^{\d\l\e\s}\left[\s_{~~~~\l}^{\a\b\g}\right]+\nonumber\\
&&+R^{\g\l\e\s}\left[\s_{~~\l}^{\a\b~\d}\right]+R^{\b\l\e\s}\left[\s_{\l}^{~\a\g\d}\right]+R^{\a\l\e\s}\left[\s_{\l}^{~\b\g\e}\right]+\nonumber\\
&&+R^{\g\l\d\s}\left[\s_{~~\l}^{\a\b~\e}\right]+R^{\b\l\d\s}\left[\s_{\l}^{~\a\g\e}\right]+R^{\a\l\d\s}\left[\s_{\l}^{~\b\g\e}\right]+R^{\b\l\g\s}\left[\s_{\l}^{~\a\d\e}\right]+\nonumber\\
&&+R^{\a\l\g\s}\left[\s_{\l}^{~\b\d\e}\right]+R^{\a\l\b\s}\left[\s_{\l}^{~\g\d\e}\right]-R^{\d\g\b\l}\left[\s_{\l}^{~\a\e\s}\right]-R^{\d\g\a\l}\left[\s_{\l}^{~\b\e\s}\right]-\nonumber\\
&&-R^{\d\b\a\l}\left[\s_{\l}^{~\g\e\s}\right]+\left[\s_{\m}^{~\b\g\d}\s^{\m\a\e\s}\right]+\left[\s_{\m}^{~\b\g\e}\s^{\m\a\d\s}\right]+\left[\s_{\m}^{~\b\g\s}\s^{\m\a\d\e}\right]+\nonumber\\
&&+\left[\s_{\m}^{~\b\d\e}\s^{\m\a\g\s}\right]+\left[\s_{\m}^{~\b\d\s}\s^{\m\a\g\e}\right]+\left[\s_{\m}^{~\b\e\s}\s^{\m\a\g\d}\right]+\nonumber\\
&&+\left[\s_{\m}^{~\g\d\e}\s^{\m\a\b\s}\right]+\left[\s_{\m}^{~\g\d\s}\s^{\m\a\b\e}\right]+
\left[\s_{\m}^{~\g\e\s}\s^{\m\a\b\d}\right]+\left[\s_{\m}^{~\d\e\s}\s^{\m\a\b\g}\right]=0\nonumber\\
\eea
and using the expression for $\left[\s_{\m\n\a\b}\right]$, \eqref{4s}, we have at coincidence 
\be\label{6Dtrace}
\boxed{\left[\Box\Box\Box \s\right]=-\frac{8}{5}\Box R+{4\over 15} R^2_{\m\n}-{4\over 15} R^2_{\m\n\a\b}}
\ee
We also need   another scalar combination of six derivatives of $\s$ to wit
\be
\boxed{\nabla_\l \Box \nabla_\l\Box \s=\frac{2}{5}\Box R-{2\over 5} R^2_{\m\n}-{4\over 15} R^2_{\m\n\a\b}}
\ee
%%%%%%%%%%%%%%%%%%%%%%%%%%%%%%%%%%%%%%%%%%%%%%%%%%%%%%%%%%%%%%%%%%%%%%%%%%%%%%%%%%%%%%%%%%%%%%%%%%%%%%%%%%%%%%%%%%%%%%%%%%%%%
\subsection{Coincidence limits of derivatives of the van Vleck determinant.}
%%%%%%%%%%%%%%%%%%%%%%%%%%%%%%%%%%%%%%%%%%%%%%%%%%%%%%%%%%%%%%%%%%%%%%%%%%%%%%%%%%%%%%%%%%%%%%%%%%%%%%%%%%%%%%%%%%%%%%%%%%%%%%%
As we shall see in a moment, another important piece in De Witt's approach is the van Vleck determinant, defined by
\be
\Delta(x,x^\prime)\equiv \text{det}~\Delta^{\a^\prime}\,_{\b^\prime}(x,x^\prime)\equiv \text{det}~\left(-g^{\a^\prime}_\a (x^\prime,x)\s^\a_{\b^\prime}(x,x^\prime)\right)
\ee
the parallel propagator is defined in terms of frames (tetrads) at both $x$ and $\xp$ as
\be
g^{\ap}\,_\a (x,x^\prime)\equiv e_a^{\ap}(x^\prime) e^a_\a(x)
\ee
in such a way  that
\be
\text{det}~\left(g^{\ap}\,_\a (x,x^\prime)\right)={e(x)\over e^\prime(x^\prime)}
\ee
Taking determinants in the definition  yields
\be
\Delta(x,x^\prime)=-{\text{det}~\left(-\s^{\a^\prime}_{~\b^\prime}(x,x^\prime)\right)\over e e^\prime}\equiv -{{\cal D}(x,x^\prime)\over e e^\prime}
\ee
It is plain that
\be
\left[\Delta^{\a^\prime}_{~\b^\prime}\right]=\d^{\a^\prime}_{~\b^\prime}
\ee
\be
\left[\Delta\right]=1
\ee

It is a fact \cite{Poisson} that the van Vleck-Morette determinant  obeys the fundamental equation
\be
\nabla_\a\left(\Delta~\s^\a\right)=\s^\a\nabla_\a \Delta+\Delta \Box \s=\s^\a\nabla_\a \Delta+\left(1+\sqrt{2 \s}~\theta\right)\Delta=n \Delta
\ee
The failure of the van Vleck determinant to be parallel propagated  is measured by the expansion of the geodesic congruence
\be
\s^\a\nabla_\a\left(\text{log}~\Delta\right)=(n-1)-\sqrt{2 \s}~\theta
\ee

Indeed, starting from
\be
\Delta^{\ap}\,_{\b^\prime}=-g^{\ap\a}\left(\s^\g\,_\a\s_{\g\b^\prime}+\s^\g\s_{\a\b^\prime\g}\right)=g^{\ap}\,_\a g^\g_{\g^\prime}\s^\a\,_\g \Delta^{\g^\prime}\,_{\b^\prime}+\nabla_\g \Delta^{\ap}_{\b^\prime}\s^\g
\ee
multiplying by the inverse matrix $\Delta^{-1}$ and taking the trace
\be
n= \Box \s+(\Delta^{-1})^{\b^\prime}\,_{\ap} \s^\g\nabla_\g\Delta^{\ap}_{\b^\prime}
\ee
which implies the desired identity.
\par
The coincidence limit of the fundamental  equation holds trivially.
%%%%%%%%%%%%%%%%%%%%%%%%%%%%%%%%%%%%%%%%%%%%%%%%%%%%%%%%%%%%%%%%%%%%%%%%%%%%%%%%%%%%%%%%%%%%%%%%%%%%%%%%%%%%%%%%%%%%%%%%%%%%%
\par 
Let us now draw some consequences for the coincidence limits of derivatives of the van Vleck determinant.

First  rewrite the fundamental equation 
\be
\nabla_\m\left(\Delta \s^\m\right)=n\Delta
\ee
like
\be \label{fundamental}
2\s^\m\Delta^{1/2}_\m+\Delta^{1/2}\Box\s=n\Delta^{1/2}
\ee
with a boundary condition $\left[\Delta\right]=1$, and taking the derivative of \eqref{fundamental}, 
\be
2\s^\m_{~\a}\Delta^{1/2}_\m+2\s^\m\Delta^{1/2}_{\m\a}+\Delta^{1/2}_{\a}\Box\s+\Delta^{1/2}\Box\s_\a=n\Delta^{1/2}_\a
\ee
At coincidence, using former results about $\s$,  it implies 
\be
\boxed{\left[\Delta^{1/2}_\r\right]=0}
\ee

The second derivative of \eqref{fundamental} leads to
\bea
&&2\s^\m_{~\a\b}\Delta^{1/2}_\m+2\s^\m_{~\a}\Delta^{1/2}_{\m\b}+2\s^\m_{~\b}\Delta^{1/2}_{\m\a}+2\s^\m\Delta^{1/2}_{\m\a\b}+\Delta^{1/2}_{\a\b}\Box\s+\Delta^{1/2}_{\a}\Box\s_\b+\nonumber\\
&&+\Delta^{1/2}_\b\Box\s_\a+\Delta^{1/2}\Box\s_{\a\b}=n\Delta^{1/2}_{\a\b}
\eea
At coincidence
\be
n\left[\Delta^{1/2}_{\r\s}\right]+\left[\nabla_\r\nabla_\s\Box \s\right]+4 \left[\Delta^{1/2}_{\r\s}\right]= n\left[\Delta^{1/2}_{\r\s}\right]
\ee
so that
\be
\boxed{\left[ \Delta^{1/2}_{\a\b}\right]={1\over 6} R_{\a\b}}
\ee
Taking the trace 
\be
\boxed{\left[\Box \Delta^{1/2}\right]={1\over 6} R}
\ee

The third derivative of \eqref{fundamental} leads to
\bea
&&2\s^\m_{~\a\b\l}\Delta^{1/2}_\m+2\s^\m_{~\a\b}\Delta^{1/2}_{\m\l}+2\s^\m_{~\a\l}\Delta^{1/2}_{\m\b}+2\s^\m_{~\a}\Delta^{1/2}_{\m\b\l}+2\s^\m_{~\b\l}\Delta^{1/2}_{\m\a}+\nonumber\\
&&+2\s^\m_{~\b}\Delta^{1/2}_{\m\a\l}+2\s^\m_{~\l}\Delta^{1/2}_{\m\a\b}+2\s^\m\Delta^{1/2}_{\m\a\b\l}+\Delta^{1/2}_{\a\b\l}\Box\s+\Delta^{1/2}_{\a\b}\Box\s_\l+\nonumber\\
&&+\Delta^{1/2}_{\a\l}\Box\s_\b+\Delta^{1/2}_{\a}\Box\s_{\b\l}+\Delta^{1/2}_{\b\l}\Box\s_\a+\Delta^{1/2}_\b\Box\s_{\a\l}+\Delta^{1/2}_\l\Box\s_{\a\b}+\nonumber\\
&&+\Delta^{1/2}\Box\s_{\a\b\l}=n\Delta^{1/2}_{\a\b\l}
\eea
At coincidence this yields
\be
4\left[\Delta^{1/2}_{\a\b\l}\right]+2\left[\Delta^{1/2}_{\l\a\b}\right]+\left[\Box\s_{\a\b\l}\right]=0
\ee
Using the commutation of derivatives
\be \left[\Delta^{1/2}_{\l\a\b}\right]=\left[\Delta^{1/2}_{\a\b\l}\right]\ee
which together with the previous result for $\left[\Box\s_{\a\b\l}\right]$, implies that
\be
\boxed{\left[\Delta^{1/2}_{\a\b\l}\right]=\frac{1}{12}\left(\nabla_\a R_{\b\l}+\nabla_\b R_{\a\l}+\nabla_\l R_{\a\b}\right)}
\ee
whose trace reads
\be
\boxed{\left[\nabla_\l\Box\Delta^{1/2}\right]=\frac{1}{6}\nabla_\l R}
\ee
Finally, the fourth derivative of \eqref{fundamental}
\bea
&&2\s^\m_{~\a\b\l\t}\Delta^{1/2}_\m+2\s^\m_{~\a\b\l}\Delta^{1/2}_{\m\t}+2\s^\m_{~\a\b\t}\Delta^{1/2}_{\m\l}+2\s^\m_{~\a\b}\Delta^{1/2}_{\m\l\t}+2\s^\m_{~\a\l\t}\Delta^{1/2}_{\m\b}+\nonumber\\
&&+2\s^\m_{~\a\l}\Delta^{1/2}_{\m\b\t}+2\s^\m_{~\a\t}\Delta^{1/2}_{\m\b\l}+2\s^\m_{~\a}\Delta^{1/2}_{\m\b\l\t}+2\s^\m_{~\b\l\t}\Delta^{1/2}_{\m\a}+2\s^\m_{~\b\l}\Delta^{1/2}_{\m\a\t}+\nonumber\\
&&+2\s^\m_{~\b\t}\Delta^{1/2}_{\m\a\l}+2\s^\m_{~\b}\Delta^{1/2}_{\m\a\l\t}+2\s^\m_{~\l\t}\Delta^{1/2}_{\m\a\b}+2\s^\m_{~\l}\Delta^{1/2}_{\m\a\b\t}+2\s^\m_{ \t}\Delta^{1/2}_{\m\a\b\l}+\nonumber\\
&&+2\s^\m\Delta^{1/2}_{\m\a\b\l\t}+\Delta^{1/2}_{\a\b\l\t}\Box\s+\Delta^{1/2}_{\a\b\l}\Box\s_\t+\Delta^{1/2}_{\a\b\t}\Box\s_\l+\Delta^{1/2}_{\a\b}\Box\s_{\l\t}+\nonumber\\
&&+\Delta^{1/2}_{\a\l\t}\Box\s_\b+\Delta^{1/2}_{\a\l}\Box\s_{\b\t}+\Delta^{1/2}_{\a\t}\Box\s_{\b\l}+\Delta^{1/2}_{\a}\Box\s_{\b\l\t}+\Delta^{1/2}_{\b\l\t}\Box\s_\a+\nonumber\\
&&+\Delta^{1/2}_{\b\l}\Box\s_{\a\t}+\Delta^{1/2}_{\b\t}\Box\s_{\a\l}+\Delta^{1/2}_\b\Box\s_{\a\l\t}+\Delta^{1/2}_{\l\t}\Box\s_{\a\b}+\Delta^{1/2}_\l\Box\s_{\a\b\t}+\nonumber\\
&&+\Delta^{1/2}_\t\Box\s_{\a\b\l}+\Delta^{1/2}\Box\s_{\a\b\l\t}=n\Delta^{1/2}_{\a\b\l\t}\eea
At coincidence reads
\bea
&&2\left[\s^\m_{~\a\b\l}\Delta^{1/2}_{\m\t}\right]+2\left[\s^\m_{~\a\b\t}\Delta^{1/2}_{\m\l}\right]+2\left[\s^\m_{~\a\l\t}\Delta^{1/2}_{\m\b}\right]+2\left[\Delta^{1/2}_{\a\b\l\t}\right]+\nonumber\\
&&+2\left[\s^\m_{~\b\l\t}\Delta^{1/2}_{\m\a}\right]+2\left[\Delta^{1/2}_{\b\a\l\t}\right]+2\left[\Delta^{1/2}_{\l\a\b\t}\right]+2\left[\Delta^{1/2}_{\t\a\b\l}\right]+\left[\Delta^{1/2}_{\a\b}\Box\s_{\l\t}\right]+\nonumber\\
&&+\left[\Delta^{1/2}_{\a\t}\Box\s_{\b\l}\right]+\left[\Delta^{1/2}_{\a\l}\Box\s_{\b\t}\right]+\left[\Delta^{1/2}_{\b\l}\Box\s_{\a\t}\right]+\left[\Delta^{1/2}_{\b\t}\Box\s_{\a\l}\right]+\nonumber\\
&&+\left[\Delta^{1/2}_{\l\t}\Box\s_{\a\b}\right]+\left[\Box\s_{\a\b\l\t}\right]=0\nonumber\\\eea
 Commutating  derivatives
\bea
\left[\Delta^{1/2}_{\l\a\b\t}\right]&&=\left[\Delta^{1/2}_{\a\b\l\t}\right]+\frac{1}{6}R_{\a\r\b\l}R^\r_{~\t}\nonumber\\
\left[\Delta^{1/2}_{\t\a\b\l}\right]&&=\left[\Delta^{1/2}_{\a\b\l\t}\right]+\frac{1}{6}R_{\b\r\l\t}R^\r_{~\a}+\frac{1}{6}R_{\a\r\l\t}R^\r_{~\b}+\frac{1}{6}R_{\a\r\b\t}R^\r_{~\l}\nonumber\\
\eea
leads to
\bea&&12\left[\Delta^{1/2}_{\a\b\l\t}\right]-\frac{2}{9}\left(R_{\a\b}R_{\l\t}+R_{\a\l}R_{\b\t}+R_{\a\t}R_{\b\l}\right)-\nonumber\\
&&-\frac{1}{9}\Big[\left(R^\m_{~\b\a\l}+R^\m_{~\l\a\b}+3R^\m_{~\a\b\l}\right)R_{\m\t}+\left(R^\m_{~\b\a\t}+R^\m_{~\t\a\b}+3R^\m_{~\a\b\t}\right)R_{\m\l}+\nonumber\\
&&+\left(R^\m_{~\l\a\t}+R^\m_{~\t\a\l}+3R^\m_{~\a\l\t}\right)R_{\m\b}+\left(R^\m_{~\l\b\t}+R^\m_{~\t\b\l}+3R^\m_{~\b\l\t}\right)R_{\m\a}\Big]+\nonumber\\
&&+\left[\Box\s_{\a\b\l\t}\right]=0\nonumber\\
\eea
 Using \eqref{6Dtrace}, we get the trace as
\be \left[\Box\Box\Delta^{1/2}\right]=\frac{1}{5}\Box R+\frac{1}{36}R^2-\frac{1}{30}R_{\m\n}^2-\frac{1}{30}R_{\m\n\a\b}^2\ee
%%%%%%%%%%%%%%%%%%%%%%%%%%%%%%%%%%%%%%%%%%%%%%%%%%%%%%%%%%%%%%%%%%%%%%%%%%%%%%%%%%%%%%%%%%%%%%%%%%%%%%%%%%%%%%%%%%%%%%%%%%%%%
\subsection{Schwinger-DeWitt coefficients.}
%%%%%%%%%%%%%%%%%%%%%%%%%%%%%%%%%%%%%%%%%%%%%%%%%%%%%%%%%%%%%%%%%%%%%%%%%%%%%%%%%%%%%%%%%%%%%%%%%%%%%%%%%%%%%%%%%%%%%%%%%5
Let us now come back towards the real thing, namely the computation of the coincidence limits of the Schwinger-DeWitt coefficients themselves. From the equation
\be\label{s0}
\s^\m \nabla_\m a_0=0
\ee
Deriving once
\be
\s^\m\,_\n\nabla_\m a_0+ \s^\m\nabla_\n\nabla_\m a_0=0
\ee
it follows
\be
\boxed{\left[\nabla_\m a_0\right]=0}
\ee
Deriving again
\be
\s^{\m}\,_{\n\l} \nabla_\m a_0+\s^\m\,_\n\nabla_\l \nabla_\m a_0+\s^\m\,_\l\nabla_\n\nabla_\m a_0+\s^\m \nabla_\l\nabla_\n\nabla_\m a_0=0
\ee
we get
\be
\left[\left(\nabla_\m\nabla_\n+\nabla_\n\nabla_\m\right) a_0\right]=0
\ee
Obviously 
\be \nabla_\m\nabla_\n a_0=\nabla_\n\nabla_\m a_0\ee
so that
\be
\boxed{\left[\nabla_\m\nabla_\n a_0\right]=0}
\ee
Whose trace is 
\be
\boxed{\left[\Box a_0\right]=0}
\ee

The third derivative of \eqref{s0}
\bea
&&\s^{\m}_{~\n\l\t} \nabla_\m a_0+\s^{\m}_{~\n\l} \nabla_\t\nabla_{\m} a_0+\s^\m_{~\n\t}\nabla_\l \nabla_\m a_0+\s^\m_{~\n}\nabla_\t\nabla_\l \nabla_\m a_0+\nonumber\\
&&+\s^\m_{~\l\t}\nabla_\n\nabla_\m a_0+\s^\m_{~\l}\nabla_\t\nabla_\n\nabla_\m a_0+\s^\m_{~\t} \nabla_\l\nabla_\n\nabla_\m a_0+\s^\m \nabla_\t\nabla_\l\nabla_\n\nabla_\m a_0=0\nonumber\\
\eea
The fourth derivative of \eqref{s0}
\bea
&&\s^{\m}_{~\n\l\t\a} \nabla_\m a_0+\s^{\m}_{~\n\l\t}\nabla_\a \nabla_\m a_0+\s^{\m}_{~\n\l\a} \nabla_\t\nabla_{\m} a_0+\s^{\m}_{~\n\l} \nabla_\a\nabla_\t\nabla_{\m} a_0+\nonumber\\
&&+\s^\m_{~\n\t\a}\nabla_\l \nabla_\m a_0+\s^\m_{~\n\t}\nabla_\a\nabla_\l \nabla_\m a_0+\s^\m_{~\n\a}\nabla_\t\nabla_\l \nabla_\m a_0+\nonumber\\
&&+\s^\m_{~\n}\nabla_\a\nabla_\t\nabla_\l \nabla_\m a_0+\s^\m_{~\l\t\a}\nabla_\n\nabla_\m a_0+\s^\m_{~\l\t}\nabla_\a\nabla_\n\nabla_\m a_0+\nonumber\\
&&+\s^\m_{~\l\a}\nabla_\t\nabla_\n\nabla_\m a_0+\s^\m_{~\l}\nabla_\a\nabla_\t\nabla_\n\nabla_\m a_0+\s^\m_{~\t\a} \nabla_\l\nabla_\n\nabla_\m a_0+\nonumber\\
&&+\s^\m_{~\t} \nabla_\a\nabla_\l\nabla_\n\nabla_\m a_0+\s^\m_{~\a} \nabla_\t\nabla_\l\nabla_\n\nabla_\m a_0+\s^\m \nabla_\a\nabla_\t\nabla_\l\nabla_\n\nabla_\m a_0=0\nonumber\\
\eea
At coincidence 
\be \left[\nabla_\m\nabla_\t\nabla_\l\nabla_\n a_0+\nabla_\m\nabla_\t\nabla_\n\nabla_\l a_0+\nabla_\m\nabla_\l\nabla_\n\nabla_\t a_0+\nabla_\t\nabla_\l\nabla_\m\nabla_\n a_0\right]=0
\ee
Ricci's identities  imply that
\bea 
&&\left[\nabla_\m\nabla_\t\nabla_\n\nabla_\l a_0\right]=\left[\nabla_\m\nabla_\t\nabla_\l\nabla_\n a_0\right]\nonumber\\
&&\left[\nabla_\m\nabla_\l\nabla_\n\nabla_\t a_0\right]=\left[\nabla_\m\nabla_\t\nabla_\l\nabla_\n a_0\right]\nonumber\\
&&\left[\nabla_\t\nabla_\l\nabla_\m\nabla_\n a_0\right]=\left[\nabla_\m\nabla_\t\nabla_\l\nabla_\n a_0\right]
\eea
then 
\be \left[\nabla_\m\nabla_\n\nabla_\a\nabla_\b a_0\right]=0\ee
so that
\be
\boxed{\left[\Box\Box a_0\right]=0}
\ee

\par
The relevant expansion has been worked out by Bryce DeWitt, \cite{DeWitt}. Let us proceed in a pedestrian way, by writing
\be
K\left(\t;x,x^\prime\right)={1\over (4\pi\t)^{n\over 2}}N(x,x^\prime)~e^{-{\s(x,x^\prime)\over 2 \t}}~\sum_{p=0}^\infty a_p (x,x^\prime) \t^p
\ee
where we have left an arbitrary  global coefficient, to be determined later, $N(x,x^\prime)$ in front of the Taylor expansion. 
The purpose here is to show that it should be equal to the van Vleck determinant. In order to do that, let us now substitute the short time expansion into the heat equation 
\bea
&&{\pd \over \pd \t}~K\left(\t;x,x^\prime\right) = {1\over (4\pi)^{n\over 2}}~N~e^{-{\s\over 2\t}}~\sum_{p=0}\left(a_p{\s\over 2} +\left(p-1-{n\over 2}\right)a_{p-1}\right)\t^{p-2-{n\over 2}}\nonumber\\
\eea
Let us do the computation for (minus) the ordinary laplacian
\bea
&&\nabla_\m K\left(\t;x,x^\prime\right)= {1\over  (4 \pi \t)^{n\over 2}}~e^{-{\s\over 2 \t}}~\sum_p\left(\nabla_\m ~N~a_p +N~\left(\nabla_\m a_p-{\s_\m\over 2\t}a_p\right)\right)\t^p\nonumber\\
\eea
\bea
&&\nabla^2 K\left(\t;x,x^\prime\right)={e^{-{\s\over 2 \t}}\over  (4 \pi )^{n\over 2}}~\sum_{p=0} ~\bigg\{(\nabla^2N) a_{p-2}+ 2 N^\m \nabla_\m a_{p-2} -\s^\m (N_\m) a_{p-1}+\nonumber\\
&&+N\Big({1\over 4} \s_\m \s^\m a_p- \s^\m \nabla_\m a_{p-1}-{1\over 2}(\Box\s) a_{p-1}+\nabla^\m\nabla_\m a_{p-2}\Big)\bigg\}\t^{p-2-{n\over 2}}
\eea
where we have defined 
\be
a_p=0
\ee
for negative values of the index $p$.

The heat kernel equation provides the recursion relation
\bea
&&(\nabla^\m\nabla_\m N) a_{p-2}+ 2 \nabla^\m N \nabla_\m a_{p-2}-\s^\m (\nabla_\m N) a_{p-1}+N\Bigg\{{1\over 4} \s_\m \s^\m a_p- \nonumber\\
&&-\s^\m \nabla_\m a_{p-1}-{1\over 2}(\Box\s) a_{p-1}+\nabla^\m\nabla_\m a_{p-2}-a_p{\s\over 2} -\left(p-1-{n\over 2}\right)a_{p-1}\Bigg\}=0
\nonumber\\\eea
We can simplify this expression with our former result like
\bea &&\s_\m \s^\m=2\s\nonumber\\ 
&&\Box\s=n\eea
and the parallel transport
\be\s^\m\nabla_\m N=0\ee
\bea
&&(\nabla^\m\nabla_\m N) a_{p-2}+ 2 \nabla^\m N \nabla_\m a_{p-2}+N\nabla^\m\nabla_\m a_{p-2}-N\s^\m \nabla_\m a_{p-1}-\nonumber\\
&&-N\left(p-1\right)a_{p-1}=0
\nonumber\\\eea
This can also be written in terms of $\Delta\equiv N^2$ as
\bea\label{rr}
&&\Box(\Delta^{1/2}a_p)-\Delta^{1/2}\s^\m \nabla_\m a_{p+1}-\Delta^{1/2}\left(p+1\right)a_{p+1}=0\eea
In the simplest case $p=0$, we have
\bea\label{p0}
&&\Box(\Delta^{1/2}a_0)-\Delta^{1/2}\s^\m \nabla_\m a_{1}-\Delta^{1/2}a_{1}=0\eea
At the coincidence limit
\bea
&&\left[\Box\Delta^{1/2}\right]-\left[a_{1}\right]=0\eea
then
\be\boxed{\left[a_{1}\right]=\frac{1}{6}R}\ee
Taking the derivative of  \eqref{p0}
\bea
&&\nabla_\a\Box(\Delta^{1/2}a_0)-\nabla_\a\Delta^{1/2}\s^\m \nabla_\m a_{1}-\Delta^{1/2}\s^\m_{~\a} \nabla_\m a_{1}-\Delta^{1/2}\s^\m \nabla_\a \nabla_\m  a_{1}-\nonumber\\
&&-\nabla_\a(\Delta^{1/2})a_{1}-\Delta^{1/2}\nabla_\a a_{1}=0\eea
At coincidence 
\bea
&&\left[\nabla_\a\Box(\Delta^{1/2}a_0)\right]-2\left[ \nabla_\a a_{1}\right]=0\eea
then
\be\left[\nabla_\a a_{1}\right]=\frac{1}{12}\nabla_\a R\ee
Deriving again
\bea
&&\nabla_\b\nabla_\a\Box(\Delta^{1/2}a_0)-\nabla_\b\nabla_\a\Delta^{1/2}\s^\m \nabla_\m a_{1}-\nabla_\a\Delta^{1/2}\s^\m_{~\b} \nabla_\m a_{1}-\nonumber\\
&&-\nabla_\a\Delta^{1/2}\s^\m \nabla_\b\nabla_\m a_{1}-\nabla_\b\Delta^{1/2}\s^\m_{~\a} \nabla_\m a_{1}-\Delta^{1/2}\s^\m_{~\a\b} \nabla_\m a_{1}-\nonumber\\
&&-\Delta^{1/2}\s^\m_{~\a} \nabla_\b\nabla_\m a_{1}-\nabla_\b\Delta^{1/2}\s^\m \nabla_\a \nabla_\m  a_{1}-\Delta^{1/2}\s^{\m}_{~\b} \nabla_\a \nabla_\m  a_{1}-\nonumber\\
&&-\Delta^{1/2}\s^\m \nabla_\b\nabla_\a \nabla_\m  a_{1}-\nabla_\b\nabla_\a(\Delta^{1/2})a_{1}-\nabla_\a(\Delta^{1/2})\nabla_\b a_{1}-\nonumber\\
&&-\nabla_\b\Delta^{1/2}\nabla_\a a_{1}-\Delta^{1/2}\nabla_\b\nabla_\a a_{1}=0\eea
At coincidence 
\bea
&&\left[\nabla_\b\nabla_\a\Box(\Delta^{1/2}a_0)\right]-3\left[\nabla_\a\nabla_\b a_{1}\right]-\left[\nabla_\b\nabla_\a(\Delta^{1/2})a_{1}\right]=0\eea
whose trace reads
\be \left[\Box a_{1}\right]=\frac{1}{15}\Box R+\frac{1}{90}R_{\m\n\a\b}^2-\frac{1}{90}R_{\m\n}^2\ee
Next, we take $p=1$ in the recursion relation
\bea\label{p1}
&&\Box(\Delta^{1/2}a_1)-\Delta^{1/2}\s^\m \nabla_\m a_{2}-2\Delta^{1/2}a_{2}=0\eea
The coincidence limit reads
\bea
&&\left[\Box(\Delta^{1/2}a_1)\right]-2\left[a_{2}\right]=0\eea
Collecting all results
\be \boxed{\left[a_2\right]=\frac{1}{30}\Box R+\frac{1}{180}R_{\m\n\a\b}^2-\frac{1}{180}R_{\m\n}^2+\frac{1}{72}R^2}
\ee
\newpage
%%%%%%%%%%%%%%%%%%%%%%%%%%%%%%%%%%%%%%%%%%%%%%%%%%%%%%%%%%%%%%%%%%%%%%%%%%%%%%%%%%%%%%%%%%%%%%%%%%%%%%%%%%%%%%%%%%%%%%%%%%%%%
\section{Recursion relations for the coefficients of the short time expansion of the heat kernel.}
%%%%%%%%%%%%%%%%%%%%%%%%%%%%%%%%%%%%%%%%%%%%%%%%%%%%%%%%%%%%%%%%%%%%%%%%%%%%%%%%%%%%%%%%%%%%%%%%%%%%%%%%%%%%%%%%%%%%%%%%%%%%%
There are some recursion relations obtained by Gilkey \cite{Gilkey} in a remarkable paper. They greatly simplify the computation of the Schwinger-DeWitt coefficients. We are refering to an operator on a riemannian manifold, $M$,
\be
\Delta\equiv -\left(g^{\m\n}\pd_\m\pd_\n+ A^\s \pd_\s+ B\right)
\ee
%Also
%\bea
%&D(\nabla,E)\equiv -\left(g^{\m\n}\nabla_\m\nabla_\n + E\right)=-\left({1\over 2} g^{\m\n}\left(\nabla_\m\nabla_\n+\nabla_\n\nabla_\m\right)+E\right)=\nonumber\\
%&=-\left({1\over 2} g^{\m\n}\left(\pd_\m\left(\pd_\n+\omega_\n\right)-\Gamma^\l_{\m\n}\left(\pd_\l+\omega_\l\right)+\omega_\m\left(\pd_\n+\omega_\n\right)+(\m\n)\right)+E\right)=\nonumber\\
%&-\bigg({1\over 2} g^{\m\n}\left(\pd_\m\pd_\n+\pd_\m\omega_\n+\omega_\n\pd_\m-\Gamma^\l_{\m\n}\pd_\l-\Gamma^\l_{\m\n}\omega_\l+\omega_\m\pd_\n+\omega_\m\omega_\n\right.+\nonumber\\
%&\left.+\pd_\m\pd_\n+\pd_\n\omega_\m+\omega_\m\pd_\n-\Gamma^\l_{\m\n}\pd_\l-\Gamma^\l_{\m\n}\omega_\l+\omega_\n\pd_\m+\omega_\n\omega_\m\right)+ E\bigg)
%\eea

%It follows that
%\bea
%&2 \omega^\l -g^{\m\n}\Gamma^\l_{\m\n}= A^\l\nonumber\\
%&\pd_\m \omega^\m-\Gamma^\l\omega_\l+\omega_\m\omega^\m+E=B
%\eea

%For example,
%\be
%D\equiv -\nabla_\m \nabla^\m=-\left(g^{\m\n}\pd_\m\pd_\n-\Gamma^\l \pd_\l\right)\Longrightarrow A^\l=-\Gamma^\l\quad B=0
%\ee
%If we define $G_\m\equiv g_{\m\l}\Gamma^\l$,
%\bea
%&D\equiv -g^{\m\n} {1\over 2}\bigg\{\left(\pd_\m-G_\m\right)\left(\pd_\n-G_\n\right)-\Gamma_{\m\n}^\l\left(\pd_\l-G_\l\right)+(\n\m)\bigg\}=\nonumber\\
%\eea
There is a unique endomorphism $E$ allowing us to rewrite the operator as
\be \Delta=-(g^{\m\n}\nabla_\m\nabla_\n+E)\ee
where
\bea
&&A^\l=2 \omega^\l -g^{\m\n}\Gamma^\l_{\m\n}\nonumber\\
&&B=\pd_\m \omega^\m-g^{\m\n}\Gamma^\l_{\m\n}\omega_\l+\omega_\m\omega^\m+E
\eea

First of all, it can be proved that the general form of the Schwinger-DeWitt coefficients is
\bea
a_0(f,\Delta)&&=\int d^n x\sqrt{g}\,\tr\left\{f(\a_0)\right\}
\nonumber\\
a_2(f,\Delta)&&={1\over 6}\int d^n x\sqrt{g}\,\tr\left\{f(\a_1 E+\a_2 R)\right\}
\nonumber\\
a_4(f,\Delta)&&={1\over 360}\int d^n x\sqrt{g}\,\tr\Big\{f(\a_3 \Box E+ \a_4 R E+\a_5 E^2+\a_6 \Box R+\nonumber\\
&&+\a_7 R^2+\a_8 R_{\m\n}^2+\a_9 R_{\m\n\r\s}^2+\a_{10} W_{\m\n}^2)\Big\}
\eea

There are three lemmas that we are going to use in the sequel in order to determine $\a_i$.

{\bf Lemma 1.}
\be
\boxed{\left.{d\over d\e}a_m\left(e^{-2 \e f}\Delta\right)\right|_{\e=0}=(n-m)a_m\left(f,\Delta\right)}\label{l1}
\ee
%where
%\bea
%&a_m(f,D)\equiv \tr\,\left(f(x) a_n(x,D)\right)\nonumber\\
%&\tr\,\left(F\, e^{- t D}\right)\equiv\sum_p a_p(F,D)\, t^{p-n\over 2}
%\eea
\begin{proof}
Consider the family
\be
\Delta(\e)\equiv e^{2 \e f} \Delta-\e F
\ee
it so happens that
\be
\left.{d\over d\e}\Delta(\e)\right|_{\e=0}=-2 f \Delta-F
\ee
now we can use the lemma in Gilkey's book \cite{Gilkey}, page 78, asserting that whenever $[P,Q]=0$,
\be
\left.{d\over d \e}\tr\, Q(\e) e^{-t P(\e)}\right|_{\e=0}=\tr\, \left(-t \dot{P} Q+ \dot{Q}\right)\,e^{-t P}
\ee
where $Q\equiv Q(0)$, $P\equiv P(0)$. It follows that
\bea
\left.{d\over d \e}\tr\, e^{-t \Delta(\e)}\right|_{\e=0}&&=t\, \tr\,\left( 2 f \Delta\, e^{-t \Delta}\right)+ t\,\tr\, \left(F e^{-t \Delta}\right)=\nonumber\\
&&=-2 t {\pd\over \pd t}\, \tr\,\left(  f \, e^{-t \Delta}\right)+ t\,\tr\, \left(F e^{-t \Delta}\right)
\eea
that is
\bea
&&\left.{d\over d \e}\sum a_p\left(e^{-2 \e f}\,\Delta-\e F\right)\right|_{\e=0}\,t^{p-n\over 2}=\nonumber\\
&&=\sum(p-n) a_p(f,\Delta)t^{p-n\over 2}+\sum a_p(F,\Delta)\,t^{p-n+2\over 2}
\eea
making $F=0$ we get the result
\end{proof}

{\bf Lemma 2.}
\be
\boxed{\left.{d\over d \e} a_p\left(\Delta-\e F\right)\right|_{\e=0}=a_{p-2}\left(F,\Delta\right)}\label{l2}
\ee
\begin{proof}
Consider now the operator
\be
\Delta_{\e,\d}\equiv e^{-2 f \e}\,\left(\Delta-\d F\right)
\ee
then
\bea
&&\pd_\e a_n\left(\Delta_{\e,\d}\right)=\pd_\d \pd_\e a_n\left(\Delta_{\e,\d}\right)=\pd_\e \pd_\d a_n\left(\Delta_{\e,\d}\right)=\nonumber\\
&&=\pd_\e a_{n-2}\left(e^{-2 f \e}F,e^{-2\e f} \Delta\right)=0
\eea
\end{proof}

{\bf Lemma 3.}
\be
\boxed{\left.{d\over d \e} a_{n-2}\left(e^{-2 f \e}F,e^{-2\e f} \Delta\right)\right|_{\e=0}=0}\label{l3}
\ee

Next, let us determine the coefficients $\a_i$
\begin{itemize}
	\item The coefficient $\a_0$ follows from the heat kernel expansion for the scalar laplacian, so that \boxed{\a_0=1}
	\item Now we use \eqref{l2} with $p=2$ 
	\be
	{1\over 6}\tr\,\left( \a_1 F\right)=\tr\, F
	\ee
	and we can directly extract \boxed{\a_1= 6}
	\item Taking now \eqref{l2} with $p=4$,
	\be
	{1\over 360}\tr\,\left(\a_4 F R+2 \a_5 F E\right)={1\over 6} \tr\,\left(\a_1 F E +\a_2 F R\right)\ee 
	from this equation we get \boxed{\a_5=180} \quad \text{and}\quad  \boxed{\a_4= 60 \a_2}	
\end{itemize}

To proceed further we have to take into account local scale transformations defined in \eqref{l1} and \eqref{l3}. A list of the relevant transformations reads 
 \bea\label{ee}
&&\left.\frac{d}{d\e}\bar{E}\right|_{\e=0}=-2f E+\frac{n-2}{2}\Box f\nonumber\\
&&\left.\frac{d}{d\e}\bar{R}\right|_{\e=0}=-2f R-2(n-1)\Box f\nonumber\\
&&\left.\frac{d}{d\e}\bar{\Box}\bar{E}\right|_{\e=0}=-4f \Box E-2E\Box f+\frac{n-2}{2}\Box^2 f-2\nabla_\m f\nabla^\m E\nonumber\\
&&\left.\frac{d}{d\e}\bar{R}\bar{E}\right|_{\e=0}=-4f R E+\frac{n-2}{2}R\Box f-2(n-1)E\Box f\nonumber\\
&&\left.\frac{d}{d\e}\bar{E}^2\right|_{\e=0}=-4f  E^2+(n-2)E\Box f\nonumber\\
&&\left.\frac{d}{d\e}\bar{\Box}\bar{R}\right|_{\e=0}=-4f\Box R -2R\Box f-2(n-1)\Box^2 f-2\nabla_\m f\nabla^\m R\nonumber\\
&&\left.\frac{d}{d\e}\bR^2\right|_{\e=0}=-4fR^2 -4(n-1)R\Box f\nonumber\\
&&\left.\frac{d}{d\e}\bR_{\m\n}^2\right|_{\e=0}=-4fR_{\m\n}^2-2R\Box f -2(n-2)\nabla_\m\nabla_\n R\nabla^\m\nabla^\n f\nonumber\\
&&\left.\frac{d}{d\e}\bR_{\m\n\r\s}^2\right|_{\e=0}=-4fR_{\m\n\r\s}^2 -8\nabla_\m\nabla_\n R\nabla^\m\nabla^\n f\nonumber\\
&&\left.\frac{d}{d\e}\bar{W}_{\m\n}^2\right|_{\e=0}=-4fW_{\m\n}^2 
\eea

Using these relations we can employ \eqref{l3} to compute the coefficients.
\begin{itemize}
	\item Applying \eqref{l3} to $(n,p)=(4,2)$ we have
	\be
\left.{d\over d\e}\,a_2\left(e^{-2\e f}\,F,e^{-2\e f}\,\Delta\right)\right|_{\e=0}=0
	\ee
	with \eqref{ee}, it follows that
	\be
	0=\tr\,\left(\a_1-6 \a_2\right)F\,\Box f
	\ee
	ergo \boxed{\a_2=1} and \boxed{\a_4=60} 
	\item Consider now a product metric $M=M_1\times M_2$ with laplacian $\Delta=\Delta_1 + \Delta_2$. It can be shown that
	\be
	a_4(\Delta)=\a_4(\Delta_1)+\a_4(\Delta_2)+\a_2(\Delta_1)\a_2(\Delta_2)
	\ee
	the only cross term comes from
	\be
	R^2(M_1 \times M_2)=R^2(M_1)+R^2(M_2)+2 R(M_1)R(M_2)
	\ee
	i.e.
	\be
	2{1\over 360}\a_7=\left({1\over 6} \a_2\right)^2
	\ee
	this means that \boxed{\a_7=5}
	\item Let us now apply again \eqref{l3} with $(n,p)=(6,4)$. It follows
	\be
	\left.\frac{d}{d\e} a_4\left(e^{-2\e f}\,F,e^{-2\e f} \Delta\right)\right|_{\e=0}=0
	\ee
	with \eqref{ee}, then
	\bea
	&0=\tr\,\bigg\{F\left(-2 \a_3-10\a_4 + 4 \a_5\right)E\Box f +\left(2 \a_3 -10\a_6\right)\Box^2 f+\nonumber\\
	&+\left(2\a_4-2 \a_6-20\a_7-2\a_8\right)R\Box f -8\left(\a_8+\a_9\right)\nabla_\m\nabla_\n R\nabla^\m\nabla^\n f\bigg\}\nonumber\\
	\eea
	we conclude 
	\boxed{\a_3=60}, \boxed{\a_6=12}, \boxed{\a_8=-2} and \boxed{\a_9=2}
\item In order to get $\a_{10}$ we shall follow Fujikawa's method as worked out by Nepomechie \cite{Nepomechie}\cite{Fujikawa}. Consider the operator {\em in flat space}
\be
\Delta\equiv -\left(\d^{\m\n} \nabla_\m\nabla_\n+E\right)
\ee
where
\be
\nabla_\m\equiv \pd_\m+A_\m
\ee

The heat kernel will be given by
\bea
K(\t;x,y)&&=\langle x\left|e^{-\t \Delta}\right|y\rangle=e^{-\t \Delta} \langle x|y\rangle=\nonumber\\
&&=\int {d^n k\over (2\pi)^n}\,e^{-\t \Delta}\langle x|k\rangle\langle k|y\rangle=\int {d^n k\over (2\pi)^n}\, e^{-i k y} e^{-\t \Delta} e^{ikx}\nonumber\\
\eea
Now it so happens that
\be
\nabla_\m e^{ikx}=e^{i kx}\,\left(ik_\m+\pd_\m+A_\m\right)=e^{i kx}\,\left(ik_\m+\nabla_\m\right)
\ee
as well as 
\bea
\nabla^\m\nabla_\m e^{ikx}&&=\nabla_\m\,e^{i kx}\,\left(ik_\m+\nabla_\m\right)=e^{ikx}\left(i k_\m+\nabla_\m\right)\left(i k^\m +\nabla^\m\right)=\nonumber\\
&&=e^{ikx}\left(- k^2+2 i k^\m\nabla_\m+\nabla^2\right)
\eea
ergo
\be
\Delta e^{ikx}=-\left(\nabla^\m\nabla_\m+E\right)e^{i k x}=e^{ikx}\left( k^2-2 i k.\nabla+\Delta\right)
\ee
Rescaling now
\be
k\rightarrow {k\over \sqrt{\t}}
\ee
we arrive at
\bea
K(\t\;x,y)&&={1\over \t^{n\over 2}}\int{ d^n k\over (2\pi)^n}\,e^{-k^2}\,e^{2 i\sqrt{\t} k.\nabla-\t \Delta} =\nonumber\\
&&={1\over \t^{n\over 2}}\int{ d^n k\over (2\pi)^n}\,e^{-k^2}\sum_{p=0}^\infty {\left(2 i\sqrt{\t} k.\nabla-\t \Delta\right)^p\over p!}
\eea

We would like to single out the power of $\t^2$ in the expansion. We notice that in the expansion of $\left(\sqrt{\t}A+ \t B\right)^p$ the coefficients of $\t^2$ come from the terms
\be
{1\over 4!} A^4+{1\over 3!}\left(A^2 B+ B A^2+ A B A\right)+ {1\over 2} B^2
\ee
where in our case
\bea
&A\equiv 2 i k^\m\nabla_\m\nonumber\\
&B\equiv -\Delta=\nabla_\m\nabla^\m+E
\eea
let us write those terms explicitly
\bea
&&\frac{2}{3}\Big[k^\m\nabla_\m k^\n\nabla_\n k^\r\nabla_\r k^\s\nabla_\s-k^\m\nabla_\m k^\n\nabla_\n\Box-k^\m\nabla_\m k^\n\nabla_\n E-\nonumber\\
&&-\Box k^\m\nabla_\m k^\n\nabla_\n-E k^\m\nabla_\m k^\n\nabla_\n-k^\m\nabla_\m \Box  k^\n\nabla_\n-k^\m\nabla_\m Ek^\n\nabla_\n\Big]+\nonumber\\
&&+\frac{1}{2}\Big[\Box^2+\Box E+e\Box+E^2\Big]
\eea
the momentum integrations are given by
\bea
&&\int {d^n k\over (2\pi)^n}e^{-k^2}=1\nonumber\\
&&\int {d^n k\over (2\pi)^n}e^{-k^2}k_\m k_\n={1\over 2} \d_{\m\n}\nonumber\\
&&\int {d^n k\over (2\pi)^n}e^{-k^2}k_\m k_\n k_\a k_\b={1\over 4}\left(\d_{\m\n}\d_{\a\b}+\d_{\m\a}\d_{\n\b}+\d_{\m\b}\d_{\n\a}\right)
\eea
therefore
\bea
a_4&&\rightarrow\frac{1}{6}\Big[\Box^2+\nabla^\m\nabla^\n\nabla_\m\nabla_\n+\nabla^\m\Box\nabla_\m\Big]-\nonumber\\
&&-\frac{1}{3}\Big[2\Box^2+\Box E+E\Box+\nabla^\m\Box\nabla_\m+\nabla_\m E\nabla^\m\Big]+\nonumber\\
&&+\frac{1}{2}\Big[\Box^2+\Box E+E\Box+E^2\Big]
\eea
ignoring the terms that can be taken as surface terms and combining the derivatives
\be
a_4\rightarrow\left({1\over 2} E^2+{1\over 12} W_{\m\n}^2+{1\over 6}\Box E\right)
\ee
which implies that $\boxed{\a_{10}=30}$.
\end{itemize}

In conclusion the heat kernel coefficients are
\bea
a_0(f,D)&&=1
\nonumber\\
a_2(f,D)&&={1\over 6}\int d^n x\sqrt{g}\,\tr\left\{6 E+ R\right\}
\nonumber\\
a_4(f,D)&&={1\over 360}\int d^n x\sqrt{g}\,\tr\Big\{60 \Box E+ 60 R E+180 E^2+12 \Box R+\nonumber\\
&&+5R^2-2 R_{\m\n}^2+2 R_{\m\n\r\s}^2+30 W_{\m\n}^2\Big\}
\eea
in flat space we recover
\bea
\left[a_0\right]&&=1\nonumber\\
\left[a_2\right]&&=-Y\nonumber\\
\left[a_4\right]&&=\frac{1}{12}W_{\m\n}^2+\frac{1}{2}Y^2-\frac{1}{6}\Box Y
\eea
where $E=-Y$.

\newpage
%%%%%%%%%%%%%%%%%%%%%%%%%%%%%%%%%%%%%%%%%%%%%%%%%%%%%%%%%%%%%%%%%%%%%%%%%%%%%%%%%%%%%%%%%%%%%%%%%%%%%%%%%%%%%%%%%%%%
\section*{Acknowledgements}
%%%%%%%%%%%%%%%%%%%%%%%%%%%%%%%%%%%%%%%%%%%%%%%%%%%%%%%%%%%%%%%%%%%%%%%%%%%%%%%%%%%%%%%%%%%%%%%%%%%%%%%%%%%%%%%%%%%%%%%%%%
We acknowledge partial financial support by the Spanish MINECO through the Centro de excelencia Severo Ochoa Program  under Grant CEX2020-001007-S  funded by MCIN/AEI/10.13039/501100011033
We also acknowledge the European Union's Horizon 2020 research and innovation programme under the Marie Sklodowska-Curie grant agreement No 860881-HIDDeN and also byGrant PID2019-108892RB-I00 funded by MCIN/AEI/ 10.13039/501100011033 and by ``ERDF A way of making Europe''

%%%%%%%%%%%%%%%%%%%%%%%%%%%%%%%%%%%%%%%%%%%%%%%%%%%%%%%%%%%%%%%%%%%%%%%%%%%%%%%%%%%%%%%%%%%%%%%%%%%%%%%%%%%%%%%%%%%%%%

\newpage
%%%%%%%%%%%%%%%%%%%%%%%%%%%%%%%%%%%%%%%%%%%%%%%%%%%%%%%%%%%%%%%%%%%%%%%%%%%%%%%%%%%%%%%%%%%%%%%%%%%%%%%%%%%%%%%%%%%%%%%%%%%%%%%
%\chapter{Windows onto the future.}
%%%%%%%%%%%%%%%%%%%%%%%%%%%%%%%%%%%%%%%%%%%%%%%%%%%%%%%%%%%%%%%%%%%%%%%%%%%%%%%%%%%%%%%%%%%%%%%%%%%%%%%%%%%%%%%%%%%%%%%%%%%%%

\newpage
\appendix
%%%%%%%%%%%%%%%%%%%%%%%%%%%%%%%%%%%%%%%%%%%%%%%%%%%%%%%%%%%%%%%%%%%%%%%%%%%%%%%%%%%%%%%%%%%%%%%%%%%%%%%%%%%%%%%%%%%%
\section{One-loop divergences in first order Einstein-Hilbert gravity}
%%%%%%%%%%%%%%%%%%%%%%%%%%%%%%%%%%%%%%%%%%%%%%%%%%%%%%%%%%%%%%%%%%%%%%%%%%%%%%%%%%%%%%%%%%%%%%%%%%%%%%%%%%%%%%%%%%%%%%%%%%
In this appendix we include a {\em final project} that all serious students should undertake. Computations are much less painful  with the use of the xTensor package of Mathematica, or any similar algebraic package. This is the renormalization of General Relativity, something  first done (by hand) by 't Hooft and Veltman in a classic paper. Here we shall follow \cite{AASG} for a first order formulation.

There are many ways to write first order actions which are fully equivalent (at least classically) to the corresponding second order ones. They are usually more elaborate than the {\em naive} first order formalism, where the metric and the connection are treated as independent fields, insofar as they often need the introduction of auxiliary fields as Lagrange multipliers. This is a general mathematical result, valid for any system of ordinary differential equations.
\par
It is well-known that when considering the Palatini version of the Einstein-Hilbert action, that is, the {\em naive} first order formalism at the classical level, the connection is required to be the Levi-Civita one once the equations of motion are imposed. However, when more general quadratic in curvature metric-affine actions are considered in the first order approach, this relationship disappears even on-shell, hence allowing for more general connections. That is, the equations of motion do not force the connection to be the Levi-Civita one. This is of particular interest when analyzing quadratic theories in first order formalism, as these theories are quadratic in the derivatives of the connection and thus, no propagator decays faster than $\frac{1}{p^\text{{\tiny 2}}}$, indicating that there is still room for the theory to be unitary. Quadratic theories of gravity are renormalizable as opposed to General Relativity (cf. \cite{review_enrique} for a general review of quantum gravity and some approaches to the problem), so the study of this kind of theories in the first order formalism could give rise to a unitary and renormalizable theory of gravity \cite{quadratic_enrique,physicalcontent}. In this context, the computation to one-loop order for the linear Einstein-Hilbert action paves the way for future studies of more complex theories. 
\par
This section aims to revisit the computation of the one-loop quantum corrections to the gravitational action in the naive first order formalism. The action is assumed to be still the Einstein-Hilbert one, with the  Riemann tensor given solely in terms of the connection field and with the addition of a minimally coupled massless scalar field. We shall find exactly 't Hoot and Veltman's counterterm \cite{tHooft} even in the presence of a scalar field, and even off-shell, which is a result stronger than the one guaranteed by the general theorems of quantum field theory that only assert on-shell equality.

\hrulefill

We consider the Einstein-Hilbert action with the metric coupled to a massless scalar field $\phi$ given by
\bea
S_{\text{\tiny{EH$\phi$}}}&&\equiv -\frac{1}{2}\int
d^n x\sqrt{|g|}g^{\m\n}R_{\m\n}+\frac{1}{2}\int
d^n x\sqrt{|g|}\frac{1}{2}g^{\m\n}\partial_\m\phi\partial_\n\phi,
\eea
where we take the gravitational coupling constant as unity, namely, $\kappa=1$.
%%%%%%%%%%%%%%%%%%%%%%%%%%%%%%%%%%%%%%%%%%%%%%%%%%%%%%%%%%%%%%%%%%%%%%%%%%%%%%%%%%%%%%%%%%%%
%\subsection{Background field expansion}
%%%%%%%%%%%%%%%%%%%%%%%%%%%%%%%%%%%%%%%%%%%%%%%%%%%%%%%%%%%%%%%%%%%%%%%%%%%%%%%%%%%%%%%%%%%%
The metric, the connection  and the scalar field are treated as independent fields and  are expanded in a background field and a perturbation as
\bea &&g_{\m\n}=\bg_{\m\n}+ h_{\m\n},\nonumber\\
&&\Gamma^\l_{\m\n}=\bar{\Gamma}^\l_{\m\n}+ A^\l_{\m\n},\nonumber\\
&&\phi=\bph+\phi.
\eea
Let us note that indices are raised with the background metric, and the quantities computed with respect to this metric have a bar. We also take as the background connection the Levi-Civita connection built from the background metric.\footnote{In this sense, the equation of motion for the connection field is already imposed from the beginning of the computation.}
As usual, linear terms cancel provided the classical fields obey the  background equations of motion  given by
\bea \label{EM}&&\frac{1}{2}\bg_{\m\n}\bR-\bR_{\m\n}-\frac{1}{4}\bg_{\m\n}\partial_\l\bph\partial^\l\bph+\frac{1}{2}\partial_\m\bph\partial_\n\bph=0,\nonumber\\ 
&&\bg^{\a\b}\bar{\nabla}_\lambda \bg_{\a\b}=0, \nonumber \\
&&\bar{\Box}\bph=0.
\eea
To take into account the one-loop effects it is enough to expand the action up to quadratic order in the perturbations so that this piece reads
\bea 
\label{BF}S_{\text{\tiny{2}}}&&=\int
d^n x~\sqrt{|\bg|}~\left\{h^{\a\b} M_{\a\b\g\e} h^{\g\e}+h^{\a\b}\vec{N}_{\a\b~\t}^{~~~\g\e}A^{\t}_{\g\e}-A^{\lambda}_{\alpha\beta}\vec{N}_{\l~\g\e}^{\a\b~~}h^{\g\e}+\right.\nonumber\\
&&\left.+A^{\l}_{\a\b}K^{\a\b~\g\e}_{~\l~~\t}A^{\t}_{\g\e}+h^{\a\b}E_{\a\b}\phi+\phi F \phi\right\}.
\eea
The symbol $\vec{N}$ is used to indicate the fact that the derivative contained in the operator acts on the right. The explicit expression for these operators is then
\bea\label{K}
M_{\a\b\g\e}&&= \frac{1}{16}\left(\bg_{\a\e}\bg_{\b\g}+\bg_{\a\g}\bg_{\b\e}-\bg_{\a\b}\bg_{\g\e}\right)\left(\bR-\frac{1}{2}\bg^{\r\s}\pd_\r\bph\pd_\s\bph\right)
+\nonumber\\
&&+\frac{1}{8}\left(\bg_{\a\b}\bR_{\g\e}+\bg_{\g\e}\bR_{\a\b}-\bg_{\a\g}\bR_{\b\e}-\bg_{\a\e}\bR_{\b\g}-\bg_{\b\g}\bR_{\a\e}-\bg_{\b\e}\bR_{\a\g}\right)-\nonumber\\
&&-\frac{1}{16}\left(\bg_{\a\b}\pd_\g\bph\pd_\e\bph+\bg_{\g\e}\pd_\a\bph\pd_\b\bph-\bg_{\a\g}\pd_\b\bph\pd_\e\bph-\bg_{\a\e}\pd_\b\bph\pd_\g\bph-\right.\nonumber\\
&&\left.-\bg_{\b\g}\pd_\a\bph\pd_\e\bph-\bg_{\b\e}\pd_\a\bph\pd_\g\bph\right)\nonumber\\
N_{\g\e~\l}^{~~\a\b}&&=\frac{1}{8}\left(\d^\a_\g\delta^{\beta}_{\e}+\d^\a_\e\delta^{\beta}_{\g}-\bg_{\g\e}\bg^{\a\b}\right)\bar{\nabla}_\lambda
+\nonumber\\
&&+\frac{1}{16}\left(\bg_{\g\e}\d^\b_\l\bar{\nabla}^\a -\d^\a_\g\delta^{\beta}_{\lambda}\bar{\nabla}_\e-\d^\a_\e\delta^{\beta}_{\lambda}\bar{\nabla}_\g
+\bg_{\g\e}\d^\a_\l\bar{\nabla}^\b -\d^\b_\g\delta^{\a}_{\lambda}\bar{\nabla}_\e-\d^\b_\e\delta^{\a}_{\lambda}\bar{\nabla}_\g\right) \nonumber\\
K^{\g\e~\a\b}_{~\t~~\l}&&= \frac{1}{8}\Big[
\d^\b_\t \d^\g_\l \bg^{\a\e}
+\d^\b_\t \d^\e_\l \bg^{\a\g}
+\d^\a_\t \d^\e_\l \bg^{\b\g}
+\d^\a_\t \d^\g_\l \bg^{\b\e}-\d^\e_\t \d^\g_\l \bg^{\a\b}-\d^\g_\t \d^\e_\l \bg^{\a\b}-\nonumber\\
&&-\d^\b_\l \d^\a_\t \bg^{\g\e} -\d^\a_\l \d^\b_\t \bg^{\g\e}\Big]
\nonumber\\
E_{\a\b}&&=\frac{1}{4}\bg_{\a\b}\bg^{\r\s}\pd_\r\bph\pd_\s-\frac{1}{4}\pd_\a\bph\pd_\b-\frac{1}{4}\pd_\b\bph\pd_\a\nonumber\\
F&&=-\frac{1}{4}\bar{\Box}\nonumber\\
\eea

In order to compute the counterterm we need to take the effective action as the starting point, which in this case depends on the three fields appearing in the theory
\bea
e^{iW \scriptstyle\left[\bg_{\m\n},\bG^\l_{\r\s}, \bph\scriptstyle\right]}&&=\int \mathcal{D}h\mathcal{D}A \mathcal{D} \phi~e^{iS_{\text{\tiny 2}}[h,A,\phi]}\eea
Taking advantage of the form of the background expansion (\ref{BF}), we can rewrite the metric and connection pieces by completing squares as
\be hMh+h\vec{N}A-A\vec{N}h+AKA
=hMh+[h\vec{N}+AK]K^{-1}[-\vec{N}h+KA]+h\vec{N}K^{-1}\vec{N}h.
\label{integrationbyparts}
\ee 
Due to the translational invariance of the integration measure, we can redefine the connection field perturbations so that the second term in \eqref{integrationbyparts} becomes quadratic in those. The integral over the quantum connection fields, ${\cal D}A$, is then a trivial gaussian integral yielding
\bea
&&e^{iW}=\int \mathcal{D}h \mathcal{D}\phi~e^{\left\{i\int
	d^n x~\sqrt{|g|}~h^{\a\b}\left(M_{\a\b\g\e}+D_{\a\b\g\e}\right)h^{\g\e}+h^{\a\b}E_{\a\b}\phi+\phi F \phi
	\right\}},\label{PI}\eea
where we have defined the new piece of the quadratic operator mediating between the metric perturbations as $D_{\a\b\g\e}\equiv \vec{N}_{\a\b~\l}^{~~~\m\n}(K^{-1})_{\m\n~\r\s}^{~\l~~\t}\vec{N}_{~\t~\g\e}^{\r\s}$.
\par
This integration before fixing the gauge is only possible if we are able to invert the operator $K$. To be specific, 
\bea
(K^{-1})_{\a\b~\g\e}^{~\l~~\t}&&=-\dfrac{2}{n-2}\left\{ \delta_{\gamma}{}^{\tau} \delta_{\epsilon}{}^{\lambda} \bg_{\alpha \beta} +  \delta_{\gamma}{}^{\lambda} \delta_{\epsilon}{}^{\tau} \bg_{\alpha \beta} + \delta_{\alpha}{}^{\tau} \delta_{\beta}{}^{\lambda} \bg_{\gamma \epsilon} +  \delta_{\alpha}{}^{\lambda} \delta_{\beta}{}^{\tau} \bg_{\gamma \epsilon} \right\}+ \nonumber\\
&& +  \delta_{\beta}{}^{\tau} \delta_{\epsilon}{}^{\lambda} \bg_{\alpha \gamma} + \delta_{\alpha}{}^{\tau} \delta_{\epsilon}{}^{\lambda} \bg_{\beta \gamma} + \delta_{\beta}{}^{\tau} \delta_{\gamma}{}^{\lambda} \bg_{\alpha \epsilon}+ \delta_{\alpha}{}^{\tau} \delta_{\gamma}{}^{\lambda} \bg_{\beta \epsilon} +  \nonumber\\&&+  \dfrac{2}{n^2-3n+2} \left\{\delta_{\beta}{}^{\lambda} \delta_{\epsilon}{}^{\tau} \bg_{\alpha \gamma} +  \delta_{\alpha}{}^{\lambda} \delta_{\epsilon}{}^{\tau} \bg_{\beta \gamma}+   \delta_{\beta}{}^{\lambda} \delta_{\gamma}{}^{\tau} \bg_{\alpha \epsilon} + \delta_{\alpha}{}^{\lambda} \delta_{\gamma}{}^{\tau} \bg_{\beta \epsilon}\right\} - \nonumber\\&&-\bg_{\alpha \epsilon} \bg_{\beta \gamma} \bg^{\lambda \tau} - \bg_{\alpha \gamma} \bg_{\beta \epsilon} \bg^{\lambda \tau}+  \dfrac{2}{n-2} \bg_{\alpha \beta} \bg_{\gamma \epsilon} \bg^{\lambda \tau},
\eea
in such a way that
\bea
D_{\a\b\g\e}&&=\frac{1}{16}(2\bg_{\a\b}\bg_{\g\e}-\bg_{\a\g}\bg_{\b\e}-\bg_{\a\e}\bg_{\b\g})\bar{\Box}-\frac{1}{8}(\bg_{\a\b}\bn_\g \bn_\e+\bg_{\g\e}\bn_\a \bn_\b)+\nonumber\\
&&+\frac{1}{16}(\bg_{\b\e}\bn_\g \bn_\a+\bg_{\a\e}\bn_\g \bn_\b+\bg_{\a\g}\bn_\e \bn_\b+\bg_{\b\g}\bn_\e \bn_\a).
\eea
Before continuing, one may ask why it is the case that we can invert the operator $K$ before gauge fixing, that is, not having to take care of any zero modes.
The quadratic action that we are left with is still invariant under the quantum gauge symmetry corresponding to diffeomorphism invariance given by the transformations
\bea
\d h_{\m\n}&&=\bn_{\m}\xi_{\n}+\bn_{\n}\xi_{\m}+\mathcal{L}_\xi h_{\m\n}\nonumber\\
\d \phi &&= \xi^\mu \bn_\mu \phi 
%\d A^\l_{\a\b}&=\bn_\a\bn_\b\xi^\l+\bR^\l_{~\b\m\a} \xi^\m+\mathcal{O}(A)
\eea
Let us  consequently define the de Donder or harmonic gauge fixing 
\be
S_{\text{\tiny{gf}}}=\frac{1}{4}\,\,\int\,d^nx\,\,\sqrt{\bg}\,\bg_{\m\n}\chi^\m\chi^\n,
\ee
where 
\be
\chi_\n=\bn^\m h_{\m\n}-\frac{1}{2}\bn_\n h-\phi\pd_\n\bp.
\ee	
The quadratic action after adding  the gauge fixing piece reads
\bea
&&S_{\text{\tiny{2+gf}}}=\frac{1}{4}\,\int\,d^nx\,\sqrt{\bg}\,\psi^A\Delta_{AB}\psi^B,
\eea 
where we have written the quadratic operator corresponding to the generalized field, $\psi^A$, defined as a vector
\be
\psi^A\equiv\begin{pmatrix}
	h^{\m\n}\\\phi
\end{pmatrix},
\ee
and the operator takes the form
\be
\Delta_{AB}=-g_{AB}\bar{\Box}+Y_{AB}.
\label{op}
\ee
Let us specify its  different pieces. The term multiplying the box operator acts as a sort of internal metric $g_{AB}$ and reads
\be
g_{AB}=\begin{pmatrix}
	C_{\a\b\m\n}&0\\0&1 \end{pmatrix},
\ee
with $C_{\m\n\r\s}=\frac{1}{4}(\bg_{\m\r}\bg_{\n\s}+\bg_{\m\s}\bg_{\n\r}-\bg_{\m\n}\bg_{\r\s})$. The components of $Y_{AB}$ are also detailed below for completeness
\bea
&&Y^{hh}_{AB}=\frac{1}{4}(\bg_{\m\r}\bg_{\n\s}+\bg_{\m\s}\bg_{\n\r}-\bg_{\m\n}\bg_{\r\s})\left(\bR-\frac{1}{2}\bg^{\a\b}\pd_\a\bph\pd_\b\bph\right)-\nonumber\\
&&-\frac{1}{2}\left(\bR_{\m\r\n\s}+\bR_{\n\r\m\s}\right)+\frac{1}{2}\left(\bg_{\m\n}\bR_{\r\s}+\bg_{\r\s}\bR_{\m\n}\right)\nonumber\\
&&-\frac{1}{4}\left(\bg_{\m\r}\bR_{\n\s}+\bg_{\m\s}\bR_{\n\r}+\bg_{\n\r}\bR_{\m\s}+\bg_{\n\s}\bR_{\m\r}\right)\nonumber\\
&&-\frac{1}{4}\left(\bg_{\m\n}\pd_\r\bph\pd_\s\bph+\bg_{\r\s}\pd_\m\bph\pd_\n\bph-\bg_{\m\r}\pd_\n\bph\pd_\s\bph-\right.\nonumber\\
&&\left.-\bg_{\m\s}\pd_\n\bph\pd_\r\bph-\bg_{\n\r}\pd_\m\bph\pd_\s\bph-\bg_{\n\s}\pd_\m\bph\pd_\r\bph\right),\nonumber\\
&&Y^{h\phi}_{AB}=Y^{\phi h}_{AB}=\frac{1}{2}\left[\partial_\a\partial_\b\bph+\partial_\b\partial_\a\bph-\bg_{\a\b}\bar{\Box}\bph\right],\nonumber\\
&&Y^{\phi\phi}_{AB}=\bg^{\r\s}\partial_\r\bph\partial_\s\bph.\eea
%For completeness, we have included a few  details of the computation of $\Delta_{AB}$ in the Appendix (\ref{B}). 
%With this, we can complete the path integral (\ref{PI})
%\be
%W[\bg_{\m\n},\bG^\l_{\r\s}, \bph]\equiv -S_0[\bg_{\m\n},\bG^\l_{\r\s}]-{1\over 2}\text{log~det}~\Delta[\bg_{\m\n},\bG^\l_{\r\s}, \bph],
%\ee
%so that the counterterm reads
%\be \Delta S[\bg_{\m\n},\bG^\l_{\r\s}]=-{1\over 2}\text{log~det}~\Delta[\bg_{\m\n},\bG^\l_{\r\s}]\ee
%As we can see, it all reduces to the computation of the determinant of the operator $\Delta$.
%We can use the heat kernel method to extract the divergent piece of the counterterm. 
The form of the usual four-dimensional one-loop counterterm is tabulated in  \cite{Barvinsky} for minimal operators of the same type as \eqref{op}, namely,
\bea\label{coe}
\Delta S&&=\frac{1}{(4\pi)^2}\frac{1}{\e}\int d^n x~\sqrt{|\bg|}\text{tr}\Big\{\frac{1}{360}\left(2\bR_{\m\n\r\s}\bR^{\m\n\r\s}-2\bR_{\m\n}\bR^{\m\n}+5\bR^2\right) \mathbb{I}+\nonumber\\
&&+\frac{1}{2}Y^2-\frac{1}{6}\bR Y+\frac{1}{12}W_{\m\n}W^{\m\n}\Big\},\eea
where the field strength is defined through
\be
[\bn_\m,\bn_\n]h^{\a\b}=W_{~~~\r\s\m\n}^{\a\b}h^{\r\s},
\ee
with
\be W_{~~~\r\s\m\n}^{\a\b}=\frac{1}{2}\left(\d^\b_\r\bR^\a_{~\s\m\n}+\d^\b_\s\bR^\a_{~\r\m\n}+\d^\a_\r\bR^\b_{~\s\m\n}+\d^\a_\s\bR^\b_{~\r\m\n}\right),\ee
which is symmetric under ${\a\b \leftrightarrow \r\s}$. Let us note that the trace in \eqref{coe} also encodes the trace of the matrix $\Delta_{AB}$. With this, only a few traces need to be computed in order to find the explicit value of the counterterm. These are given by
\bea
&&\text{tr}~ \mathbb{I} =\frac{n(n+1)}{2}+1\nonumber\\
&&\text{tr}~Y=g^{AB}Y_{AB}
=\frac{n(n-1)}{2}\bR+\frac{8+3n-n^2}{4}\bg^{\r\s}\partial_\r\bph\partial_\s\bph\nonumber\\
&&\text{tr}~Y^2=Y_{AB}\,g^{BC}\,Y_{CD}\,g^{DA}=3\bR_{\m\n\r\s}\bR^{\m\n\r\s}+\frac{n^2-8n+4}{n-2}\bR_{\m\n}\bR^{\m\n}+\nonumber\\
&&+\frac{n^3-5n^2+8n+4}{2(n-2)}\bR^2-\left[\frac{2n(n-4)}{ (n-2)}+4\right]\bR^{\m\n}\partial_\m\bph\partial_\n\bp+\nonumber\\
&&+\frac{n^3-7n^2+10n+8}{2(2-n)}\bR\bg^{\r\s}\partial_\r\bph\partial_\s\bph +2\bar{\Box}\bph\bar{\Box}\bph\nonumber\\
&&+\frac{n^3-n^2+14n-40}{8(n-2)}\left(\bg^{\r\s}\pd_\r\bp\pd_\s\bp\right)^2\nonumber\\
&&\text{tr}~W_{\m\n}W^{\m\n}=-(n+2)\bR_{\m\n\r\s}\bR^{\m\n\r\s}
\eea
%Please note that $\bn^\n\bn^\m\bph\bn_\m\bn_\n\bph+\bn^\m\bn^\n\bph\bn_\m\bn_\n\bph+2\bR^{\m\n}\bn_\m\bph\bn_\n\bph=2\bar{\Box}\bph\bar{\Box}\bph$.
Using expression (\ref{coe}), the full gauge, gravitational and scalar field contributions to the one-loop counterterm are given by
\bea\label{a2}
\Delta S_{\text{\tiny{2+gf}}}&&=\frac{1}{(4\pi)^2}\frac{1}{\e}\int d^n x~\sqrt{|\bg|}\frac{1}{360}\Bigg\{\left(482-29n+n^2\right)\bR_{\m\n\r\s}\bR^{\m\n\r\s}+\nonumber\\
&&+\frac{724-1440n+181n^2-n^3}{n-2}\bR_{\m\n}\bR^{\m\n}+\nonumber\\
&&+
\frac{5(140+264n-145n^2+25n^3)}{2(n-2)}\bR^2-\nonumber\\
&&-\frac{360(-4-2n+n^2)}{ n-2}\bR^{\m\n}\partial_\m\bph\partial_\n\bph-\nonumber\\
&&-\frac{15(32+62n-37n^2+5n^3)}{n-2}\bR\bg^{\r\s}\partial_\r\bph\partial_\s\bph+\nonumber\\
&&+\frac{45(n^3-n^2+14n-40)}{2(n-2)}\left(\bg^{\r\s}\partial_\r\bph\partial_\s\bph\right)^2+360\bar{\Box}\bph\bar{\Box}\bph\Bigg\}.
\eea
\par
Finally, the contribution coming from the ghost loops is also needed. The quadratic ghost Lagrangian  reads
\be
S_{\text{\tiny{gh}}}=\frac{1}{4}\,\,\int\,d^nx\,\,\sqrt{\bg}\,V^*_\m\left(-\bg^{\m\n}\bar{\Box}-\bR^{\m\n}+\bn^\m\bph\bn^\n\bph\right)V_\n,
\ee
where operators cubic in the fluctuations are not taken into account because they are irrelevant at one-loop (the ghosts being quantum fields do not appear as external states). The corresponding Laplacian operator is simply given by
\be \Delta_{\m\n}=-\bg_{\m\n}\bar{\Box}+Y_{\m\n},\ee 
with
\be Y_{\m\n}=-\bR_{\m\n}+\bn_\m\bph\bn_\n\bph.\ee
We can compute the traces in the same way as before so that
%\bea
%\text{tr}\, \mathbb{I}&=n,\nonumber\\
%\text{tr}\,Y&=-\bR+\bg^{\r\s}\partial_\r\bph\partial_\s\bph,\nonumber\\
%\text{tr}\,Y^2&=\bR_{\m\n}\bR^{\m\n}-2\bR^{\m\n}\partial_\m\bph\partial_\n\bph+
%\left(\bg^{\r\s}\partial_\r\bph\partial_\s\bph\right)^2,\nonumber\\
%\text{tr}\, W_{\m\n}W^{\m\n}&=-\bR_{\m\n\r\s}\bR^{\m\n\r\s},
%\eea
%where the field strength in this case is defined as
%\be [\bn_\m,\bn_\n]V^\r=W^\r_{~\l\m\n}V^\l.\ee
the complete ghost contribution can be read from (\ref{coe}) yielding
\bea\label{a2g}
\Delta S_{\text{\tiny{gh}}}&&=\frac{1}{(4\pi)^2}\frac{1}{\e}\int d^n x~\sqrt{|\bg|}\frac{1}{360}\left[(2n-30)\bR_{\m\n\r\s}\bR^{\m\n\r\s}+\right.\nonumber\\
&&\left.+(180-2n)\bR_{\m\n}\bR^{\m\n}+(5n+60)\bR^2-360 \bR^{\m\n}\partial_\m\bph\partial_\n\bph-\right.\nonumber\\
&&\left.-60\bR \bg^{\r\s}\partial_\r\bph\partial_\s\bph+180\left(\bg^{\r\s}\partial_\r\bph\partial_\s\bph\right)^2\right].\nonumber\\
\eea
Adding the two pieces (\ref{a2}) and (\ref{a2g}) (note the factor and the sign of the ghost contribution), and specifying the result to $n=4$, the full one-loop counterterm reads 
\bea\label{countfull}
\Delta S_{\text{\tiny{EH$\phi$}}}&&=\frac{1}{(4\pi)^2}\frac{1}{\e}\int d^n x~\sqrt{|\bg|}\left(\text{tr}\, a_2 \left(x,x\right)-2\text{tr}\, a^{\text{gh}}_2 \left(x,x\right)\right)=\nonumber\\&&=\frac{1}{(4\pi)^2}\frac{1}{\e}\int d^n x~\sqrt{|\bg|}\Big\{\frac{43}{60}\bR_{\m\n}\bR^{\m\n}+\frac{1}{40}\bR^2+\frac{1}{6}\bR \bg^{\r\s}\pd_\r\bph\pd_\s\bph+\nonumber\\
&&+\left(\bg^{\r\s}\pd_\r\bp\pd_\s\bp\right)^2+\bar{\Box}\bph\bar{\Box}\bph\Big\},
\eea
where the well-known four-dimensional Gauss-Bonnet identity has been used, namely,
\be 
\label{GB}\bR_{\m\n\r\s}\bR^{\m\n\r\s}-4\bR_{\m\n}\bR^{\m\n}+\bR^2=\text{total derivative}.
\ee
Substituting the background equations of motion, the {\em on-shell} effective action is finally obtained
\bea
\label{count}\Delta S_{\text{\tiny{EH$\phi$}}}&&=\frac{1}{(4\pi)^2}\frac{1}{\e}\int d^n x~\sqrt{|\bg|}\frac{203}{40}\bR^2,
\eea
which exactly matches 't Hooft and Veltman's result \cite{tHooft}. Obviously when setting $\bph$ equal to zero, we recover the off-shell counterterm for pure gravity given by
\bea
\Delta S_{\text{\tiny{EH}}}&&=\frac{1}{(4\pi)^2}\frac{1}{\e}\int d^n x~\sqrt{|g|}\left(\frac{7}{10}\bR_{\m\n}\bR^{\m\n}+\frac{1}{60}\bR^2\right).\eea
%which again agrees with \cite{tHooft}. 
.
%%%%%%%%%%%%%%%%%%%%%%%%%%%%%%%%%%%%%%%%%%%%%%%%%%%%%%%%%%%%%%%%%%%%%%%%%%%%%%%%%%%%%%%%%%%%%%%%%%%%%%%%%%%%%%%%%%%%%%%%%%%%%
               
\end{document}